\documentclass[11pt]{article}
\usepackage[utf8]{inputenc}
\usepackage[T1]{fontenc}
\usepackage[toc,page]{appendix}
\usepackage{enumerate}
\usepackage{fixltx2e}
\usepackage{graphicx}
\usepackage{longtable}
\usepackage{float}
\usepackage{setspace}
\usepackage{wrapfig}
\usepackage{soul}
\usepackage{amsmath}
\usepackage{mathtools}
\usepackage{mathrsfs}
\usepackage{textcomp}
\usepackage{marvosym}
\usepackage{wasysym}
\usepackage{latexsym}
\usepackage{amssymb}
\usepackage{hyperref}
\usepackage{graphicx}
\usepackage{subfig}
\usepackage{longtable}
\usepackage{float}
\usepackage{array}
\usepackage[ruled]{algorithm}
\usepackage[noend]{algpseudocode}
\usepackage[table]{xcolor}
\usepackage{arydshln}
\usepackage{cancel}
\usepackage{url}
\usepackage[round, colon]{natbib}
\usepackage[margin=1in]{geometry}
\usepackage[affil-it]{authblk}

\usepackage{color}

\input{./Definitions}

\renewcommand{\mb}{\mathbf{m}}

\newcommand{\inid}{\emph{inid}}
\newcommand{\iid}{\emph{iid}}

\newcommand{\f}{\boldsymbol{f}}
\newcommand{\bl}{\boldsymbol{l}}
\newcommand{\bu}{\boldsymbol{u}}
\newcommand{\bp}{\boldsymbol{p}}
\newcommand{\bby}{\boldsymbol{y}}

\allowdisplaybreaks

\title{Scalable Bayes via Barycenter in Wasserstein Space}

\author[1]{Sanvesh Srivastava \thanks{\url{sanvesh-srivastava@uiowa.edu}}}
\author[2]{Cheng Li \thanks{\url{stalic@nus.edu.sg}}}
\author[3]{David B. Dunson \thanks{\url{dunson@duke.edu}}}

\affil[1]{Department of Statistics and Actuarial Science, The University of Iowa, Iowa City, Iowa, USA}
\affil[2]{Department of Statistics and Applied Probability, National University of Singapore, Singapore}
\affil[3]{Department of Statistical Science, Duke University, Durham, North Carolina, USA}

\date{\today}

\begin{document}

\maketitle

\begin{abstract}
  Divide-and-conquer based methods for Bayesian inference provide a general approach for tractable posterior inference when the sample size is large. These methods divide the data into smaller subsets,  sample from the posterior distribution of parameters in parallel on all the subsets, and combine posterior samples from all the subsets to approximate the full data posterior distribution.  The smaller size of any subset compared to the full data implies that posterior sampling on any subset is computationally more efficient than sampling from the true posterior distribution. Since the combination step takes negligible time relative to sampling, posterior computations can be scaled to massive data by dividing the full data into sufficiently large number of data subsets. One such approach relies on the geometry of posterior distributions estimated across different subsets and combines them through their barycenter in a Wasserstein space of probability measures. We provide theoretical guarantees on the accuracy of approximation that are valid in many applications. We show that the geometric method approximates the full data posterior distribution better than its competitors across diverse simulations and reproduces known results when applied to a movie ratings database.
\end{abstract}


\section{Introduction}
\label{sec:intro}

Developing efficient sampling algorithms is an active area of research motivated by tractable Bayesian inference in large sample settings. Sampling remains a primary tool for inference in Bayesian models, with Markov chain Monte Carlo (MCMC) and sequential Monte Carlo (SMC) providing two broad classes of algorithms that are routinely used.  Most MCMC and SMC algorithms face problems in scaling up to massive data settings due to memory and computational bottlenecks that arise; this has motivated a rich literature in recent years proposing a variety of strategies to enable better performance in such settings. Our focus is on proposing a very general divide-and-conquer technique, which is designed to combine results from any posterior sampling algorithm applied in parallel using subsets of the data.

Massive data pose major problems for existing sampling algorithms. First, if full data require multiple machines for storage, then {a sampler has access to only a small fraction of the full data stored on the machine where it runs. Posterior sampling given the full data is expensive due to network latency and extensive communication among machines.} Second, with sample size $n$, sampling in hierarchical Bayesian models requires generation of $O(n)$ latent variables, which becomes inefficient as $n$ increases. {Finally, even if full data are available to the sampler, sampling can be infeasible in practice because computation of Hessians and acceptance ratios can scale as $O(n^3)$ in some nonparametric models based on Gaussian process priors \citep{RasWil06}.} A variety of methods exist to address these issues using optimization and sampling.

Optimization-based methods for Bayesian inference obtain an analytic approximation of the full data posterior distribution. The two most common techniques are polynomial approximation \citep{Rueetal09} and projection of the full data posterior distribution on a class of distributions with analytically tractable posterior densities, which includes variational Bayes and expectation propagation \citep{WaiJor08, Geletal14}. Both techniques estimate parameters of the approximate distribution using a variety of optimization algorithms \citep{TanNot13,Kucetal15,RezMoh15,Ragetal16}. Stochastic approximation significantly improves the efficiency of estimation by accessing the data in small batches and updating the parameter estimates sequentially \citep{Broetal13, Hofetal13}; however, optimization can be nontrivial for complex likelihoods frequently used in hierarchical models. Furthermore, variational Bayes and expectation propagation often have excellent predictive performance but can be highly biased in estimation of posterior uncertainty and dependence \citep{Giaetal17}.

There is extensive work in sampling-based methods for Bayesian inference. The three main techniques used  are as follows. First, subsampling-based methods obtain posterior samples conditioned on a small fraction of the data \citep{MacAda15}. Coupling of subsampling with modified Hamiltonian or Langevin dynamics improves posterior exploration and convergence to the stationary distribution \citep{WelTeh11,AhnKorWel12,Cheetal14,KorCheWel13,Lanetal14,Shaetal14}; see \citet{Baretal15} for a review. Second, the exact transition kernel in posterior sampling is replaced by an approximation that significantly reduces the time required to finish an iteration of the sampler \citep{Johetal15,Alqetal16}. Finally, divide-and-conquer approaches first divide the data into smaller subsets and sample in parallel across subsets, and then combine the posterior samples from all the subsets. Our focus is on scalable Bayesian methods based on the divide-and-conquer technique. These methods have two subgroups that differ mainly in their sampling scheme for every subset and their method for combining posterior samples obtained from all the subsets.

The first subgroup modifies the prior to sample from the posterior distribution of the parameter conditioned on a data subset. Let $k$ be the number of subsets, $\pi(\theta)$ be the prior density of parameter $\theta$, and $l_i(\theta)$ be the likelihood for subset $i$ ($i=1, \ldots, k$). Samples from subset posterior distribution $i$ are obtained using $l_i(\theta)$ and $\pi(\theta)^{1/k}$ as the likelihood and prior. {Consensus Monte Carlo combines subset posterior samples by averaging, which has been generalized in many ways \citep{Rabetal15,Scoetal16}}. This relies heavily on the normality assumption, which is relaxed using a combination based on kernel density estimation \citep{NeiWanXin13}. Both methods perform poorly if the supports of subset posteriors are different, which motivates the combination using the Weierstrass transform and random partition trees \citep{WanDun13, Wanetal15}. These methods offer simple approaches for combining samples from subset posterior distributions {but have a major limitation that the sampling algorithm depends on the model parameterization.}

The second subgroup modifies the subset likelihood to sample from a subset posterior distribution and combines samples from subset posterior distributions through their geometric center. These methods modify the likelihood to $l_i(\theta)^{k}$ and use prior $\pi(\theta)$ to sample from subset posterior distribution $i$ ($i=1, \ldots, k$). M-Posterior combines subset posterior distributions through their median in the Wasserstein space of order 1 \citep{Minetal14, Minetal14b}. The robustness of the median implies that it could ignore valuable information in some subset posterior distributions, which motivates combination through the mean in the Wasserstein space of order 2 called Wasserstein Posterior (WASP) \citep{Srietal15}. The WASP approach strikes a balance between the generality of sampling and the efficiency of optimization. While WASP can be applied to any data or Bayesian model, its computations are developed for independent identically distributed (\iid) data and its theoretical properties are unknown. 

Our main goal is to study {the} theoretical properties of WASP and apply WASP in a variety of practical problems. The \iid\ assumption of WASP rules out many important practical problems, including regression and classification, where the data are independent and non-identically distributed (\inid). {We relax this assumption and our theoretical results are applicable to \inid\ data}. Second, we show that if the number of subsets are chosen appropriately, then the WASP achieves almost the same rate of convergence as that of the full data posterior distribution. {For linear models with error distribution in the location-scale family, we strengthen this result and show that the WASP and the full data posterior distribution have the same asymptotic mean and asymptotic variance.} This implies that WASP can be used as an efficient alternative to the full data posterior distribution in massive data settings. Third, we show that the method for estimating WASP is independent of the form of the model, which implies that WASP is very general and can be easily used for estimating posterior summaries for any function of the model parameters. We emphasize that WASP is not a new sampling algorithm but a general approach to easily extend any existing sampling algorithms for massive data applications.

\section{Preliminaries}

\subsection{Wasserstein space, Wasserstein distance, and Wasserstein barycenter}

We recall elementary properties and definitions related to the Wasserstein space of probability measures.  Let $(\Theta, \rho)$ be a complete separable metric space and $\Pcal(\Theta)$ be the space of all probability measures on $\Theta$. The Wasserstein space of order $2$ is defined as
\begin{align}
  \Pcal_2(\Theta) := \bigg\{\mu \in \Pcal(\Theta): \int_{\Theta} \rho^2(\theta_0, \theta) \mu(d \theta) < \infty \bigg\}, \label{eqn:wass}
\end{align}
where $\theta_0 \in \Theta$ is arbitrary and $\Pcal_2(\Theta) $ does not depend on the choice of $\theta_0$. The space $\Pcal_2(\Theta)$ is equipped with a natural distance between its elements. Let $\mu, \nu \in \Pcal_2(\Theta)$ and $\Pi (\mu, \nu)$ be the set of all probability measures on $\Theta \times \Theta$ with marginals $\mu$ and $\nu$, then the Wasserstein distance of order $2$ between $\mu$ and $\nu$ is defined as
\begin{align}
  W_2(\mu, \nu) &= \bigg( \underset{\pi \in \Pi (\mu, \nu)} {\mathrm{inf}} \int_{\Theta \times \Theta} \rho^2(x, y) \, d \pi(x, y) \bigg)^{\frac{1}{2}}. \label{eqn:wass2}
\end{align}
In our applications $\rho$ is the Euclidean metric and we refer to $\Pcal_2(\Theta)$ and $W_2$ as the Wasserstein space and the Wasserstein distance without explicitly mentioning their order. If $\Pi_1, \ldots, \Pi_k$ are a collection of probability measures in $\Pcal_2(\Theta)$, then their barycenter in $\Pcal_2(\Theta)$ is defined as
\begin{align}
  \overline \Pi = \underset{ \Pi \in \Pcal_2(\Theta)} {\argmin} \sum_{j=1}^{k} \frac{1}{k}  W_2^2(\Pi, \Pi_j). \label{eqn:wasser-bary}
\end{align}
This generalizes the Euclidean barycenter, which is the sample mean, to $\Pcal_2(\Theta)$ \citep{AguCar11}. The barycenter $\overline \Pi$ is analytically intractable, except in few special cases. Let $\delta_a(x) = 1$ if $a = x$ and 0 otherwise. If $X_{j1}, \ldots, X_{jm}$ are samples from $\Pi_j$ ($j=1, \ldots, k$), then $\widehat \Pi_j(\cdot) = \sum_{i=1}^m \delta_{X_{ji}}(\cdot) / m$ is an empirical measure that approximates $\Pi_j$ ($j=1, \ldots, k$). If $\overline \Pi$ is assumed to be an empirical measure, then the optimization problem in  \eqref{eqn:wasser-bary} reduces to a linear program; see  \citet{CutDou14}, \citet{Caretal15}, and \citet{Srietal15} for different algorithms to solve this linear program.

\subsection{Stochastic approximation and subset posterior density}
\label{sec:sub-post}

Consider a general set-up for \inid\ data. Let $Y^{(n)} = (Y_1, \ldots, Y_n)$ be $n$ observations and the distribution of $Y_i$ is $P_{\theta, i}$, $i=1, \ldots, n$, where $\theta$ lies in the parameter space $\Theta \subset \mathbb{R}^p$. Assume that $P_{\theta, i}$ has density $p_{i}(\cdot |\theta)$ with respect to the Lebesgue measure, so $dP_{\theta, i}(y_{i}) = p_{i}(y_{i}|\theta) d y_{i}$ and the likelihood given $Y^{(n)}$ is $l(\theta) = \prod_{i=1}^n p_{i}(y_{i}|\theta)$. Given a prior distribution $\Pi$ on $\Theta$ that has density $\pi$ with respect to the Lebesgue measure, the posterior density of $\theta$ given $Y^{(n)}$ using Bayes theorem is
\begin{align}
  \pi(\theta \mid Y^{(n)}) = \frac{\prod_{i = 1}^n p_{i}(y_i \mid \theta) \pi(\theta) }{\int_{\Theta} \prod_{i = 1}^n p_{i}(y_i \mid \theta) \pi(\theta) d \theta } = \frac{l(\theta) \pi(\theta) }{\int_{\Theta} l(\theta) \pi(\theta) d \theta }. \label{eqn:post}
\end{align}
In most cases $\pi(\theta \mid Y^{(n)})$ is analytically intractable, and accurate approximations of $\pi(\theta \mid Y^{(n)})$ are obtained using Monte Carlo methods, such as importance sampling and MCMC, and deterministic approximations, such as Laplace's method and variational Bayes. For example, in the context of logistic regression,  $P_{\theta, i}$ is the Bernoulli distribution with mean $1 / \left\{ 1 + \exp ( -x_i^T \theta ) \right\}$, where $x_i^T$ is the $i$th row of the design matrix $X  \in \RR^{n \times p} $ and  $\Theta=\RR^p$. The posterior density of $\theta$ is analytically intractable, and it is typical to rely on Gibbs samplers based on data augmentation {\citep{Bis06}}. These samplers introduce latent variables $\{ z_i, i=1,\ldots, n\}$ and alternately sample the latent variables and the parameters from their full conditional distributions.  Related algorithms are very common and are computationally prohibitive for large $n$ because they require repeated passes through the whole data.

Divide-and-conquer-type methods resolve this problem by partitioning the data into smaller subsets. Let $k$ be the number of subsets. The default strategy is to randomly allocate samples to subsets. Let $Y_{[j]} \equiv Y^{(m_j)}_j= (Y_{j1}, \ldots, Y_{jm_j})$ denote data on the $j$th subset, where $m_j$ is the size of the $j$th subset and $\sum_{j=1}^{k}m_j = n$. We assume that $m_j = m$ ($j=1, \ldots, k$) for ease of presentation, so $n = km$, the likelihood given $Y_{[j]}$ is $l_j(\theta) = \prod_{i=1}^{m} p_{ji}(y_{ji}|\theta)$, and $l(\theta)$ in \eqref{eqn:post} equals $\prod_{j=1}^k l_j(\theta)$. Define subset posterior density $j$ given $Y_{[j]}$ as
\begin{align}
  \pi_{m}(\theta \mid Y_{[j]}) = \frac{ \{\prod_{i=1}^{m} p_{ji}(y_{ji}|\theta)\}^{\gamma} \pi(\theta) }{\int_{\Theta} \{\prod_{i=1}^{m} p_{ji}(y_{ji}|\theta)\}^{\gamma} \pi(\theta) d \theta } = \frac{ l_j(\theta)^{\gamma} \pi(\theta)}{\int_{\Theta} l_j(\theta)^{\gamma} \pi(\theta) d \theta }, \label{eqn:wasp-post}
\end{align}
where $\gamma$ is a positive real number such that $g_1\gamma m \leq n \leq g_2\gamma m$ for some $g_1, g_2 > 0$. In the present context, we assume that $\gamma = k$ with $g_1=g_2=1$ following \citet{Minetal14}; more general conditions on $\gamma$ are defined later in Section \ref{wasp-theory}. This modified form of subset posterior compensates for the fact that $j$th subset has access to only $(m/n)$-fraction of the full data and ensures that $\pi_{m}(\theta \mid Y_{[j]})$ and $\pi_n(\theta \mid Y^{(n)})$ in \eqref{eqn:post} have variances of the same order. \citet{Minetal14} refer to this as \emph{stochastic approximation} because raising $l_j(\theta)$ ($j=1, \ldots, k$) to the power $\gamma$ is equivalent to replicating every $X_{ji}$ ($i=1, \ldots, m$) $\gamma$-times so that
$\pi_{m}(\theta \mid Y_{[j]})$ ($j=1, \ldots, k$) are noisy approximations of $\pi(\theta \mid Y^{(n)})$.

One advantage of using stochastic approximation to define $\pi_{m}(\theta \mid Y_{[j]})$ in \eqref{eqn:wasp-post} is that off-the-shelf sampling algorithms can be used directly even when the prior density is the form of a discrete mixture. Consider a simple example of univariate density estimation using Dirichlet process (DP) mixtures of Gaussians. Let $X_i$ ($i=1, \ldots, n$) be \iid\ samples from a distribution $P_0$ with density $p_0$. The data are randomly split into $k$ subsets of equal size $m$. The truncated stick-breaking representation of DP implies that the prior distribution $\Pi$ on $\Pcal$ has a finite mixture representation, where $\Pcal$ is the set of probability distributions that have a density. We show in the Appendix that modification of the likelihood using stochastic approximation {leads to nearly identical subset and full data posterior computations. }

Stochastic approximation does not add any extra burden to the computations required for sampling from the subset posterior distribution of $\theta$ conditioned on $m$ observations. We raise the likelihood in every subset to the power $\gamma$. This is equivalent to replicating observations $\gamma$-times, which seems to offset the benefits of partitioning. However, the replication of observation is not required in implementation of the sampler; we simply modify the likelihood in the full data sampler by raising it to the power $\gamma$. For example, stochastic approximation is easily implemented using the \texttt{increment\_log\_prob} function in Stan \citep{Sta14}. We provide more examples for a variety of models in Section \ref{experiments}.

\begin{figure}[!t]
  \centering
    \includegraphics[scale=0.3]{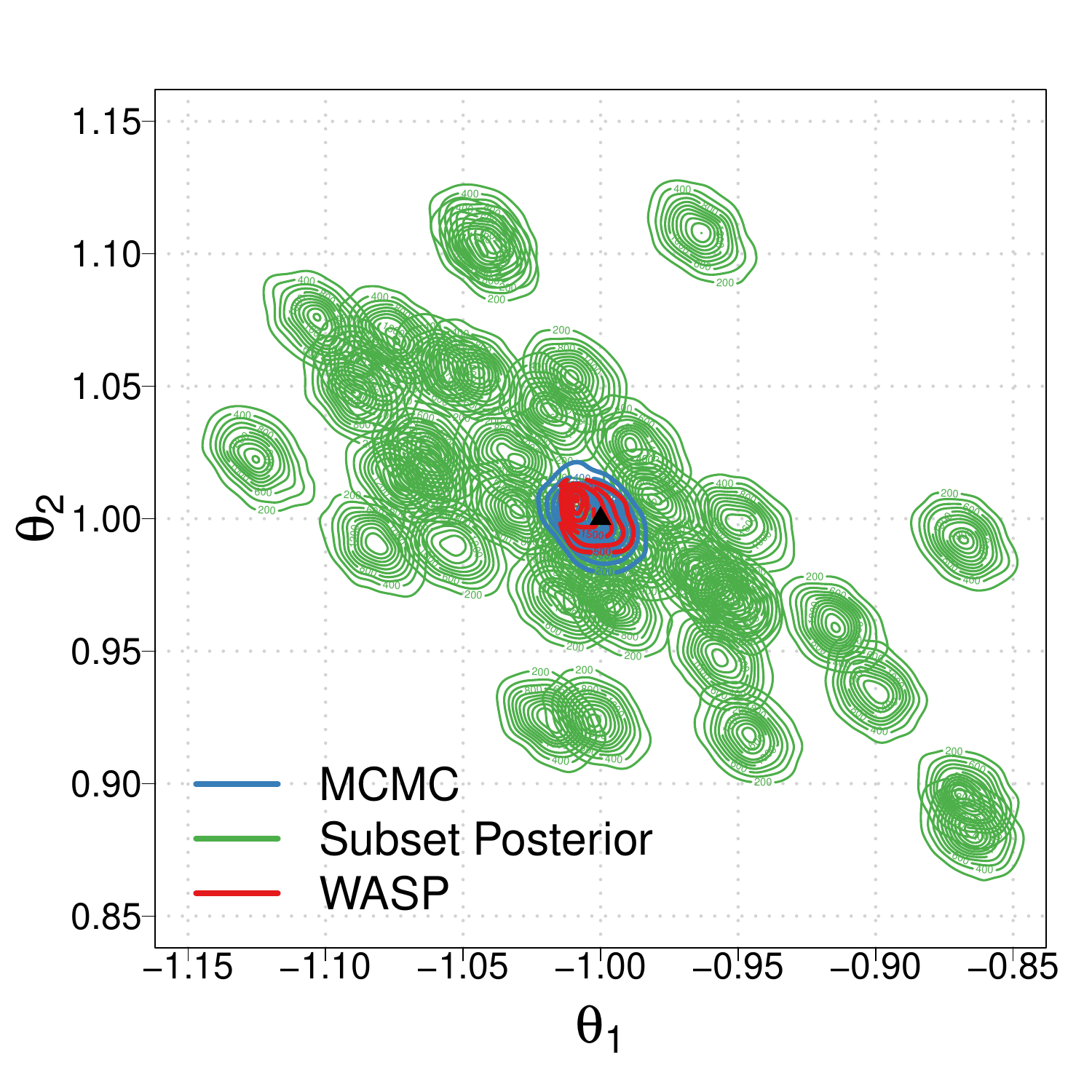}
  \caption{Binned kernel density estimates of full data posterior distribution, subset posterior distributions, and WASP for coefficients $(\theta_1, \theta_2)$ in logistic regression. The x and y axes represent posterior samples for $\theta_1$ and $\theta_2$. The true values of $\theta_1$ and $\theta_2$ are $-1$ and $1$ (black triangle). }
  \label{fig:demo1}
\end{figure}

A simple logistic regression example demonstrates that $\pi_{m}(\theta \mid Y_{[j]})$ in \eqref{eqn:wasp-post} is a noisy approximation of $\pi(\theta \mid Y^{(n)})$ in \eqref{eqn:post}. We simulated data for logistic regression with $n=10^5$, $p=2$, $\theta = (-1, 1)^T$, and entries of $X$ randomly set to $\pm 1$ (Figure \ref{fig:demo1}). We set $\gamma=k=40$ and obtained samples of $\theta$ from $\pi(\theta \mid Y^{(n)})$ and from $\pi_{m}(\theta \mid Y_{[j]})$ $(j=1, \ldots, k)$ using the Stan's HMC sampling algorithm. The contours for the  subset and full data posterior densities are very similar, indicating all densities have similar spreads.  We also notice that subset posteriors are noisy approximations of the full data posterior in that most of them have a bias and do not concentrate at the true $\theta$.

\section{Wasserstein Posterior (WASP): The general framework}
\label{sec:wasp:-wass-post}

\subsection{Definition and estimation of the WASP}

  \begin{enumerate}[]
  \begin{algorithm}[t]
    \caption{Estimation of the WASP for $f(\theta)$ given samples of $\theta$ from $k$ subset posteriors} \label{algo:wasp}
    {\scriptsize
  \item {\bf Input}: Samples from $k$ subset posteriors, $\{\theta_{ji} : \theta_{ji} \sim \Pi_m(\cdot \mid Y_{[j]})$, $i=1, \ldots, s_j$, $j = 1, \ldots, k\}$; mesh size $\epsilon > 0$.
  \item {\bf Do}:
    \begin{enumerate}[1.]
    \item Define $ \phi_{i}^j = (\phi_{i1}^j, \ldots, \phi_{iq}^j) = f(\theta_{ji})$ ($i=1, \ldots, s_j$; $j = 1, \ldots, k$), the matrix of atoms of subset posterior $j$, $\Phi_j \in \RR^{s_j \times q}$, with $\phi_{i}^j$ as row $i$ ($i=1, \ldots, s_j$). For $r=1,\ldots,q$, let $\phi_{\min} = (\phi_{\min \, 1}, \ldots, \phi_{\min \, q})$ with $\phi_{\min \, r} = \underset{j\, i}{\min} \, \phi^j_{i r}$, and $\phi_{\max} = (\phi_{\max \, 1}, \ldots, \phi_{\max \, q})$ with $\phi_{\max \, r} = \underset{j\, i}{\max} \, \phi^j_{i r}$.
    \item Set the number of atoms in the empirical approximation for the WASP $g = g_1 \times \ldots \times  g_q$, where  $g_r = \big \lceil {\tfrac{\phi_{\max \, r} - \phi_{\min \, r}}{\epsilon}} \big \rceil$ ($r = 1, \ldots, q$).
   \item Define the matrix of WASP atoms $\overline \Phi \in \RR^{g \times q}$ with rows formed by stacking vectors
      \begin{align*}
         \left\{ \phi_{\min \, 1} + \tfrac{i_1} {g_1} \left( \phi_{\max \, 1} - \phi_{\min \, 1}\right), \ldots, \phi_{\min \, q} + \tfrac{i_q} {g_q} \left( \phi_{\max \, q} - \phi_{\min \, q}\right) \right\}, \quad (i_r = 1, \ldots, g_r; \; r =1, \ldots, q).
      \end{align*}
    \item Set the distance matrix between the atoms of WASP and the $j$th subset posterior, $D_j \in \RR_{+}^{g \times s_j}$,  as
      \begin{align*}
        (D_j)_{uv} =  \sum_{r = 1}^q (\overline \phi_{ur} - \phi^j_{vr})^2, \quad (u = 1, \ldots, g;\; v = 1, \ldots, s_j; \;j = 1, \ldots, k),
      \end{align*}
      where $\overline \phi_{ur}$ is the $(u,r)$-entry of $\overline \Phi$.
    \item Estimate $\hat a_1, \ldots, \hat a_g$ by solving the linear program \eqref{lpwasp} in Appendix \ref{sec:linear-program}.
    \end{enumerate}
  \item {\bf Return}: $\hat{ \overline{f\sharp\Pi}}(\cdot \mid Y^{(n)}) = \sum_{i=1}^g \hat a_{i} \delta_{\overline \phi_i}(\cdot)$, the atomic approximation of $\overline{ f\sharp\Pi}_n (\cdot \mid Y^{(n)})$.}%
  \end{algorithm}
\end{enumerate}

The WASP approach combines subset posterior distributions $\Pi_{m}(\cdot \mid Y_{[j]}) $ ($j = 1, \ldots, k$) through their barycenter in $\Pcal_2(\Theta)$, where the density of $\Pi_{m}(\cdot \mid Y_{[j]}) $ is $\pi_{m}(\cdot \mid Y_{[j]})$ in \eqref{eqn:wasp-post}. The barycenter represents a  geometric center of a collection of probability distributions that can be efficiently computed using a linear program. Motivated by this, \citet{Srietal15} proposed to combine a collection of subset posterior distributions through their barycenter in the Wasserstein space called \emph{WASP}. Assuming that subset posterior distributions $\Pi_m( \cdot \mid Y_{[j]})$ ($j=1, \ldots, k$) have finite second moments, the WASP is defined  using \eqref{eqn:wasser-bary} as
\begin{align}
  \overline \Pi_{n}( \cdot \mid Y^{(n)}) = \underset{ \Pi \in \Pcal_2(\Theta)} {\argmin} \sum_{j=1}^{k} \frac{1}{k} W_2^2\{\Pi, \Pi_{m}( \cdot \mid Y_{[j]})\}. \label{eqn:was-bary}
\end{align}

{Consider the following Gaussian example where the WASP is analytically tractable. Assume that the subset posterior distributions, $\Pi_1, \ldots, \Pi_k$, are Gaussian with means $\mu_1, \ldots, \mu_k$ and covariance matrices $\Sigma_1, \ldots, \Sigma_k$. If we fix $\rho$ to be the Euclidean metric and $\Theta = \RR^d$ in \eqref{eqn:wass2}, then \eqref{eqn:wasser-bary} implies that $\overline \Pi_n$ is Gaussian with mean $\overline \mu$ and covariance matrix $\overline \Sigma$, where
\begin{align}
  \label{eq:mult-gaus}
  \overline \mu = \frac{1}{k} \sum_{j=1}^k \mu_{j} \text{  and  } \overline \Sigma \text{  is such that  } \frac{1}{k}\sum_{j=1}^k \left( \overline \Sigma^{1/2} \Sigma_j \overline \Sigma^{1/2} \right)^{1/2} =  \overline \Sigma,
\end{align}
where $A^{1/2}$ is the symmetric square root of $A$ \citep{AguCar11}. If $\theta$ is one dimensional, then \eqref{eq:mult-gaus} says that the standard deviation of WASP is the average of standard deviations of subset posteriors; therefore, the variance of WASP is typically about the same order as that of any subset posterior distribution. A similar relation also holds in higher dimensions and for a large class of posterior distributions, including elliptical distributions \citep{Alvetal16}. }

The WASP is analytically tractable only in special cases, but it can be estimated using a linear program if the subset posterior distributions have an atomic form. Let $\{\theta_{j1}, \ldots, \theta_{jS}\}$ be
the $\theta$ samples obtained from subset posterior density $j$ in \eqref{eqn:was-bary} using a sampling algorithm, including HMC, MCMC, SMC, or importance sampling. Approximate $j$th subset posterior distribution $\Pi_{m}( \cdot \mid Y_{[j]})$ using the empirical measure
\begin{align}
  \label{eq:emp-approx-sub}
  \hat \Pi_{m}(\cdot \mid Y_{[j]}) = \sum_{i=1}^S \frac{1}{S} \delta_{\theta_{ji}}(\cdot) \quad (j=1, \ldots, k).
\end{align}
\citet{Srietal15} approximate the WASP as
\begin{align}
  \label{eq:emp-approx-wasp}
  \hat{\overline{\Pi}}_{n}(\cdot \mid Y^{(n)}) = \sum_{j=1}^k \sum_{i=1}^S  a_{ji} \delta_{\theta_{ji}}(\cdot), \quad 0 \leq a_{ji} \leq 1, \quad \sum_{j=1}^k \sum_{i=1}^S a_{ji} =1,
\end{align}
where $a_{ji}$ $(j=1,\ldots, k; \; i=1, \ldots, S)$ are unknown weights of the atoms. There are many specialized algorithms to estimate the WASP that exploit the structure of the linear program in \eqref{eqn:was-bary} when $\Pi_{m}( \cdot \mid Y_{[j]})$ and $\overline \Pi_{n}( \cdot \mid Y^{(n)}) $ are restricted to have atomic forms in \eqref{eq:emp-approx-sub} and \eqref{eq:emp-approx-wasp}, respectively; for example, \citet{CutDou14} extend the Sinkhorn algorithm using entropy-smoothed sub-gradient methods, \citet{Caretal15} develop a non-smooth optimization algorithm, and \citet{Srietal15} propose an efficient linear program that exploits the sparsity of constraints to solve \eqref{eqn:was-bary}. {A simple and efficient algorithm to find the WASP of a given function of parameters is summarized in Algorithm \ref{algo:wasp}.} \\

\subsection{Theoretical properties of the WASP}
\label{wasp-theory}

The WASP, denoted as $\overline \Pi_n$, replaces the full data posterior distribution, denoted as $\Pi_n$, for inference and prediction in massive data applications where $n$ is large. In motivating applications, computation of $\Pi_n$ is inefficient, and dividing the data into smaller subsets and performing posterior computations in parallel leads to massive speed-ups. A formal asymptotic justification for using $\overline \Pi_n$ to approximate $\Pi_n$ would ideally show that the distance between $\overline \Pi_n$ and $\Pi_n$ tends to 0 as the full data size $n$ increases to infinity. We will illustrate this using a linear model example in Section \ref{wlm-bvm}, where we show that $n^{1/2} W_2(\overline \Pi_n, \Pi_n) \to 0$ as $n \to \infty$. Since both $\overline \Pi_n$ and $\Pi_n$ have variances of order $n^{-1}$, our result implies that the mean and the variance of WASP match those of the full data posterior distribution.

A general theoretical justification for using $\overline \Pi_n$ in the place of $\Pi_n$ for a multivariate $\theta$ given \inid\ data is technically much more challenging. If the data are \iid\ and $\theta$ is one-dimensional, then \citet{Lietal16} proves that $n^{1/2}W_2(\overline \Pi_n, \Pi_n) \to 0$ as $n \to \infty$ for regular parametric models. The proof in \citet{Lietal16} relies heavily on the Bernstein-von Mises theorem (BvM) for \iid\ data and the one-dimensional quantile representation of Wasserstein distance. Unlike the \iid\ case, a BvM-type theorem is generally unavailable if the data are \inid\ or the model is non-regular \citep{IbrHas81}. In Section \ref{convergence-rate}, we show that the WASP $\overline \Pi_n$ converges to the true parameter value at almost the same rate as $\Pi_n$ when the number of subset $k$ increases slowly with $n$. The previous theoretical justification of WASP in \citet{Srietal15} only includes posterior consistency under the stronger \emph{iid} assumption without characterizing the convergence rate. Relaxing these limitations, we provide the convergence rate for the WASP in the \emph{inid} case, including the convergence rate for WASP of general functionals of the original parameters.

\subsubsection{Approximation Error of WASP for inid data: Weighted linear model example} \label{wlm-bvm}

We use a weighted linear model example to illustrate the theoretical approximation accuracy of WASP to the true posterior under the \emph{inid} setup. For $i=1,\ldots,n$, let $y_i$ be a scalar response, $x_i$ be a $p \times 1$ vector of predictors, and $\epsilon_i$ be the idiosyncratic error in $y_i$. Let $\theta = (\theta_1, \ldots, \theta_p)^T$ be the $p \times 1$ regression coefficients vector. Let $y = (y_1, \ldots, y_n)^T$, $X = [x_1, \ldots, x_n]^T$, and $\epsilon = (\epsilon_1, \ldots, \epsilon_n)^T$ be the $n \times 1$ response vector, the $n \times p$ design matrix, and the $n \times 1$ error vector, respectively. If $\Sigma$ is a known diagonal matrix with positive elements and $\text{cov}(\epsilon) = \Sigma$, then the weighted linear regression model of $y$ on $X$ with a flat prior on $\theta$ assumes that
\begin{align}
  y = X \theta + \epsilon, \quad \epsilon \sim N_n(0, \Sigma), \quad \Sigma = \diag(\sigma_1^2, \ldots, \sigma_n^2), \quad \pi(\theta) \propto 1, \label{wlin-reg}
\end{align}
where $\pi(\theta)$ is the flat prior on $\theta$ and $N_n(0, \Sigma)$ is a $n$-variate Gaussian distribution with $n \times 1$ mean $0$ and covariance $\Sigma$. In this case, the data are \emph{inid} since the distribution of $y_i$ depends on the value of $x_i$. Since $\Sigma$ is assumed to be known, the posterior distribution of $\theta$ is normal with mean $\mu = (X^T \Sigma^{-1} X)^{-1} X^T \Sigma^{-1} y$ and covariance matrix $V = (X^T \Sigma^{-1} X)^{-1}$. Although the posterior of $\theta$ has a closed form in this example, the computational complexity of finding $\mu$ and $V$ is $O(n^2)$, which becomes inefficient as the size of the data $n$ increases.

The WASP of $\theta$ in \eqref{wlin-reg} is analytically tractable.  The computation of WASP has three steps.
First, the training data are randomly split into $k$ subsets. Let $y_j$, $X_j$, and $\Sigma_j$ be the response vector, design matrix, and error covariance matrix specific to subset $j$ ($j=1, \ldots, k$). Second, we compute the subset posterior distributions after stochastic approximation on the $k$ subsets in parallel as in \eqref{eqn:wasp-post} with $\gamma=k$. The $j$th subset posterior distribution of $\theta$ is $N_p(\mu_j, V_j)$, where $\mu_j = (X_j^T \Sigma_j^{-1} X_j)^{-1} X_j^T \Sigma^{-1}_j y_j$ and $V_j = k^{-1}(X_j^T \Sigma_j^{-1} X_j)^{-1}$. Third, \eqref{eq:mult-gaus} implies that the WASP of $\theta$ is also Gaussian with mean vector $\overline \mu$ and covariance matrix $\overline V$, where $\overline \mu = k^{-1} \sum_{j=1}^k \mu_j$ and $\overline V$ satisfies $\overline V = k^{-1}\sum_{j=1}^k (\overline V^{1/2} V_j \overline V^{1/2} )^{1/2}$.

The WASP and full data posterior distributions lead to the same posterior inference on $\theta$ up to $o(n^{-1})$ terms. Let $\overline \Pi_n = N_p(\overline \mu, \overline V)$ and $\Pi_n = N_p(\mu, V)$ be the WASP and full data posterior distributions for $\theta$. Based on the divide-and-conquer technique, the computational complexity of $\overline \Pi_n$ is $O(k m^2)$, which is smaller than that of $\Pi_n$ by a factor of $k$. The true distribution of $y$, denoted as $P_{\theta_0}^{(n)}$, in \eqref{wlin-reg} is $N_n(X \theta_0, \Sigma)$. If uncertainty quantification using $\overline \Pi_n$ and $\Pi_n$ is the same, then it suffices to show that the difference in the second moments of $\overline \Pi_n$ and $\Pi_n$ is $o(n^{-1})$ in ${P_{\theta_0}^{(n)}}$-probability because the variances $\overline V$ and $V$ are both of order $n^{-1}$.  This is equivalent to showing that the $W_2$ distance between $\overline \Pi_n$ and $\Pi_n$ is $o(n^{-1})$ in ${P_{\theta_0}^{(n)}}$-probability, which is proved in the next theorem. {In the statement of the theorem, we denote $A \prec B$ for positive definite matrices $A$ and $B$ if $B - A$  is also positive definite.
\begin{theorem}\label{lin-mdl-uq}
    Assume that there exist $a_n = o(1)$, $b_m = o(1)$ such that $\Omega_0 - a_n I_p \prec \tfrac{1}{n} X^T \Sigma^{-1} X \prec \Omega_0 + a_n I_p$ and $\Omega_0 - b_m I_p \prec \tfrac{1}{m} X_j^T \Sigma_j^{-1} X_j \prec \Omega_0 + b_m I_p$ for all $j=1, \ldots, k$, where $I_p$, $\Omega_0$ are $p\times p$ identity and constant positive definite matrices. Then, 
  \begin{align*}
    &E_{P_{\theta_0}^{(n)}} \|\overline \mu - \mu\|^2_2 = o\left(n^{-1}\right),\quad
      \tr \left( \overline V - V \right) = o\left(n^{-1}\right), \quad
      E_{P_{\theta_0}^{(n)}} W_2^2(\overline \Pi_n, \Pi_n) =  o\left(n^{-1}\right).
  \end{align*}
\end{theorem}
The proof of this theorem is in the appendix along with other proofs. }

Theorem \ref{lin-mdl-uq} shows that the uncertainty quantification of $\Pi_n$ and $\overline \Pi_n$ are the same in $P_{\theta_0}^{(n)}$-probability for the data following the model in \eqref{wlin-reg}. Essentially, the WASP and the true posterior have the same posterior mean and posterior variance, and their differences are only in high order of the full data size $n$. Furthermore, Theorem \ref{lin-mdl-uq} is valid for any block diagonal $\Sigma$ as long as the data that belong to a particular diagonal block of $\Sigma$ also belong to the same partition. In other words, Theorem \ref{lin-mdl-uq} even holds for dependent data in which the dependence can be expressed as a block diagonal $\Sigma$ in \eqref{wlin-reg}. Finally, Theorem \ref{lin-mdl-uq} is in fact true for any error distribution satisfying $E(\epsilon) = 0$ and $\text{cov}(\epsilon) = \Sigma$, which includes the Gaussian distribution; see Definition 2.1 and Theorem 2.3 in \citet{Alvetal16}.

\subsubsection{General convergence rates of WASP for inid data} \label{convergence-rate}

For general non-iid data, the standard Bayesian asymptotic theory for posterior convergence rates has been established in \citet{GhoVan07}, which also includes our \emph{inid} setup. We follow the theoretical framework of \citet{GhoVan07} and develop the corresponding theory for divide-and-conquer Bayesian inference using the WASP.

We start with two definitions required to state the assumptions of our theoretical setup.
\begin{definition}[Pseudo Hellinger distance] \label{def:prod-hell-wass}
The pseudo Hellinger distance between probability measures $P_{\theta_1}^{(m)}, P_{\theta_2}^{(m)} \in  \{\otimes_{i=1}^m P_{\theta, j,i}: \theta \in \Theta, \; d P_{\theta, j,i}(y) = p_{ji}(y \mid \theta) dy  \}$ is $h_{mj}^2(\theta_1, \theta_2) = \frac{1}{m} \sum_{i=1}^m h^2 \left\{ p_{ji}(\cdot \mid \theta_1), p_{ji}(\cdot \mid \theta_2) \right\}$,
where $h(p_1,p_2)= [\int \{\sqrt{p_1(y)}-\sqrt{p_2(y)}\}^2 dy ]^{1/2}$ is the Hellinger distance between two generic densities $p_1,p_2$.
\end{definition}
This definition generalizes the usual Hellinger distance to account for the \emph{inid} data generating mechanism. The space $\left(\{\otimes_{i=1}^m P_{\theta, j,i}: \theta \in \Theta \}, h_{mj}\right)$ is a metric space.

\begin{definition}[Generalized bracketing entropy] \label{def:brac-ent}
Let $\Xi$ be a fixed subset of $\Theta$. For an $m$-dimensional random vector $Z=(Z_1,\ldots,Z_m)^T$, denote its $L_q$ norm as $|Z|_q = \left[\frac{1}{m} \sum_{i=1}^m E\left(|Z_i|^q\right) \right]^{1/q}$ and use $\|Z\|$ to represent $|Z|_2$. For a fixed $j\in \{1,\ldots, k\}$, let $$\mathcal{P}_{j}(\Xi)=\left\{\bp_{j}(\bby|\theta)=(p_{j1}(y_{1}|\theta),\ldots,p_{jm}(y_{m}|\theta))^T:\bby=(y_{1},\ldots,y_{m})^T\in
\otimes_{i=1}^m \Ycal_{ji},\theta\in \Xi \right\}$$
be the class of $m$-dimensional functions indexed by $\theta$. For a given $\delta>0$, let
\begin{align*}
\mathcal{B}\left(\delta,\mathcal{P}_{j}(\Xi)\right)=&\Big\{[\bl_s,\bu_s]: \bl_s(\bby)=(l_{s1}(y_{1}),\ldots,l_{sm}(y_{m}))^T,\bu_s(\bby)=(u_{s1}(y_{1}),\ldots,u_{sm}(y_{m}))^T, \\
&\bby=(y_{1},\ldots,y_{m})^T\in
\otimes_{i=1}^m \Ycal_{ji}, s=1,\ldots,N \Big\}
\end{align*}
be the {\it generalized bracketing set} of $\mathcal{P}_{j}(\Xi)$ with cardinality $N$, such that for any $\bp_{j}(\bby|\theta) \in \mathcal{P}_{j}(\Xi)$, there exists a pair of functions $[\bl_s,\bu_s]\in \mathcal{B}\left(\delta,\mathcal{P}_{j}(\Xi)\right)$, such that
\begin{align*}
& l_{si}(y_{i}) \leq p_{ji}(y_{i}) \leq u_{si}(y_{i}), \text{ for all } \bby \in \otimes_{i=1}^m \Ycal_{ji}, \text{ and all } i=1,\ldots, m\\
\text{ and } & \|\sqrt{\bu_s}-\sqrt{\bl_s}\| \leq \delta.
\end{align*}
The $h_{mj}$-bracketing number of $\mathcal{P}_{j}(\Xi)$, $N_{[]}\left(\delta, \mathcal{P}_{j}(\Xi), h_{mj}\right)$, is defined as the smallest cardinality of the generalized bracketing set $\mathcal{B}\left(\delta,\mathcal{P}_{j}(\Xi)\right)$. The $h_{mj}$-bracketing entropy of $\mathcal{P}_{j}(\Xi)$ is defined as $H_{[]}\left(\delta, \mathcal{P}_{j}(\Xi), h_{mj}\right) = \log \left(1+N_{[]}\left(\delta, \mathcal{P}_{j}(\Xi), h_{mj}\right)\right)$.
\end{definition}
Again, this definition generalizes the usual bracketing entropy to the \emph{inid} cases. If the data are indeed \emph{iid}, then Definition \ref{def:brac-ent} coincides with that of the usual bracketing entropy.

Our theory for the convergence rate of WASP is built on the following assumptions.
\begin{enumerate}
\item[(A1)]  $\Theta$ is a compact space in $\rho$ metric,  $\theta_0$ is an interior point of $\Theta$, and $g_1\gamma m\leq n\leq g_2\gamma m$ for some constants $g_1,g_2>0$.
\item[(A2)] For any $\theta, \theta^{'} \in \Theta$ and $j = 1, \ldots, m$, there exist positive constants $\alpha$ and $C_L$ such that $h^2_{mj}(\theta,\theta^{'}) \geq C_L \rho^{2\alpha}(\theta, \theta^{'})$,
where $h^2_{mj}$ is the pseudo Hellinger distance in Definition \ref{def:prod-hell-wass}.
\item[(A3)] (Entropy Condition) There exist constants $D_1>0$, $0<D_2<D_1^2/2^{12}$, a function $\Psi(u,r)\geq 0$ that is nonincreasing in $u\in\mathbb{R}^+$ and nondecreasing in $r\in \mathbb{R}^+$, such that for all $j=1,\ldots,k$, for any $u,r>0$ and for all sufficiently large $m$,
\begin{align*}
& H_{[]}\left(u, \left\{\bp_j(\bby|\theta):\theta\in\Theta,h_{mj}(\theta,\theta_0)\leq r\right\}, h_{mj}\right) \leq \Psi(u,r) \text{ for all } j=1,\ldots,k;\\
\text{ and } & \int_{D_1r^2/2^{12}}^{D_1r} \sqrt{\Psi(u,r)} d u < D_2\sqrt{m}r^2,
\end{align*}
where $\bp_j(\bby|\theta) = \{p_{j1} (y_{j1} \mid \theta), \ldots, p_{jm} (y_{jm} \mid \theta)\}^T$ and $H_{[]}$ is the $h_{mj}$-bracketing entropy of the set $\left\{\bp_j(\bby|\theta):\theta\in\Theta,h_{mj}(\theta,\theta_0)\leq r\right\}$ in Definition \ref{def:brac-ent}.
\item [(A4)] (Prior Thickness) There exist positive constants $\kappa$ and $c_{\pi}$, such that uniformly over all $j=1,\ldots,k$,
\begin{align*}
  &\Pi\left(\theta\in \Theta: \frac{1}{m}\sum_{i=1}^m E_{P_{\theta_0}}\exp\left(\kappa \log_+\frac{p_{ji}(Y_{ji}|\theta_0)}{p_{ji}(Y_{ji}|\theta)} \right) -1 \leq \frac{\log^2 m}{m}\right) \geq \exp(-c_{\pi}k\log^2 m)
\end{align*}
where $\log_+ x = \max(\log x,0)$ for $x>0$.
\item [(A5)] The metric $\rho$ satisfies $\rho (\sum_{i=1}^N w_i \theta_i,\theta') \leq \sum_{i=1}^N w_i \rho(\theta_i,\theta')$ for any $N \in \{1, 2, \ldots\}$, $\theta_1,\ldots,\theta_{N},\theta' \in \Theta$ and non-negative weights $\sum_{i=1}^{N} w_i =1$.
\end{enumerate}

Our assumptions above are based on the standard assumptions in Bayesian asymptotic theory. Similar to Theorem 10 in \citet{GhoVan07}, we have assumed a compact support in (A1) and lower bounded pseudo Hellinger distance in (A2).  Typically, $\alpha=1$ for most regular models, such as generalized linear models. If the model is non-regular, then  $\alpha$ can be less than 1; for example, the densities may have discontinuities depending on the parameter \citep[Chapters V, VI]{IbrHas81}. Assumption (A3) parallels the entropy condition used in Theorem 1 of \citet{WonShe95}, which has been adapted here for the \inid\ setup using the generalized bracketing entropy, and will simplify to a similar entropy condition to that in Theorem 1 of \citet{WonShe95} if the data are \iid. Assumption (A4) is crucial in providing a stronger control over the tail probability as the posterior probability mass moves away from the true parameter $\theta_0$, typically with an exponentially decaying rate. The convexity property of $\rho$ in (A5) is mainly used to establish an averaging inequality under $W_2$ distance  and is satisfied by, for example, the Euclidean metric and $L_q$ metric with $q\geq 1$.

The posterior risks of $\Pi_n$ and $\overline \Pi_n$ in the $\rho$ metric is directly related to the $W_2$ distance based on the $\rho$ metric. If $\theta_0$ denotes the true parameter value from which the data are generated, then the posterior risk of $\Pi_n$ in the estimation of $\theta_0$ is
\begin{align}
  \label{eq:post-risk-1}
  \int_{\Ycal^{(n)}} \int_{\Theta} \rho^2(\theta, \theta_0) d \Pi_n(\theta \mid Y^{(n)}) d P^{(n)}_{\theta_0}(y_1, \ldots, y_n) =
  E_{P^{(n)}_{\theta_0}} \left[ W_2^2 \left\{ \Pi_n (\cdot \mid Y^{(n)}) , \delta_{\theta_0}(\cdot)  \right\} \right] .
\end{align}
The classical result says that the posterior risk \eqref{eq:post-risk-1} in regular parametric models converges to zero at the $n^{-1}$ rate under assumptions similar to (A2)--(A4), with $m$ replaced by $n$ \citep{Van00}. The next theorem shows that the same posterior risk of the WASP converges at a similar rate to that of the true posterior $\Pi_n$, which mainly depends on the size of subsets $m$, and can be made close to the standard $n^{-1}$ rate up to some logarithmic factors for regular parametric models.

\begin{theorem}\label{unionbound}
If Assumptions (A1)-(A4) hold for the $j$th subset posterior $\Pi_m(\cdot \mid Y_{[j]})$ ($j=1,\ldots,k$), then there exists a constants universal $C_1>0$ independent of $j$, such that as $m\to \infty$,
\begin{align}\label{jbound}
E_{P_{\theta_0}^{(m)}} \left[ W_2^2 \left\{ \Pi_m (\cdot \mid Y_{[j]}) , \delta_{\theta_0}(\cdot)  \right\} \right]  \leq C_1 \left(\frac{\log^2 m}{m}\right)^{\frac{1}{\alpha}},  \quad j = 1, \ldots, k.
\end{align}
Additionally, if Assumption (A5) holds, then as $m\to \infty$,
\begin{align}\label{waspbound}
E_{P_{\theta_0}^{(n)}} \left[ W^2_2 \left\{ \overline \Pi_n(\cdot \mid Y^{(n)}),\delta_{\theta_0}(\cdot)  \right\} \right] \leq C_1 \left(\frac{\log^2 m}{m}\right)^{\frac{1}{\alpha}}.
\end{align}
\end{theorem}

Theorem \ref{unionbound} proves posterior convergence in expectation, which is stronger than the commonly studied posterior convergence in probability. We present our results using the $W_2$ distance in order to account for the fact that the $k$ subset posteriors sit on a common parameter space. Alternatively, from \eqref{eq:post-risk-1}, the convergence rates in \eqref{jbound} and \eqref{waspbound} are also the rates of posterior risks for the subset posterior distributions and the WASP. For regular models with $\alpha=1$, if the number of subsets $k$ increases slowly with $n$ (e.g., $k=O(\log^c n)$ for some constant $c>0$), then Theorem \ref{unionbound} implies that the WASP converges in $W_2$ distance at a near optimal convergence rate $O_p(n^{-1/2}\log^{c/2+1} n)$ to $\delta_{\theta_0}$. In this case, the standard parametric convergence rate of $\Pi_n$ is $O_p(n^{-1/2})$, so the WASP attains the optimal convergence rate up to the $\log^{c/2+1} n$ factor. Equivalently, using \eqref{eq:post-risk-1}, the posterior risk of the WASP converges to zero at the near optimal rate $O_p(n^{-1}\log^{c+2} n)$, compared to the $O_p(n^{-1})$ posterior risk of the true posterior $\Pi_n$.

In most applications, the interest also lies in functions of $\theta$. Suppose $f:\Theta \mapsto \RR^q$ is a function that maps $\theta$ to $\{f_1(\theta), \ldots, f_q(\theta)\}$, where $q \geq 1$ is a positive integer. A direct application of Lemma 8.5 in \citet{BicFre81} gives the following corollary about the WASP of a function of $\theta$. As long as the function is bounded almost linearly by the $\rho$ metric in \eqref{eqn:wass}, its WASP possesses the same posterior convergence rate as in Theorem \ref{unionbound}.
\begin{corollary}\label{funcrate}
Suppose $f(\cdot) = \{f_1(\cdot), \ldots, f_q(\cdot)\}$ is a function that maps $\Theta \mapsto \RR^q$ such that $| f(\theta) |^2 = \sum_{i=1}^q \{f_i(\theta)\}^2 \leq C_f \{1 + \rho^2(\theta, \theta_0)\} $, where $C_f>0$ is a fixed constant. If the conditions in Theorem \ref{unionbound} hold and $\overline{ f\sharp\Pi}_n (\cdot \mid Y^{(n)})$ represents the WASP of the subset posterior distributions for $f(\theta)$, then as $m\to \infty$,
$$ W_2 \left\{ \overline{ f\sharp\Pi}_n (\cdot \mid Y^{(n)}),\delta_{f(\theta_0)}(\cdot) \right\}  =O_{P_{\theta_0}^{(n)}}\left(\sqrt{\frac{\log ^{2/\alpha} m}{ m^{1/\alpha}}}\right).$$
\end{corollary}

Corollary \ref{funcrate} is very useful in applications because it says that the combination step in the WASP is independent of the model parametrization. Let
$f\sharp\Pi_m(\cdot \mid Y_{[j]})$ be the $j$th subset posterior distribution for $f(\theta)$ ($j=1, \ldots, k$), then the WASP of $k$ subset posterior distributions converges to $f(\theta_0)$ at the rate obtained in Theorem \ref{unionbound}. In practice, we have $S_j$ posterior samples of $\theta$ obtained from subset posterior $j$ denoted as $\theta_{ji}$ ($i=1, \ldots, s_j$; $j=1, \ldots, k$). Algorithm \ref{algo:wasp} estimates an atomic approximation of $\overline{ f\sharp\Pi}_n (\cdot \mid Y^{(n)})$, denoted as  $\hat {\overline{f\sharp\Pi}}_n (\cdot \mid Y^{(n)})$, based on the subset posterior samples $f(\theta_{ji})$ ($i=1, \ldots, s_j$; $j=1, \ldots, k$). The atomic form of the WASP is supported on a grid with mesh-size $\epsilon$ estimated from the subset posterior samples of $f(\theta)$. Algorithm \ref{algo:wasp} estimates the weights of the atoms located on the grid by solving a discrete version of \eqref{eqn:was-bary}.  The theoretical properties of discrete barycenters imply that $\hat {\overline{f\sharp\Pi}}_n (\cdot \mid Y^{(n)})$ is supported only on $O(k)$ elements of the grid; see Theorem 2 in \citet{Andetal16}. We exploit this sparsity by adapting the algorithm in \citet{Srietal15} and by using Gurobi \citep{Gur14}.

{A key assumption in Theorem \ref{unionbound} and Corollary \ref{funcrate} is that the subset posterior distributions provide a noisy approximation of the full data posterior distribution. This is stated precisely in \eqref{jbound}, which shows that the convergence rate of a subset posterior distribution in $W_2$ distance is obtained by using $m$ as the sample size instead of $n$. If any of the assumptions (A1)--(A4) fail, then the subset posterior distributions may approximate the full data posterior distribution poorly, which could possibly lead to poor approximation quality for the WASP.}

\begin{figure}[!t]
  \centering
    \includegraphics[width=\textwidth]{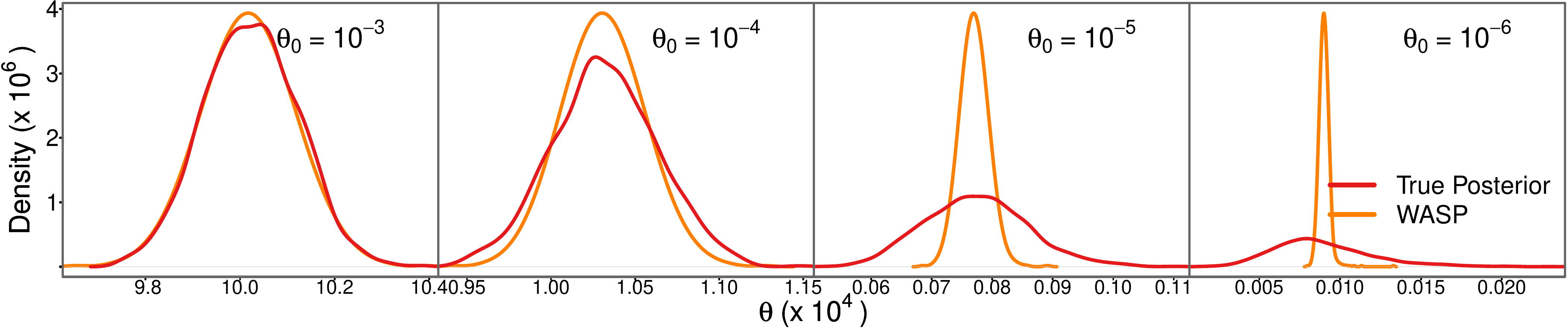}
    \caption{Kernel density estimates of the posterior densities of $\theta$ in the rare events example where assumption (A1) fails to hold for $\theta_0 = 10^{-5}, 10^{-6}$.}
  \label{fig:path_ex}
\end{figure}

{A simple example based on rare events demonstrates this phenomenon. Let $Y_1, \ldots, Y_n$ be \emph{iid} Bernoulli random variables with unknown success probability $\theta \in (0, 1)$. The assumption (A1) is violated if the true parameter $\theta_0$ is very close to 0; that is, observing 1 is a rare event. In our simulation example, we set $n = 10^7$ and $\theta_0 = 10^{-a}$ for $a = 3, 4, 5, 6$ so that as $a$ increases, $s = \sum_{i=1}^n Y_i$ decreases and $\theta_0$ gets closer and closer to the boundary of the parameter space. The standard Bayesian approach is to put Jefferys' prior Beta(0.5, 0.5) on $\theta$ and perform inference on $\theta$ using Beta($s + 0.5$, $n - s + 0.5$), which leads to a full data posterior that concentrates around the correct value of $\theta_0$ even if $\theta_0$ is small (Figure \ref{fig:path_ex}). However, if the data are randomly divided into $k=100$ subsets, then a majority of the subsets contain only 0s as $\theta_0$ decreases. As a result, a majority of the subset posterior distributions differ significantly in shape from the full data posterior distribution, leading to a failure of the WASP in approximating the full data posterior distribution because the assumption (A1) is severely violated for $\theta_0 = 10^{-5}, 10^{-6}$ (Figure \ref{fig:path_ex}).
}

\section{Experiments}
\label{experiments}

\subsection{Setup}

We compared WASP with consensus Monte Carlo (CMC) \citep{Scoetal16}, semiparametric density product (SDP) \citep{NeiWanXin13}, and variational Bayes (VB). The sample sizes and the number of parameters in our experiments were chosen such that sampling from the full data posterior distribution was computationally feasible. Every sampling algorithm ran for 10,000 iterations. We discarded the first 5,000 samples as burn-in and thinned the chain by collecting every fifth sample. Convergence of the chains to their stationary distributions was confirmed using trace plots. All experiments ran on an Oracle Grid Engine cluster with 2.6GHz 16 core compute nodes. Full data posterior computations were allotted memory resources of 64GB, and all other methods were allotted memory resources of 16GB.

The sampling algorithm for the full data posterior was modified to obtain samples from the subset posteriors in CMC, SDP, and WASP. The sampling algorithms for subset posteriors in CMC and SDP were the same and were based on Equation (2) in \citet{Scoetal16}. The sampling algorithm for subset posteriors in WASP was based on \eqref{eqn:wasp-post}. Samples from the approximate posterior distributions of $\theta$ in CMC, SDP, and WASP were obtained in two steps. First, samples from subset posteriors of $\theta$ were obtained in parallel across $k$ subsets. Second, the samples of $\theta$ from all the subsets were combined using implementations of CMC and SDP in \texttt{parallelMCMC} package \citep{MirCon14} and using Algorithm \ref{algo:wasp} for the WASP.

The full data posterior distribution obtained using MCMC served as the benchmark in all our comparisons. Let $\pi(\theta \mid  Y^{(n)})$ be the density of the full data posterior distribution for $\theta$ estimated using sampling and $\hat \pi (\theta \mid  Y^{(n)})$ be the density of an approximate posterior distribution for $\theta$ estimated using the WASP or its competitors. We used the following metric based on the total variation distance to compare the accuracy $\hat \pi (\theta \mid  Y^{(n)})$ in approximating $\pi(\theta \mid  Y^{(n)})$
\begin{align}
  \text{accuracy} \left\{ \hat \pi (\theta\mid  Y^{(n)})\right\} = 1 - \frac{1}{2} \int_{\Theta} \left|\hat \pi(\theta \mid Y^{(n)}) - \pi(\theta \mid  Y^{(n)})\right| \dd \theta. \label{acc}
\end{align}
The accuracy metric lies in $[0, 1]$ \citep{Faeetal12}. The approximation of full data posterior density by $\hat \pi$ is poor or excellent if the accuracy metric is close to 0 or 1, respectively. In our experiments, we computed the kernel density estimates of $\hat \pi$ and $\pi$ from the posterior samples of $\theta$ using R package \texttt{KernSmooth} \citep{Wan15}
and calculated the integral in \eqref{acc} using numerical approximation.

\subsection{Simulated data: finite mixture of Gaussians}
\label{gmm}

Finite mixture of Gaussians are widely used for model-based classification, clustering, and density estimation \citep{FraRaf02}. Let $n$, $p$, and $L$ be the sample size, the dimension of observations, and the number of mixture components. If $\yb_i \in \RR^p$ is the $i$th observation ($i=1, \ldots, n$), then the mixture of $L$ Gaussians assumes that any $\yb \in \{\yb_1, \ldots, \yb_n\}$ is generated from the density
\begin{align}
  f_{\text{mix}}(\yb \mid \theta) = \sum_{l=1}^L \pi_l \Ncal_p(\yb \mid \mub_l, \Sigma_l), \label{mix-1}
\end{align}
where $\pib = (\pi_1, \ldots, \pi_L)$ lies in a $(L-1)$-simplex, $\mu_l$ and $\Sigma_l$ ($l=1, \ldots, L$) are the mean and covariance parameters of a $p$-variate Gaussian distribution, and $\theta = \{\pib, \mub_1, \ldots, \mub_L, \Sigma_1, \ldots, \Sigma_L\}$. We set $L=2$ and $p=2$ and simulated data from \eqref{mix-1} using $\pib=(0.3, 0.7)$, $\mu_1= (1, 2)^T$, $\mu_2 = (7, 8)^T$, and $\Sigma_l=\Sigma$ ($l=1, 2$), where $\Sigma_{12} = 0.5$, $\Sigma_{11} = 1$, and $\Sigma_{22} = 2$.  We performed 10 simulation replications.

This simple example demonstrated the generality of WASP in estimating the posterior distribution of functions of $\theta$ as described in Corollary \ref{funcrate}. We defined two nonlinear functions of $\theta$ as
\begin{align}
  \label{eq:mix-2}
  \rho_{l} = (\Sigma_l)_{12} / \left\{ (\Sigma_l)_{11} (\Sigma_l)_{22} \right\}^{1/2} \quad  l=1, 2, \quad
  g(x)  = f_{\text{mix}}\{(x, x)^T \} \quad x \in \RR,
\end{align}
where $\rho_{l}$ is the correlation of $l$th mixture component and $g(x)$ is the value of density $f_{\text{mix}}$  in \eqref{mix-1} when $\yb = (x, x)^T$. Our simulation setup implied that $\rho_1 = \rho_2$ and $g(x)$ was bimodal for $x \in \RR$. We completed the hierarchical model in \eqref{mix-1} by specifying independent conjugate priors on $\pib$ and $(\mu_l, \Sigma_l)$ ($l=1, 2$) as
\begin{align}
  \label{eq:1}
  \pib \sim \text{Dirichlet}(1/2, 1/2), \quad \mub_l \mid \Sigma_l \sim \Ncal_2(\zero, 100 \Sigma_l), \quad \Sigma_l \sim \text{Inverse-Wishart}(2, 4 I_2) ,
\end{align}
where 2 is the prior degrees of freedom and $4I_p$ is the scale matrix of the Inverse-Wishart distribution. The posterior samples of $\theta$ were obtained using Gibbs sampling \citep{Bis06}, which were used to obtain posterior samples for  $\rho_1$, $\rho_2$, and $g$.

We compared WASP with the posterior distributions estimated using CMC, Gibbs sampling, SDP, and VB. We used the VB algorithm developed in \citet{Bis06}. Two values of $k \in \{5, 10\}$ were used for CMC, SDP, and WASP and full data were partitioned into $k$ subsets such that the mixture proportions were preserved in every subset. The approximate posterior distributions of $\rho_1$, $\rho_2$, and $g(x), x \in \RR,$ under each method were estimated using the subset posterior samples obtained after modifying the original Gibbs sampler. The sampling algorithm for WASP is described in Section 2.1 of Supplementary Material.

We compared the accuracy \eqref{acc} of CMC, SDP, VB, and WASP in approximating the full data posterior distributions of $\rho_1$, $\rho_2$, and point-wise $90\%$ credible bands of $g(x)$ for $x \in \RR$, denoted as $g_{0.05}(x)$ and $g_{0.95}(x)$. CMC, SDP, and WASP accurately approximated the full data posterior distributions of $\rho_1$ and $\rho_2$ for both $k$s, but VB underestimated the posterior uncertainty in $\rho_1$ and $\rho_2$. CMC, VB, and WASP were very accurate in estimating $g_{0.05}(x)$ and $g_{0.95}(x)$ for $x \in \RR$, whereas the application of SDP failed due to a numerical error in matrix inversion (Table \ref{tbl:acc_mix}). This provides an empirical verification of Corollary \ref{funcrate}, showing that the accuracy of the WASP was unaffected by the form of the parameters in the combination step. Theoretical guarantees similar to Corollary \ref{funcrate} were unavailable for CMC or SDP, but our numerical results illustrated that a similar result might also hold for these methods in mixture models.

\begin{table}[t]
  \caption{Accuracies of the approximate posteriors for $\rho_{1}$, $\rho_2$, and $g_{0.05}(x)$ and $g_{0.95}(x)$ for $x \in \RR$. The accuracies are averaged over 10 simulation replications. Monte Carlo errors are in parenthesis. CMC, consensus Monte Carlo; SDP, semiparametric density product; VB, variational Bayes; WASP, Wasserstein posterior
  }
  \label{tbl:acc_mix}
  \centering
  {\tiny
  \begin{tabular}{rcccccccc}
    \hline
    & \multicolumn{2}{c}{$\rho_1$} & \multicolumn{2}{c} {$\rho_2$} & \multicolumn{2}{c} {$g_{0.05}$} & \multicolumn{2}{c} {$g_{0.95}$} \\
    \hline
    VB & \multicolumn{2}{c}{0.77 (0.31)} & \multicolumn{2}{c} {0.76 (0.29)} & \multicolumn{2}{c} {0.99 (0.00)} & \multicolumn{2}{c} {0.99 (0.00)}  \\
    \hline
    & $k=5$ & $k=10$ & $k=5$ & $k=10$ & $k=5$ & $k=10$ & $k=5$ & $k=10$\\
    \hline
    CMC & 0.97 (0.01) & 0.96 (0.01) & 0.96 (0.01) & 0.96 (0.01) & 0.99 (0.00) & 0.99 (0.00) & 0.99 (0.00) & 0.99 (0.00) \\
    SDP & 0.97 (0.01) & 0.96 (0.01) & 0.95 (0.01) & 0.96 (0.01) & - & - & - & - \\
    WASP & 0.97 (0.01) & 0.95 (0.01) & 0.97 (0.01) & 0.96 (0.01) & 0.99 (0.00) & 0.99 (0.00) & 0.99 (0.00) & 0.99 (0.00) \\
    \hline
  \end{tabular}
  }%
\end{table}

\subsection{Simulated data: Linear mixed effects model}
\label{llmm}

Linear mixed effects models are extensively used in extending linear regression to account for longitudinal and nested dependence structures. Let $n$, $s$, and $s_i$ be the sample size, total number of observations, and total number of observations for sample $i$ ($i=1, \ldots, n$) so that $s = \sum_{i=1}^n s_i$. Suppose $X_i \in \RR^{s_i \times p}$ and $Z_i \in \RR^{s_i \times r}$ include predictors in the fixed and random effects components, respectively.  Letting $\yb_i \in \RR^{s_i}$ be the response for sample $i$, the linear mixed effects model assumes that
\begin{align}
  \label{eq:dat1}
  \yb_i \mid \betab, \ub_i, \tau^2 \sim \Ncal_{s_i}(X_i \betab + Z_i \ub_i, \tau^2 I_{n_i}), \quad \ub_i \sim \Ncal_r(\zero, \Sigma), \quad (i = 1, \ldots, n),
\end{align}
where $\ub_i \in \RR^r$ is the random effect for sample $i$ with mean $\zero$ and $r \times r$ covariance $\Sigma$, $\betab \in \RR^{p}$ denotes the fixed effects, and $\tau^2$ is the error variance. The model parameters are $\theta = \{\betab, \Sigma, \tau^2\}$.

We simulated data for a fixed $n$ and $s$ and varying $p$ and $r$. We chose two values of $(p, r)\in \{(4, 3), (80, 6)\}$, fixed $n$ and $s$ to be $6000$ and 100,000, and randomly assigned the $s$ observations to $n$ samples. The two choices of $(p, r)$ ensured that the number of unknown parameters in $\betab$ and $\Sigma$ was 10 and 100 in the former and latter cases. The entries of $X_i$ and $Z_i$ were set to $1$ or $-1$ with equal probability for every $i$. We fixed $\betab$ entries as $-2$ and 2 alternately and $\tau^2=1$. {The random effects covariance matrix $\Sigma = \diag(\sqrt{1}, \ldots, \sqrt{r})R\diag(\sqrt{1}, \ldots, \sqrt{r})$, where $\diag(\ab)$ is a diagonal matrix with $\ab$ along the diagonal and $R$ is a correlation matrix with 1 along the diagonal. We set $R = R_1$ if $r = 3$ and $R = \text{bdiag}(R_1, R_1)$ if $r = 6$, where $\text{bdiag}(A, B)$ is a block-diagonal matrix with $A, B$ along the diagonal,  $(R_1)_{ii} = 1$ ($i=1, 2, 3$), $R_{12}=-0.40$, $R_{13}=0.30$, and $R_{23}=0.001$. The matrix $R_1$ included negative, positive, and small to moderate strength correlations \citep{Kimetal13}. }We used this setup to simulate data from \eqref{eq:dat1} and performed 10 replications.

We used the HMC algorithm in Stan for sampling from the full data and subset posterior distributions. The full data posterior computations were feasible for the chosen values of $n$ and $s$ and posterior samples were obtained after completing the hierarchical model in \eqref{eq:dat1} by using the default weakly informative priors for $\betab$, $\Sigma$, and $\tau^2$ in Stan. Two values of $k \in \{10, 20\}$ were used for CMC, SDP, and WASP, and the $n$ samples were randomly partitioned into $k$ subsets. The sampling algorithms for subset posterior distributions for the three methods were implemented in Stan and posterior samples of $\theta$ were obtained in parallel across $k$ subsets. This was followed by a combination step to estimate the approximate posterior distributions for the three methods. The sampling algorithm for WASP is described in Section 2.2 of Supplementary Material. {Stochastic gradient Langevin dynamics (SGLD; \citealt{WelTeh11})  has proven to be a successful stochastic version of MCMC in mixture and regression models but has not been extensively tested on linear mixed effects models in which multiple observations are available on a subject}. {We compared Stan's HMC and SGLD with batch sizes 2000, 4000, step sizes $10^{-4}, 10^{-5}$ and {$10^4$ iterations}.}

We compared the accuracy \eqref{acc} of CMC, SDP, SGLD, VB, and WASP in approximating the marginal posterior distributions of fixed effects, variances and covariances of random effects, and the joint posterior distributions of three pairs of covariances of random effects. We used the streamlined algorithm (SA; \citealt{LeeWan16}) and automatic differentiation variational inference in Stan (ADVI; \citealt{Kucetal15}) for estimating the VB posteriors for $\betab$ and $\Sigma$ . {All methods except SGLD were significantly faster than the full data posterior distribution, with SA being the fastest.} CMC, SA, SDP, and WASP provided accurate approximations of the marginal posterior distributions of fixed effects and covariances of random effects. {Unlike Stan's HMC, SGLD's performance was sensitive to the choices of step size and batch size. SGLD failed to converge for all batch sizes when the step size was $10^{-4}$, and its accuracy increased with batch size.} The performance of ADVI {and SGLD} deteriorated quickly as $r$ increased from $3$ to $6$. The accuracy of CMC and SDP in approximating the marginal posterior distributions of variances of random effects depended on $k$ and $r$. ADVI and SA provided a poor approximation for the posterior variances of random effects. In all these cases, the accuracy of WASP was stable for every $k$ and $r$ (Tables \ref{tbl:acc_varef} and \ref{tbl:acc_covef}). {All methods except SGLD showed similar accuracies in approximating the true joint posterior distributions of three pairs of covariances of random effects.} The differences in accuracies  of CMC, SA, SDP, and WASP for different values of $k$ and $r$ were due to the differences in numerical approximation of \eqref{acc} (Tables \ref{tbl:acc_ranef1_cov2d} and \ref{tbl:acc_ranef2_cov2d} and Figures \ref{fig:cov_2d_q3} and \ref{fig:cov_2d_q6}); see Table 1 in the Supplementary Material.

The accuracy of CMC, SDP, and WASP decreased when $k$ increased from $10$ to $20$ because subset posterior distributions conditioned on a smaller fraction of the data. This provided an empirical verification of Theorem \ref{unionbound} for the WASP. Our numerical results illustrated that a similar result might also hold for CMC and SDP. The stable performance of WASP compared to that of CMC and SDP in the approximation of the posterior distributions of variances of random effects showed that the validity of the normal approximation for subset posterior distributions was crucial in obtaining accurate approximations of full data posterior using CMC and SDP. On the other hand, WASP results were free of any such assumptions and were valid for any nonlinear function of $\mu$ and $\Sigma$; see Corollary \ref{funcrate}.

\begin{table}[t]
  \caption{Accuracies of the approximate posteriors for variances in \eqref{eq:dat1}. The accuracies are averaged over 10 simulation replications and across all diagonal elements of $\Sigma$. Monte Carlo errors are in parenthesis. ADVI, automatic differentiation variational inference; SA, streamlined algorithm; SGLD, stochastic gradient Langevin dynamics with batch size in parenthesis; CMC, consensus Monte Carlo; SDP, semiparametric density product; WASP, Wasserstein posterior
  }
  \label{tbl:acc_varef}
  \centering
  {\tiny
    \begin{tabular}{rcccc}
      \hline
      & \multicolumn{2}{c}{$r = 3$} & \multicolumn{2}{c}{$r = 6$} \\
      \hline
      ADVI & \multicolumn{2}{c}{0.48 (0.31)} & \multicolumn{2}{c}{0.09 (0.23)} \\
      SA & \multicolumn{2}{c}{0.26 (0.19)} & \multicolumn{2}{c}{0.34 (0.22)} \\
      SGLD (2000) & \multicolumn{2}{c}{0.68 (0.08)} & \multicolumn{2}{c}{0.73 (0.08)} \\
      SGLD (4000) & \multicolumn{2}{c}{0.69 (0.09)} & \multicolumn{2}{c}{0.72 (0.08)} \\      
      \hline
      & $k = 10$ & $k = 20$ & $k=10$ & $k=20$ \\
      \hline
      CMC & 0.93 (0.03) & 0.91 (0.05) & 0.89 (0.05) & 0.80 (0.08) \\
      SDP & 0.92 (0.06) & 0.86 (0.07) & 0.84 (0.10) & 0.77 (0.14) \\
      WASP & 0.97 (0.01) & 0.97 (0.01) & 0.97 (0.01) & 0.97 (0.01) \\
      \hline
    \end{tabular}
  }
\end{table}

\begin{table}[t]
  \caption{Accuracies of the approximate posteriors for covariances in \eqref{eq:dat1}. The accuracies are averaged over 10 simulation replications and across all off-diagonal elements of $\Sigma$. Monte Carlo errors are in parenthesis. ADVI, automatic differentiation variational inference; SA, streamlined algorithm; SGLD, stochastic gradient Langevin dynamics with batch size in parenthesis; CMC, consensus Monte Carlo; SDP, semiparametric density product; WASP, Wasserstein posterior
  }
  \label{tbl:acc_covef}
  \centering
  {\tiny
    \begin{tabular}{rcccc}
      \hline
      & \multicolumn{2}{c}{$r = 3$} & \multicolumn{2}{c}{$r = 6$} \\
      \hline
      ADVI & \multicolumn{2}{c}{0.69 (0.23)} & \multicolumn{2}{c}{0.49 (0.29)} \\
      SA & \multicolumn{2}{c}{0.94 (0.02)} & \multicolumn{2}{c}{0.94 (0.02)} \\
      SGLD (2000) & \multicolumn{2}{c}{0.07 (0.11)} & \multicolumn{2}{c}{0.13 (0.09)} \\
      SGLD (4000) & \multicolumn{2}{c}{0.07 (0.11)} & \multicolumn{2}{c}{0.12 (0.09)} \\            
      \hline
      & $k = 10$ & $k = 20$ & $k=10$ & $k=20$ \\
      \hline
      CMC & 0.94 (0.03) & 0.91 (0.05) & 0.94 (0.03) & 0.92 (0.05) \\
      SDP & 0.92 (0.04) & 0.89 (0.06) & 0.89 (0.07) & 0.87 (0.10) \\
      WASP & 0.97 (0.01) & 0.97 (0.01) & 0.97 (0.01) & 0.96 (0.01) \\
      \hline
    \end{tabular}
  }
\end{table}

\begin{table}[t]
  \caption{Accuracies of the approximate two-dimensional joint posteriors for the covariances of random effects when $r = 3$ in \eqref{eq:dat1}. The accuracies are averaged over 10 simulation replications. Monte Carlo errors are in parenthesis. ADVI, automatic differentiation variational inference; SA, streamlined algorithm; SGLD, stochastic gradient Langevin dynamics with batch size in parenthesis; CMC, consensus Monte Carlo; SDP, semiparametric density product; WASP, Wasserstein posterior
  }
  \label{tbl:acc_ranef1_cov2d}
  \centering
  {\tiny
    \begin{tabular}{rcccccc}
      \hline
      & \multicolumn{2}{c}{$(\sigma_{12}, \sigma_{13})$} & \multicolumn{2}{c}{$(\sigma_{12}, \sigma_{23})$} & \multicolumn{2}{c}{$(\sigma_{13}, \sigma_{3 2})$} \\
      \hline
      ADVI & \multicolumn{2}{c}{0.53 (0.28)} & \multicolumn{2}{c}{0.62 (0.14)} & \multicolumn{2}{c}{0.49 (0.25)} \\
      SA & \multicolumn{2}{c}{0.91 (0.01)} & \multicolumn{2}{c}{0.91 (0.01)} & \multicolumn{2}{c}{0.92 (0.01)} \\
      SGLD (2000) & \multicolumn{2}{c}{0.03 (0.01)} & \multicolumn{2}{c}{0.01 (0.00)} & \multicolumn{2}{c}{0.02 (0.01)} \\
      SGLD (4000) & \multicolumn{2}{c}{0.03 (0.01)} & \multicolumn{2}{c}{0.01 (0.00)} & \multicolumn{2}{c}{0.02 (0.01)} \\            
      \hline
      &  $k = 10$ & $k = 20$ & $k=10$ & $k=20$ & $k = 10$ & $k = 20$ \\
      \hline
      CMC & 0.88 (0.05) & 0.79 (0.06) & 0.88 (0.04) & 0.82 (0.07) & 0.91 (0.02) & 0.85 (0.04) \\
      SDP & 0.90 (0.03) & 0.89 (0.03) & 0.90 (0.03) & 0.87 (0.05) & 0.92 (0.02) & 0.89 (0.04) \\
      WASP & 0.93 (0.01) & 0.94 (0.01) & 0.93 (0.01) & 0.94 (0.01) & 0.94 (0.01) & 0.94 (0.01) \\
      \hline
    \end{tabular}
  }%
\end{table}

\begin{table}[t]
  \caption{Accuracies of the approximate two-dimensional joint posteriors for the covariances of random effects when $r = 6$ in \eqref{eq:dat1}. The accuracies are averaged over 10 simulation replications. Monte Carlo errors are in parenthesis. ADVI, automatic differentiation variational inference; SA, streamlined algorithm; SGLD, stochastic gradient Langevin dynamics with batch size in parenthesis; CMC, consensus Monte Carlo; SDP, semiparametric density product; WASP, Wasserstein posterior
  }
  \label{tbl:acc_ranef2_cov2d}
  \centering
  {\tiny
    \begin{tabular}{rcccccc}
      \hline
      & \multicolumn{2}{c}{$(\sigma_{12}, \sigma_{13})$} & \multicolumn{2}{c}{$(\sigma_{12}, \sigma_{2 3})$} & \multicolumn{2}{c}{$(\sigma_{1 3}, \sigma_{3 2})$} \\
      \hline
      ADVI & \multicolumn{2}{c}{0.06 (0.16)} & \multicolumn{2}{c}{0.08 (0.22)} & \multicolumn{2}{c}{0.08 (0.17)} \\
      SA & \multicolumn{2}{c}{0.89 (0.02)} & \multicolumn{2}{c}{0.90 (0.02)} & \multicolumn{2}{c}{0.91 (0.02)} \\
      SGLD (2000) & \multicolumn{2}{c}{0.02 (0.01)} & \multicolumn{2}{c}{0.01 (0.01)} & \multicolumn{2}{c}{0.01 (0.01)} \\
      SGLD (4000) & \multicolumn{2}{c}{0.02 (0.01)} & \multicolumn{2}{c}{0.01 (0.01)} & \multicolumn{2}{c}{0.01 (0.01)} \\            
      \hline
      &  $k = 10$ & $k = 20$ & $k=10$ & $k=20$ & $k = 10$ & $k = 20$ \\
      \hline
      CMC & 0.88 (0.05) & 0.76 (0.10) & 0.88 (0.04) & 0.78 (0.07) & 0.90 (0.04) & 0.83 (0.07) \\
      SDP & 0.90 (0.03) & 0.86 (0.05) & 0.90 (0.04) & 0.86 (0.04) & 0.90 (0.03) & 0.87 (0.04) \\
      WASP & 0.93 (0.02) & 0.94 (0.01) & 0.94 (0.01) & 0.94 (0.01) & 0.94 (0.01) & 0.94 (0.02) \\
      \hline
    \end{tabular}
  }%
\end{table}

\begin{figure}[!t]
  \centering
    \includegraphics[width=\textwidth]{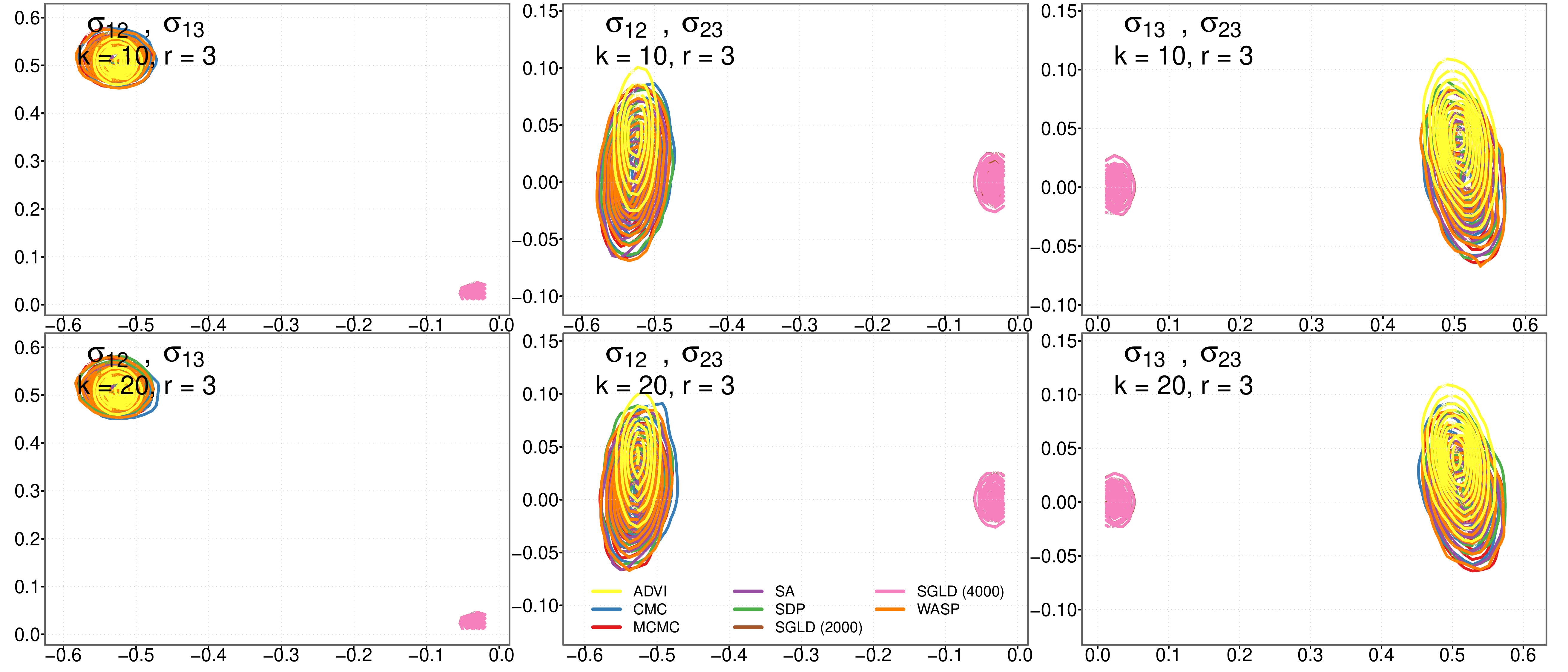}
  \caption{Kernel density estimates of the posterior densities of three covariance pairs when $r = 3$  in \eqref{eq:dat1}, where $\sigma_{a b}, \sigma_{c d}$ on every panel represents the two-dimensional posterior density of $(\sigma_{a b}, \sigma_{cd})$. ADVI, automatic differentiation variational inference; SGLD, stochastic gradient Langevin dynamics with batch size in parenthesis; CMC, consensus Monte Carlo; MCMC, Markov chain Monte Carlo; SA, streamlined algorithm; SDP, semiparametric density product; WASP, Wasserstein posterior.}
  \label{fig:cov_2d_q3}
\end{figure}

\begin{figure}[!t]
  \centering
    \includegraphics[width=\textwidth]{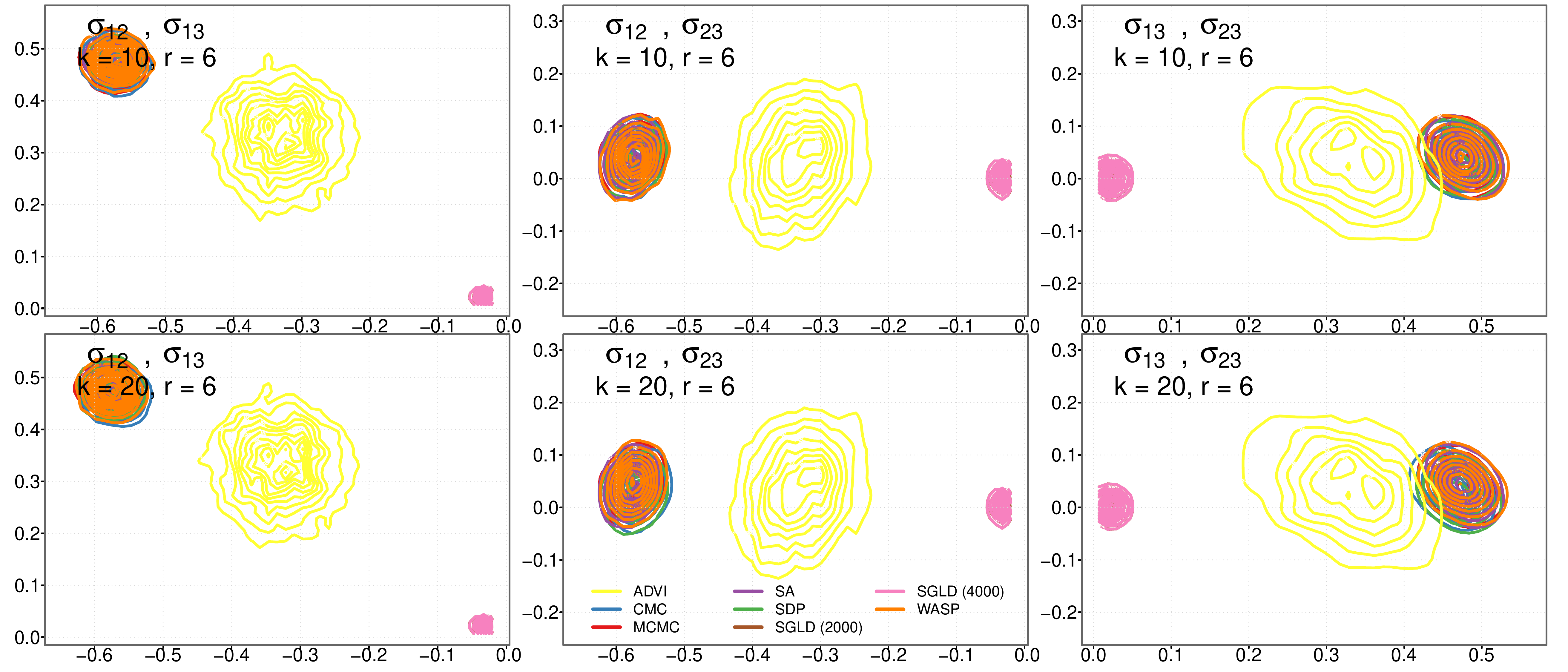}
  \caption{Kernel density estimates of the posterior densities of three covariance pairs when $r = 6$  in \eqref{eq:dat1}, where $\sigma_{a b}, \sigma_{c d}$ on every panel represents the two-dimensional posterior density of $(\sigma_{a b}, \sigma_{cd})$. ADVI, automatic differentiation variational inference; SGLD, stochastic gradient Langevin dynamics with batch size in parenthesis; CMC, consensus Monte Carlo; MCMC, Markov chain Monte Carlo; SA, streamlined algorithm; SDP, semiparametric density product; WASP, Wasserstein posterior.}
  \label{fig:cov_2d_q6}
\end{figure}

\subsection{Simulated data analysis: Probablistic parafac model} %
\label{ppm}

\begin{figure}[!t]
  \centering
  \includegraphics[width=\textwidth]{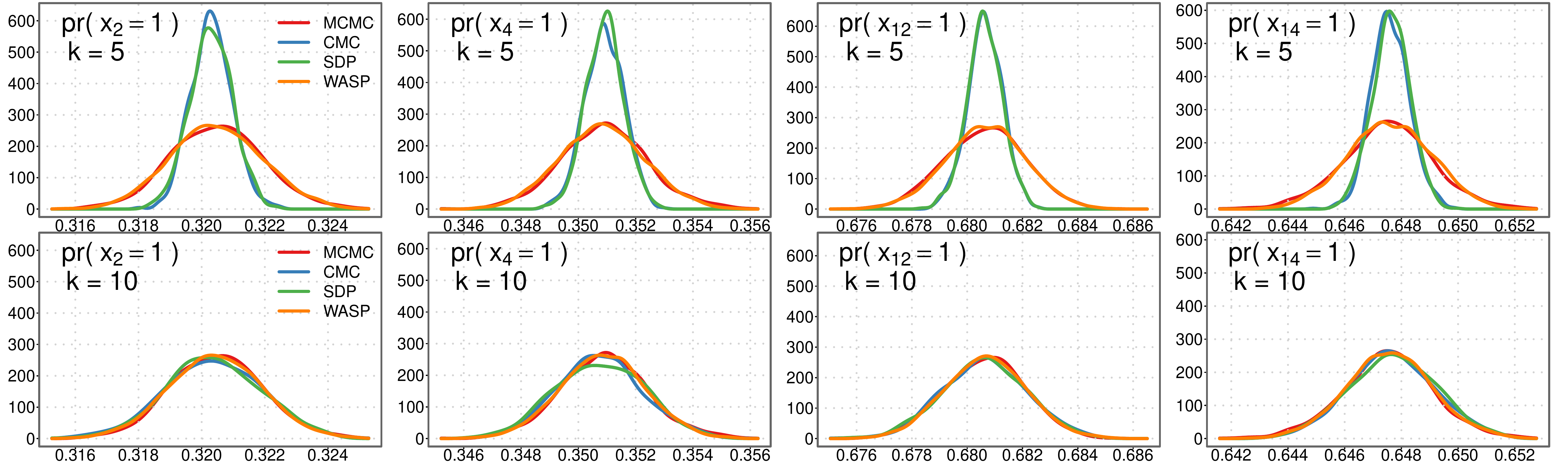}
  \caption{Kernel density estimates of the marginal posterior densities for dimensions 2, 4, 12, and 14. MCMC, Gibbs sampling algorithm of \citet{DunXin09}; CMC, consensus Monte Carlo; SDP, semiparametric density product; VB, variational Bayes; WASP, Wasserstein posterior}
  \label{fig:para}
\end{figure}

We use probabilistic parafac model as a representative example for nonparametric density estimation using WASP.  Probabilistic parafac is an approach for nonparametric Bayes modeling of joint dependence in multivariate categorical data \citep{DunXin09}. Let $\xb_i = (x_{i1}, \ldots, x_{ij}, \ldots, x_{ip})$ be the data from sample $i$, where $x_{ij}$ has $d_j$ possible categorical values in $ \{1, \ldots, d_j\}$ ($j = 1, \ldots, p$). The hierarchical model for $x_{ij}$ ($i=1, \ldots, n$; $j=1, \ldots, p$) is
\begin{align}
  \label{eq:dx}
  x_{ij} \mid \left( \psi^{(j)}_{h 1} \right)_{h=1}^{\infty}, \ldots, \left( \psi^{(j)}_{h d_j} \right)_{h=1}^{\infty}, z_{i}  &\sim \text{Multinomial}(\{1, \ldots, d_j\}, \psi^{(j)}_{z_i 1}, \ldots, \psi^{(j)}_{z_i d_j}), \nonumber \\
  z_i &\sim \sum_{h=1}^{\infty} V_h \prod_{l < h} (1 - V_l) \delta_h \equiv \sum_{h=1}^{\infty} \nu_h \delta_h, \quad V_h \sim \text{Beta}(1, \alpha), \nonumber\\
  \psi^{(j)}_{h}   &\sim \text{Dirichlet}(a_{j1}, \ldots, a_{j d_j}), \quad \alpha \sim \text{Gamma}(a_{\alpha}, b_{\alpha}),
\end{align}
where $\alpha$ has prior mean $a_{\alpha} / b_{\alpha}$. The hierarchical model for probabilistic parafac implies that
\begin{align}
  \label{eq:jt-prob-para}
  \text{pr}(x_{i1}=c_1, \ldots, x_{ij}= c_j, \ldots, x_{ip} = c_p) = \pi_{c_1, \ldots, c_p} = \sum_{h=1}^{\infty} \nu_h \prod_{j=1}^p \psi_{h c_j}^{(j)}.
\end{align}
The $x_{ij}$s are sampled independently given the latent class $z_i$ and probability vectors $\psib^{(j)}_h$ ($h=1, \ldots, \infty$). The latent class for every sample is generated using the stick breaking representation of Dirichlet processes. The Gibbs sampling algorithm developed in \citet{DunXin09} is very slow even for moderate sample sizes. This example demonstrates that WASP can easily scale existing sampling algorithms to massive data, even when efficient VB alternatives are unavailable.

We followed the simulation setup in \citet{DunXin09}, except with a much larger sample size. We fixed the sample size, number of dimensions, and number of categories in each dimension at $n=10^5$, $p=20$, and $d_j=2$ ($j= 1, \ldots, p$), respectively. These choices of $n$, $p$, and $d_j$s ensured that computations for sampling from the full data posterior were tractable. Data were simulated as a mixture of two populations such that any sample belonged to the two populations with equal probability. The two categories in every dimension excluding $2, 4, 12,$ and $14$ were simulated from a discrete uniform in both populations. The dependence across dimensions $2, 4, 12,$ and $14$ was induced as follows. The probabilities $\pi_2, \pi_4, \pi_{12},$ and $\pi_{14}$ were set to (0.20, 0.80), (0.25, 0.75), (0.80, 0.20), and (0.75, 0.25) in the first population and to (0.80, 0.20), (0.75, 0.25), (0.20, 0.80), and (0.25, 0.75) in the second population. The simulation setup was replicated 10 times.

We used CMC, SDP, and WASP to approximate the full data posterior distributions for $\text{pr}(x_{i}= 1)$, where $i \in \{2, 4, 12, 14\}$. Two values of $k \in \{5, 10\}$ were used for CMC, SDP, and WASP. The full data were randomly partitioned into $k$ subsets and subset posterior samples for WASP were obtained after modifying the Gibbs sampling algorithm in \citet{DunXin09} using \eqref{eqn:wasp-post}. Examples for the application of CMC and SDP were unavailable for Dirchlet process mixtures, and it was unclear how to raise the prior density to the power $1/k$ when the prior distribution has an atomic form similar to that in \eqref{eq:dx}; therefore, we did not raise the prior to a power of $1/k$ for sampling from the subset posterior distributions in CMC and SDP. The sampling algorithm for WASP based on stochastic approximation is summarized in Section 2.3 of Supplementary Material. Subset posterior samples for $\text{pr}(x_{2}= 1)$, $\text{pr}(x_{4}= 1)$, $\text{pr}(x_{12}= 1)$, and $\text{pr}(x_{14}= 1)$ were combined to obtain their approximate posterior distributions using CMC, SDP, and WASP.

The accuracy \eqref{acc} of CMC and SDP in approximating the full data marginal posterior distribution depended on $k$, with WASP outperforming CMC and SDP when $k=5$ (Table \ref{tbl:acc_para}). The approximate and full data posterior distributions were centered at the same value across all dimensions and replications, but the  posterior densities for CMC and SDP were highly concentrated compared to the full data posterior density when $k=5$ (Figure \ref{fig:para}). The accuracy of WASP remained stable with varying $k$, providing an empirical verification of Theorem \ref{unionbound} in cases where our theory is not applicable. The time spent in combining subset posterior samples was negligible compared to the time spent in sampling; therefore, WASP could be used for data with much larger sample size by choosing $k$ large enough such that sampling was efficient across all the data subsets.

\begin{table}[t]
  \caption{Accuracies of the approximate marginal posterior distributions for dimensions 2, 4, 12, and 14 in \eqref{eq:dx}. The accuracies are averaged over 10 simulation replications. Monte Carlo errors are in parenthesis. CMC, consensus Monte Carlo; SDP, semiparametric density product; WASP, Wasserstein posterior}
  \label{tbl:acc_para}
  \centering
  {\tiny
    \begin{tabular}{rcccccc}
      \hline
      & \multicolumn{3}{c}{$k = 5$} & \multicolumn{3}{c}{$k = 10$} \\
      \hline
      & CMC & SDP & WASP & CMC & SDP & WASP \\
      \hline
      $\text{pr}(x_{2}= 1)$ & 0.63 (0.02) & 0.62 (0.02) & 0.97 (0.01) & 0.96 (0.02) & 0.95 (0.01) & 0.97 (0.01) \\
      $\text{pr}(x_{4}= 1)$ & 0.63 (0.02) & 0.62 (0.02) & 0.97 (0.01) & 0.96 (0.01) & 0.95 (0.02) & 0.97 (0.01) \\
      $\text{pr}(x_{12}= 1)$ & 0.62 (0.02) & 0.62 (0.02) & 0.97 (0.01) & 0.95 (0.01) & 0.96 (0.02) & 0.97 (0.01) \\
      $\text{pr}(x_{14}= 1)$ & 0.64 (0.01) & 0.63 (0.01) & 0.97 (0.01) & 0.96 (0.02) & 0.95 (0.02) & 0.97 (0.01) \\
      \hline
    \end{tabular}
  }%
\end{table}

\subsection{Real data analysis: MovieLens ratings data}
\label{ml}

We used MovieLens data to illustrate the application of WASP to large-scale ratings data. MovieLens data are one of the largest publicly available ratings data with about 10 million ratings from about 72 thousand users of the MovieLens recommender system. Each observation in the database consists of a user, movie, rating of the movie from 0.5 to 5 in increments of 0.5, and the time of rating. Every movie is also classified into at least one of the 19 genres. We fit a linear mixed effects model \eqref{eq:dat1} using movie- and user-specific information as predictors and the ratings as responses.

We generated three new predictors for accurate modeling of ratings following \cite{Per15}. First, movie genres were grouped into \emph{movie categories} to reduce the number of genres from 19 to four: \emph{Action} category included Action, Adventure, Fantasy, Horror, Sci-Fi, and Thriller genres; \emph{Children} category included Animation and Children genres; \emph{Comedy} category included Comedy genre; and \emph{Drama} category included Crime, Documentary, Drama, Film-Noir, Musical, Mystery, Romance, War, and Western genres. If a movie belonged to multiple genres, then movie category scores were fractions proportional to the number of genres in the respective categories.  Second, \emph{popularity} predictor was defined as $\mathrm{logit}\{(l + 0.5) / (n + 1.0) \}$, where $l$ and $n$ respectively were the number of users who liked and rated the movie in 30 most recent observations for the movie and $\mathrm{logit}(x) = \log \tfrac{x}{1-x}$. Third, \emph{previous} predictor was defined to be 1 if the user liked the previous movie and 0 otherwise. We used \emph{Action}, \emph{Children} $-$ \emph{Action}, \emph{Comedy} $-$ \emph{Action}, \emph{Drama} $-$ \emph{Action}, \emph{popularity}, and \emph{previous} as the fixed and random effects in \eqref{eq:dat1}.

Following the setup in Section \ref{llmm}, we compared the performance of WASP with ADVI, CMC, SA, {SGLD with batch sizes 2000, 4000, step size 10$^{-5}$ and $10^4$ iterations}, and SDP using the full data posterior distribution as the benchmark. Sampling using the HMC algorithm in Stan was prohibitively slow for the full data posterior distribution, so we first randomly selected 5000 users and then randomly selected 20 ratings for every user. This resulted in a data set with 100,000 ratings. We randomly split the users into 10 training data sets such that ratings for any user belonged to the same training data set. To compute the approximate posteriors using CMC, SDP, and WASP, we set $k=10$ and randomly partitioned the users into $k$ subsets such that each subset contained all the ratings for a user. This setup was replicated for every training data.

\begin{figure}[!t]
  \centering
  \includegraphics[width=\textwidth]{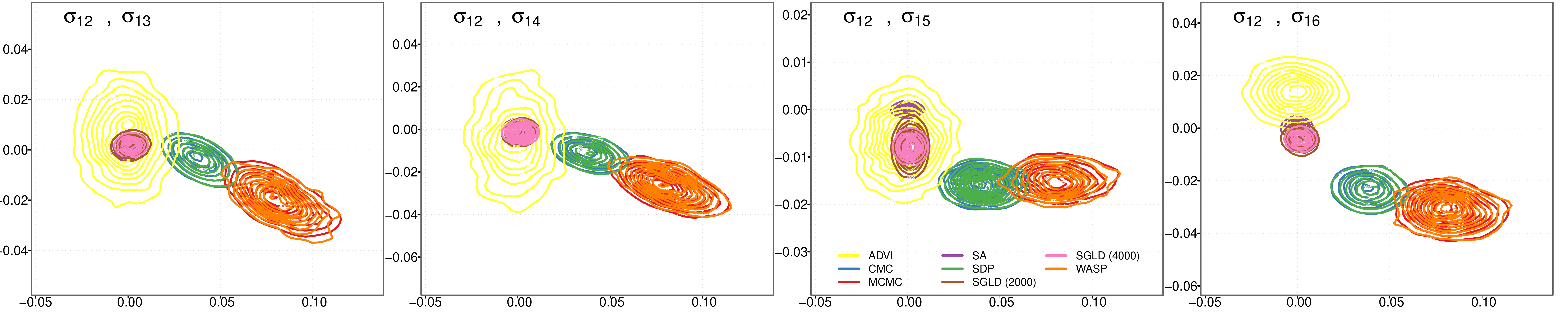}
  \caption{Kernel density estimates of the posterior densities of four covariance pairs, where $\sigma_{a b}, \sigma_{c d}$ on every panel represents the two-dimensional posterior density of $(\sigma_{ab}, \sigma_{cd})$. ADVI, automatic differentiation variational inference; SGLD, stochastic gradient Langevin dynamics with batch size in parenthesis; CMC, consensus Monte Carlo; MCMC, Markov chain Monte Carlo; SA, streamlined algorithm; SDP, semiparametric density product; WASP, Wasserstein posterior.}
  \label{fig:cov_2d_ml}
\end{figure}

WASP performed better than its competitors in approximating the full data posterior distributions for variances and covariances of the random effects. Similar to the simulation results in Section \ref{llmm}, ADVI, CMC, SA, SDP, and WASP were significantly faster than the full data posterior distribution, with SA being the fastest, and SGLD was the slowest. CMC, SDP, and WASP showed excellent performed in approximating the full data posterior distributions for the fixed effects. WASP outperformed its competitors in approximating the full data posterior distributions for variances, covariances, and pairs of covariances of the random effects (Tables \ref{tbl:acc_ml_var}, \ref{tbl:acc_ml_cov}, and \ref{tbl:acc_ml_2d}). ADVI, SA, and SGLD significantly under-performed in the estimation of the posterior distribution for the fixed effects and covariance matrix of the random effects.
The accuracy of marginals in CMC and SDP depended on the magnitude of covariances, with both methods showing excellent accuracy for covariances with low magnitude. The accuracies of the two-dimensional joint distributions in CMC and SDP were poor because the full data posteriors concentrated at different locations (Figure \ref{fig:cov_2d_ml}). {Except for the poor performance of CMC, SA, and SDP in approximating the posterior distribution of variances and covariances of the random effects, our real data results agreed with our simulation results. We concluded that WASP performed better than its competitors in MovieLens data analysis.}

\begin{table}[t]
  \caption{Accuracies of the approximate posteriors for variances in \eqref{eq:dat1}. The accuracies are averaged over 10 replications. Monte Carlo errors are in parenthesis. ADVI, automatic differentiation variational inference; SA, streamlined algorithm; SGLD, stochastic gradient Langevin dynamics with batch size in parenthesis; CMC, consensus Monte Carlo; SDP, semiparametric density product; WASP, Wasserstein posterior}
  \label{tbl:acc_ml_var}
  \centering
  {\tiny
    \begin{tabular}{rcccccc}
      \hline
      & $\sigma^2_{\text{Action}}$ & $\sigma^2_{\text{Children $-$ Action}}$ & $\sigma^2_{\text{Comedy $-$ Action}}$ & $\sigma^2_{\text{Drama $-$ Action}}$ & $\sigma^2_{\text{Popularity}}$ & $\sigma^2_{\text{Previous}}$ \\
      \hline
      ADVI & 0.06 (0.14) & 0.33 (0.30) & 0.16 (0.23) & 0.00 (0.00) & 0.00 (0.00) & 0.00 (0.00) \\
      SA & 0.00 (0.00) & 0.00 (0.00) & 0.00 (0.00) & 0.00 (0.00) & 0.00 (0.00) & 0.00 (0.00) \\
      SGLD (2000) & 0.10 (0.06) & 0.06 (0.03) & 0.05 (0.05) & 0.08 (0.04) & 0.10 (0.00) & 0.10 (0.07) \\ 
      SGLD (4000) & 0.07 (0.06) & 0.06 (0.03) & 0.02 (0.06) & 0.08 (0.04) & 0.10 (0.00) & 0.08 (0.07) \\ 
      CMC & 0.28 (0.13) & 0.01 (0.01) & 0.01 (0.01) & 0.14 (0.09) & 0.74 (0.10) & 0.22 (0.10) \\
      SDP & 0.05 (0.03) & 0.00 (0.00) & 0.00 (0.00) & 0.01 (0.01) & 0.35 (0.10) & 0.03 (0.03) \\
      WASP & 0.92 (0.04) & 0.93 (0.02) & 0.87 (0.06) & 0.85 (0.08) & 0.92 (0.03) & 0.93 (0.05) \\
      \hline
    \end{tabular}
  }%
\end{table}

\begin{table}[t]
  \caption{Accuracies of the approximate posteriors for covariances in \eqref{eq:dat1}. The accuracies are averaged over 10 replications. Monte Carlo errors are in parenthesis. The subscripts $1, \ldots, 6$ are used for predictors \emph{Action}, \emph{Children} $-$ \emph{Action}, \emph{Comedy} $-$ \emph{Action}, \emph{Drama} $-$ \emph{Action}, \emph{popularity}, and \emph{previous}. ADVI, automatic differentiation variational inference; SA, streamlined algorithm; SGLD, stochastic gradient Langevin dynamics with batch size in parenthesis; CMC, consensus Monte Carlo; SDP, semiparametric density product; WASP, Wasserstein posterior}
  \label{tbl:acc_ml_cov}
  \centering
  {\tiny
    \begin{tabular}{rcccccccc}
      \hline
      & $\sigma_{1 2}$ & $\sigma_{1 3}$ & $\sigma_{1 4}$ & $\sigma_{1 5}$ & $\sigma_{1 6}$ & $\sigma_{2 3}$ & $\sigma_{2 4}$ &  $\sigma_{2 5}$ \\
      \hline
      ADVI& 0.15 (0.30) & 0.25 (0.26) & 0.14 (0.16) & 0.32 (0.12) & 0.06 (0.09) & 0.00 (0.00) & 0.18 (0.20) & 0.66 (0.15)\\
      SA & 0.00 (0.00) & 0.00 (0.00) & 0.00 (0.00) & 0.00 (0.00) & 0.00 (0.00) & 0.00 (0.00) & 0.00 (0.00) & 0.00 (0.00) \\
      SGLD (2000) & 0.08 (0.03) & 0.19 (0.10) & 0.18 (0.08) & 0.18 (0.11) & 0.23 (0.09) & 0.14 (0.00) & 0.14 (0.01) & 0.14 (0.10)\\
      SGLD (4000) & 0.08 (0.02) & 0.16 (0.10) & 0.14 (0.08) & 0.12 (0.08) & 0.20 (0.08) & 0.14 (0.00) & 0.13 (0.01) & 0.11 (0.10) \\
      CMC & 0.06 (0.03) & 0.16 (0.04) & 0.18 (0.04) & 0.83 (0.07) & 0.33 (0.13) & 0.01 (0.01) & 0.07 (0.02) & 0.80 (0.04) \\
      SDP & 0.01 (0.01) & 0.08 (0.03) & 0.07 (0.02) & 0.75 (0.06) & 0.14 (0.09) & 0.00 (0.00) & 0.02 (0.01) & 0.73 (0.08) \\
      WASP & 0.95 (0.02) & 0.91 (0.04) & 0.91 (0.05) & 0.94 (0.03) & 0.90 (0.07) & 0.89 (0.07) & 0.85 (0.08) & 0.93 (0.03) \\
      \hline
      & $\sigma_{2 6}$ & $\sigma_{3 4}$ & $\sigma_{3 5}$ & $\sigma_{3 6}$ & $\sigma_{4 5}$ & $\sigma_{4 6}$ & $\sigma_{5 6}$ &  \\
      \hline
      ADVI & 0.47 (0.22) & 0.50 (0.22) & 0.64 (0.11) & 0.62 (0.23) & 0.64 (0.18) & 0.49 (0.29) & 0.42 (0.11) \\
      SA & 0.01 (0.00) & 0.01 (0.00) & 0.01 (0.00) & 0.01 (0.00) & 0.00 (0.00) & 0.00 (0.00) & 0.00 (0.00) \\
      SGLD (2000) & 0.11 (0.10) & 0.14 (0.10) & 0.16 (0.07) & 0.10 (0.11) & 0.14 (0.12) & 0.12 (0.10) & 0.15 (0.09) \\
      SGLD (4000) & 0.07 (0.10) & 0.11 (0.09) & 0.14 (0.07) & 0.03 (0.11) & 0.10 (0.11) & 0.08 (0.10) & 0.14 (0.08) \\
      CMC & 0.66 (0.09) & 0.65 (0.07) & 0.76 (0.08) & 0.71 (0.05) & 0.82 (0.04) & 0.61 (0.11) & 0.55 (0.09) \\
      SDP & 0.59 (0.11) & 0.62 (0.06) & 0.64 (0.09) & 0.66 (0.08) & 0.66 (0.09) & 0.56 (0.14) & 0.55 (0.13) \\
      WASP & 0.91 (0.05) & 0.94 (0.05) & 0.93 (0.03) & 0.91 (0.04) & 0.93 (0.04) & 0.93 (0.04) & 0.94 (0.04) \\
      \hline
    \end{tabular}
  }%
\end{table}

\begin{table}[t]
  \caption{Accuracies of the approximate two-dimensional joint posteriors for the covariances of random effects. The accuracies are averaged over 10 replications. Monte Carlo errors are in parenthesis. The subscripts $1, \ldots, 6$ are used for predictors \emph{Action}, \emph{Children} $-$ \emph{Action}, \emph{Comedy} $-$ \emph{Action}, \emph{Drama} $-$ \emph{Action}, \emph{popularity}, and \emph{previous}. ADVI, automatic differentiation variational inference; SA, streamlined algorithm; SGLD, stochastic gradient Langevin dynamics with batch size in parenthesis; CMC, consensus Monte Carlo; SDP, semiparametric density product; WASP, Wasserstein posterior}
  \label{tbl:acc_ml_2d}
  \centering
  {\tiny
  \begin{tabular}{rllll}
    \hline
    & $\left(\sigma_{1 2}, \sigma_{1 3} \right)$ & $\left(\sigma_{1 2}, \sigma_{1 4} \right)$ & $\left(\sigma_{1 2}, \sigma_{1 5} \right)$  & $\left(\sigma_{1 2}, \sigma_{1 6} \right)$  \\
    \hline
    ADVI & 0.03 (0.06) & 0.03 (0.07) & 0.02 (0.06) & 0.05 (0.11) \\
    SA & 0.18 (0.04) & 0.22 (0.07) & 0.31 (0.03) & 0.31 (0.02) \\
    SGLD (2000) & 0.01 (0.02) & 0.01 (0.02) & 0.01 (0.01) & 0.01 (0.01) \\
    SGLD (4000) & 0.01 (0.02) & 0.01 (0.02) & 0.01 (0.01) & 0.01 (0.01) \\
    CMC & 0.05 (0.02) & 0.04 (0.02) & 0.06 (0.03) & 0.05 (0.02) \\
    SDP & 0.05 (0.02) & 0.04 (0.02) & 0.06 (0.03) & 0.05 (0.02) \\
    WASP & 0.88 (0.03) & 0.88 (0.03) & 0.88 (0.02) & 0.86 (0.06) \\
    \hline
  \end{tabular}
  }
\end{table}

\section{Discussion}
\label{sec:discussion}

We have presented WASP as an approach for computationally efficient approximation of the posterior distributions of parameters and their functions when the sample size is large.
WASP allows extensions of existing samplers to massive data with minimal modifications and is easily implemented using probabilistic programming languages, such as Stan.  {Theoretically, we have showed that the rate of convergence of WASP to the Dirac measure centered at the true parameter value in $W_2$ distance matches the optimal parametric rate up to a logarithmic factor if the number of subsets increases slowly with the size of the full data set.} Empirically, we demonstrated that results from WASP and MCMC agree closely in several widely different examples, while WASP enables massive speed-ups in computational time.

We plan to explore several extensions of WASP in the future. First, the combination of subset posterior distributions using WASP and the {proof of the convergence rate for the WASP} in Theorem \ref{unionbound} are valid even if the data in different subsets are dependent; however, independence assumption within each subset is required in {the proof of \eqref{jbound} in Theorem} \ref{unionbound} and in our justification of stochastic approximation. Currently, it is unclear how to extend  stochastic approximation to cases where the likelihood is unavailable in a product form. This extension in crucial for proper uncertainty quantification outside of settings in which the observations are conditionally independent given latent variables. Second, it is unclear how to optimally choose $k$ in practice; larger $k$ improves computational time when abundant processors are available but choosing $k$ too large may lead to increasing statistical errors (refer to Theorem \ref{unionbound}).  Our numerical experiments show that the accuracy of WASP is robust to the choice of $k$ if all the subset sizes are moderately large relative to the number of parameters. In addition, it is of interest to study more deeply the impact of the partitioning schemes and attempt to develop approaches that deal with not only large sample sizes but also high-dimensional data. A possibility in this regard is to combine WASP with approximate MCMC \citep{Johetal15}.

\section*{Acknowledgement}
{Volkan Cevher and Quoc Tran-Dinh proposed and implemented the linear program for calculating Wasserstein barycenter described in \citet{Srietal15}. All experiments were based on a modified version of Tran-Dinh's Matlab and Gurobi code for estimating Wasserstein barycenter. Jack Baker provided extensive help in implementing the SGLD algorithm. The code used in the experiments is available at \url{https://github.com/blayes/WASP}. Cheng Li's work was partially supported by National University of Singapore start-up grant R155000172133.}

\newpage

\appendix
\section{Proofs of Theorems}

{\subsection{Proof of Theorem \ref{lin-mdl-uq}}

If $E_{P_{\theta_0}^{(n)}}$ represents the expectation with respect to $P_{\theta_0}^{(n)}$, then
\begin{align}
  \label{eq:wass-full-wasp}
  E_{P_{\theta_0}^{(n)}} \left[ W_2^2\{N_p(\mu, V), N_p(\overline \mu, \overline V)\} \right]  = E_{P_{\theta_0}^{(n)}} \| \mu - \overline \mu \|_2^2 + \tr \left\{ V + \overline V - 2 ( \overline V^{1/2} V \overline V ^{1/2})^{1/2} \right\}.
\end{align}
First, we find the asymptotic order of $E_{P_{\theta_0}^{(n)}} \| \mu_1 - \mu_2 \|_2^2$ in \eqref{eq:wass-full-wasp}. Define
\begin{align*}
  A = (X^T \Sigma^{-1} X)^{-1} X^T \Sigma^{-1}, \quad B = k^{-1} \left[ (X_1^T \Sigma_1^{-1} X_1)^{-1}X_1^T \Sigma_1^{-1}, \cdots , (X_k^T \Sigma_k^{-1} X_k)^{-1}X_k^T \Sigma_k^{-1} \right],
\end{align*}
and $C = A - B$. After some algebra, we have that $AX = I_p$, $BX = I_p$, where $I_p$ is a $p \times p$ identity matrix, and
\begin{align*}
  \| \mu - \overline \mu \|_2^2 =  \| C y \|_2^2, \quad E_{P_{\theta_0}^{(n)}}\| \mu - \overline \mu \|_2^2 = E_{P_{\theta_0}^{(n)}}(y^T) C^T C E_{P_{\theta_0}^{(n)}}(y) +  \tr(C \Sigma C^T ).
\end{align*}
Since $E_{P_{\theta_0}^{(n)}}(y) = X \theta_0$ and $CX = AX - BX = I_p - I_p = 0$, $E_{P_{\theta_0}^{(n)}}\| \mu - \overline \mu \|_2^2= \tr(C \Sigma C^T)$. Expanding $C \Sigma C^T$, we get
\begin{align*}
  C &= (X^T \Sigma^{-1} X)^{-1} X^T \Sigma^{-1} - k^{-1} \left[ (X_1^T \Sigma^{-1}_1 X_1)^{-1}X_1^T\Sigma_1^{-1}, \cdots , (X_k^T \Sigma_k^{-1} X_k)^{-1}X_k^T \Sigma^{-1}_k \right], \\
  C^T &= \Sigma^{-1} X (X^T \Sigma^{-1} X)^{-1} - k^{-1}
        \begin{bmatrix}
          \Sigma^{-1}_1 X_1 (X_1^T \Sigma^{-1}_1 X_1)^{-1} \\
          \vdots\\
          \Sigma^{-1}_k  X_k(X_k^T \Sigma^{-1}_k  X_k)^{-1}
        \end{bmatrix},\\
  \tr(C \Sigma C^T) &=  \tr\{(X^T \Sigma X)^{-1}\} +k^{-2} \sum_{j=1}^k \tr \left\{(X_j^T \Sigma_j X_j)^{-1} \right \} - 2 \tr (D),
\end{align*}
where
\begin{align*}
  D &= k^{-1} \left[ (X_1^T \Sigma_1 X_1)^{-1}X_1^T , \cdots , (X_k^T  \Sigma_k X_k)^{-1}X_k^T \right]  \Sigma^{-1} X (X^T \Sigma^{-1} X)^{-1} \\
    &= \left\{ k^{-1} \sum_{j=1}^k (X_j^T \Sigma^{-1}_j X_j)^{-1}X_j^T \Sigma^{-1}_j X_j  \right\}  (X^T \Sigma^{-1} X)^{-1} = (X^T \Sigma^{-1} X)^{-1}
\end{align*}
because $\Sigma$ is diagonal. We use the above display to obtain that
\begin{align*}
  E_{P_{\theta_0}^{(n)}} \| \mu - \overline \mu \|_2^2 &= \tr(C \Sigma C^T) =  \frac{1}{k^2}\sum_{j=1}^k \tr \left\{ (X_j^T \Sigma_j^{-1} X_j)^{-1} \right\} -  \tr \left\{(X^T \Sigma^{-1} X)^{-1} \right\},  \\
                                                       &=  \frac{1}{km} \tr \Big\{ \tfrac{1}{k} \sum_{j=1}^k \left(\tfrac{1}{m}X_j^T \Sigma^{-1}_j X_j\right)^{-1} \Big\} -  \frac{1}{n} \tr \left\{\left(\tfrac{1}{n}X^T\Sigma^{-1} X\right)^{-1} \right\}.
\end{align*}
{Our assumptions and continuity of the matrix inverse for positive definite matrices imply that there are exist positive $a'_n = o(1)$, $b'_m = o(1)$, such that
  \begin{align*}
    \Omega_0^{-1} - a'_n I_p \prec \left( \tfrac{1}{n} X^T \Sigma^{-1} X \right)^{-1}  \prec \Omega_0^{-1} + a'_n I_p, \\
    \Omega_0^{-1} - b'_m I_p \prec \left( \tfrac{1}{m} X_j^T \Sigma_j^{-1} X_j \right)^{-1}  \prec \Omega_0^{-1} + b'_m I_p    .
  \end{align*}
This implies that the previous display reduces to
\begin{align} \label{mnbd}
   E_{P_{\theta_0}^{(n)}} \| \mu - \overline \mu \|_2^2  \leq p(b'_m + a'_n) / n = o(n^{-1}),
\end{align}
where the equality follow because $p$ is fixed.}

We now find the asymptotic order of the traces of the covariance matrices in \eqref{eq:wass-full-wasp}. Following  the same arguments used to derive \eqref{mnbd}, the full data and $j$th subset posterior covariance matrices satisfy
\begin{align}
  \frac{1}{n} \left( \Omega_0^{-1} - a_n' I_p \right) \prec V  &= \frac{1}{n} \left(\frac{1}{n} X^T \Sigma^{-1} X \right)^{-1} \prec  \frac{1}{n} \left( \Omega_0^{-1} + a_n' I_p \right), \nonumber\\
  \frac{1}{n}  \left( \Omega_0^{-1} - b_m' I_p \right) \prec V_j &= \frac{1}{km} \left(\frac{1}{m} X_j^T \Sigma_j^{-1} X_j \right)^{-1} \prec  \frac{1}{n} \left( \Omega_0^{-1} + b_m' I_p \right). \label{cov-ord}
\end{align}
Let $M_j = \left\{\overline V^{1/2} \frac{1}{km} \left(\tfrac{1}{m}X_j^T \Sigma_j^{-1} X_j \right)^{-1} \overline V^{1/2} \right\}^{1/2} $. Then \eqref{cov-ord} implies that
\begin{align}\label{Mjtwoside}
& - b_m'\overline V \prec nM_j^2 - \overline V^{1/2} \Omega_0^{-1} \overline V^{1/2} = n\overline V^{1/2} \left(V_j - n^{-1}\Omega_0^{-1}\right) \overline V^{1/2} \prec b_m'\overline V.
\end{align}
From the first inequality in \eqref{Mjtwoside}, we have
\begin{align*}
& \left(\overline V^{1/2} \Omega_0^{-1} \overline V^{1/2} \right)^{1/2} \prec \left(nM_j^2 + b_m'\overline V\right)^{1/2} \prec n^{1/2} M_j +  b_m^{' 1/2} \overline V^{1/2}.
\end{align*}
And similarly the second inequality in \eqref{Mjtwoside} implies that
\begin{align*}
& n^{1/2} M_j \prec \left(\overline V^{1/2} \Omega_0^{-1} \overline V^{1/2} + b_m'\overline V\right)^{1/2} \prec \left(\overline V^{1/2} \Omega_0^{-1} \overline V^{1/2} \right)^{1/2} + b_m^{' 1/2} \overline V^{1/2}.
\end{align*}
Therefore
\begin{align*}
& \left(\overline V^{1/2} \Omega_0^{-1} \overline V^{1/2} \right)^{1/2} - b_m^{' 1/2} \overline V^{1/2} \prec n^{1/2} M_j \prec \left(\overline V^{1/2} \Omega_0^{-1} \overline V^{1/2} \right)^{1/2} + b_m^{' 1/2} \overline V^{1/2}.
\end{align*}
Using this relation and the definition of $\overline V$, we have that
\begin{align}\label{nvbar}
\left( \overline V^{1/2} \Omega_0^{-1} \overline V^{1/2}  \right)^{1/2} - b_m^{' 1/2} \overline V^{1/2} \prec n^{1/2} \overline V = \frac{1}{k} \sum_{j=1}^k n^{1/2} M_j \prec \left( \overline V^{1/2} \Omega_0^{-1} \overline V^{1/2}  \right)^{1/2} + b_m^{' 1/2} \overline V^{1/2}.
\end{align}
In \eqref{nvbar}, we take the square of $n^{1/2} \overline V$, apply the inequality $(A_1+A_2)^2\prec 2(A_1^2+A_2^2)$ for two generic positive definite matrices $A_1,A_2$, and obtain that
\begin{align*}
n \overline V^2 &\prec 2\overline V^{1/2} \Omega_0^{-1} \overline V^{1/2}   +2 b'_m \overline V ,\\
n \overline V^2 &\succ \frac{1}{2}\overline V^{1/2} \Omega_0^{-1} \overline V^{1/2}   - b'_m \overline V.
\end{align*}
Multiplying by $\overline V^{-1/2}$ on the left and right hand sides yields,
\begin{align}
n \overline V &\prec 2\Omega_0^{-1}  + 2b'_m I_p, \nonumber \\
n \overline V &\succ \frac{1}{2}\Omega_0^{-1}  - b'_m I_p. \label{ev-wcov1}
\end{align}
Notice that $b_m' = o(1)$, $\Omega_0$ is a constant positive definite matrix, and $\overline V$ is a positive definite matrix.  Clearly, \eqref{ev-wcov1} forces $n \overline V$ to be an order-1 matrix. Now we take the square of $n^{1/2} \overline V$ in \eqref{nvbar} again and obtain that 
\begin{align*}
n \overline V^2 &\prec \overline V^{1/2} \Omega_0^{-1} \overline V^{1/2}   + b'_m \overline V + b_m^{'1/2} \left( \overline V^{1/2} \Omega_0^{-1} \overline V^{1/2}  \right)^{1/2} \overline V^{1/2} + b_m^{'1/2} \overline V^{1/2} \left( \overline V^{1/2} \Omega_0^{-1} \overline V^{1/2}  \right)^{1/2},\\
n \overline V^2 &\succ \overline V^{1/2} \Omega_0^{-1} \overline V^{1/2}   + b'_m \overline V - b_m^{'1/2} \left( \overline V^{1/2} \Omega_0^{-1} \overline V^{1/2}  \right)^{1/2} \overline V^{1/2} - b_m^{'1/2} \overline V^{1/2} \left( \overline V^{1/2} \Omega_0^{-1} \overline V^{1/2}  \right)^{1/2}.
\end{align*}
Multiplying by $\overline V^{-1/2}$ on the left and right hand sides yields,
\begin{align}
n \overline V &\prec \Omega_0^{-1}  + b'_m I_p + b_m^{'1/2} \overline V^{-1/2}  \left( \overline V^{1/2} \Omega_0^{-1} \overline V^{1/2}  \right)^{1/2}  + b_m^{'1/2}  \left( \overline V^{1/2} \Omega_0^{-1} \overline V^{1/2}  \right)^{1/2}\overline V^{-1/2}, \nonumber \\
n \overline V &\succ \Omega_0^{-1}  + b'_m I_p - b_m^{'1/2} \overline V^{-1/2}   \left( \overline V^{1/2} \Omega_0^{-1} \overline V^{1/2}  \right)^{1/2} - b_m^{'1/2} \left( \overline V^{1/2} \Omega_0^{-1} \overline V^{1/2}  \right)^{1/2} \overline V^{-1/2} . \label{ev-wcov}
\end{align}
Since $n \overline V$ is an order-1 matrix, we have that $b_m' \overline V^{-1/2}   \left( \overline V^{1/2} \Omega_0^{-1} \overline V^{1/2}  \right)^{1/2} = o(1)$, $b_m'\left( \overline V^{1/2} \Omega_0^{-1} \overline V^{1/2}  \right)^{1/2} \overline V^{-1/2} = o(1)$. Hence \eqref{cov-ord} and \eqref{ev-wcov} reduce to
\begin{align*}
  \frac{1}{n} \{\Omega_0^{-1}  - o(1) I_p\} \prec  V_j \prec \frac{1}{n} \{\Omega_0^{-1} + o(1) I_p\}, \quad  \frac{1}{n}(\Omega_0^{-1}  - o(1) I_p)  \prec \overline V \prec \frac{1}{n} \{\Omega_0^{-1} + o(1) I_p\}.
\end{align*}
This implies that
\begin{align}  \label{eq:wcov2}
\tr(\overline V - V) = o(n^{-1}),
\end{align}
where the last equality follows because $p$ is fixed.

Finally, we find the asymptotic order of the variance term in \eqref{eq:wass-full-wasp}. The display before \eqref{eq:wcov2} implies that for some positive $c_n=o(1)$,
\begin{align*}
  \overline V^{1/2} V \overline V ^{1/2} &\prec  \frac{1}{n^2} \{\Omega_0^{-1/2}  + o(1) I_p\} \{\Omega_0^{-1}  + o(1) I_p\}\{\Omega_0^{-1/2}  + o(1) I_p\} \\
                                        &\prec \frac{1}{n^2} [\Omega_0^{-2} +  c_n I_p ], \\
  \overline V^{1/2} V \overline V ^{1/2} &\succ \frac{1}{n^2} \{\Omega_0^{-1/2}  - o(1) I_p\} \{\Omega_0^{-1}  - o(1) I_p\}\{\Omega_0^{-1/2}  - o(1) I_p\} \\
                                        &\succ \frac{1}{n^2} [\Omega_0^{-2} - c_n I_p ].
\end{align*}
Therefore, $\tr \{( \overline V^{1/2} V \overline V ^{1/2} )^{1/2}\}  = n^{-1} \tr(\Omega_0^{-1}) + o(n^{-1})$ since $p$ is fixed. Using this and \eqref{cov-ord} for the variance term in \eqref{eq:wass-full-wasp} gives
\begin{align}
& \tr \left\{V + \overline V - 2 \left( \overline V^{1/2} V \overline V ^{1/2}\right)^{1/2} \right\}  \nonumber \\
&= \{n^{-1} \tr(\Omega_0^{-1}) + o(n^{-1}) \} + \{n^{-1} \tr(\Omega_0^{-1}) + o(n^{-1}) \} - \{2 n^{-1} \tr(\Omega_0^{-1}) + 2o(n^{-1})\} \nonumber \\
&= o\left(n^{-1}\right).\label{covmbd}
\end{align}
Combining the asymptotic expressions for the mean and variance terms in \eqref{mnbd} and \eqref{covmbd},   \eqref{eq:wass-full-wasp} reduces to
\begin{align*}
E_{P_{\theta_0}^{(n)}} \left[ W_2^2\left\{N(\overline \mu, \overline V), N(\mu, V)\right\}  \right]= o\left(n^{-1}\right),
\end{align*}
which completes the proof. $\Box$}

\subsection{Proof of Theorem \ref{unionbound}}
Let $\epsilon_m= \left( \frac{m}{\log^2 m}  \right)^{-1/(2\alpha)}$. For ease of notation, in all the following proofs, we will sometimes write $p(y_{ji} \mid \theta) \equiv p_{ji}(y_{ji} \mid \theta)$.

\label{sec:proof-main-theorem}
Due to the compactness of $\Theta$ in (A1), we assume that $\rho(\theta,\theta_0)\leq M_0$ for a large finite constant $M_0$. We start with a decomposition of the $W_2$ distance from the $j$th subset posterior $ \Pi_m (\cdot \mid Y_{[j]})$ to the Dirac measure at the true parameter $\theta_0$:
\begin{align}\label{J12}
  & E_{P_{\theta_0}} W_2^2 \left( \Pi_m (\cdot \mid Y_{[j]}) , \delta_{\theta_0}(\cdot) \right) = E_{P_{\theta_0}} \int_{\Theta} \rho^2(\theta,\theta_0)  \Pi_m ( d\theta \mid Y_{[j]}) \nonumber \\
  \leq & E_{P_{\theta_0}} \int_{\left \{\theta:\rho(\theta,\theta_0)\leq c_1 \epsilon_m \right\}} \rho^2(\theta,\theta_0)  \Pi_m ( d\theta \mid Y_{[j]}) +
      E_{P_{\theta_0}} \int_{\left \{\theta: \rho(\theta,\theta_0) > c_1 \epsilon_m \right\}} \rho^2(\theta,\theta_0)  \Pi_m ( d\theta \mid Y_{[j]}) \nonumber \\
  \leq & (c_1 \epsilon_m)^{2} + M_0^2 E_{P_{\theta_0}} \Pi_m \left(\rho(\theta,\theta_0)> c_1 \epsilon_m \mid Y_{[j]} \right).
\end{align}
We will choose the constant $c_1$ as $c_1=\left(\frac{2r_1g_2}{q_1C_L}\right)^{1/(2\alpha)}$, where $g_1,C_L,q_1,r_1$ are the constants in (A1), (A2), and Lemma 5 and Lemma 6 in Supplementary Material.

The following proofs are similar to the proofs of Theorem 1, 4, and 10 in \citet{GhoVan07}. The main difference is that our likelihood has been raised to the power $\gamma$. Using condition (A2), we can further replace the $\rho$ metric by the pseudo Hellinger distance:
\begin{align}\label{J22}
  & \Pi_m \left(\theta\in \Theta: \rho(\theta,\theta_0)>c_1 \epsilon_m  \mid  Y_{[j]}\right) \nonumber \\
\leq{}& \Pi_m \left(\theta\in \Theta: h_{mj}(P_{\theta,j},P_{\theta_0,j})> \sqrt{C_L} (c_1\epsilon_m)^\alpha \mid Y_{[j]}\right) \nonumber \\
={}&  \int_{ \big\{ \theta\in \Theta: h_{mj}(\theta,\theta_0)> \sqrt{\frac{2r_1g_2}{q_1}} \epsilon_m^\alpha \big\} } \frac{\prod_{i=1}^{m} \left[\frac{p(Y_{ji} \mid \theta)}{p(Y_{ji} \mid \theta_0)}\right]^{\gamma} \Pi(d \theta)}{\int_{\Theta} \prod_{i=1}^{m} \left[\frac{p(Y_{ji} \mid \theta)}{p(Y_{ji} \mid \theta_0)}\right]^{\gamma} \Pi(d \theta)}.
\end{align}
For the denominator in \eqref{J22}, by Condition (A4) and Lemma 6, for $m$ sufficiently large, with probability at least $1-\exp(-r_2 m \epsilon_m^{2\alpha})$
\begin{align}\label{denomJ}
  \int_{\Theta}\prod_{i=1}^m \frac{p(Y_{ji}|\theta)^{\gamma}}{p(Y_{ji}|\theta_0)^{\gamma}} \Pi(d \theta) > \exp(- r_1 n \epsilon_m^{2\alpha}).
\end{align}

For the numerator in \eqref{J22}, by Condition (A3) and Lemma 5, we set $\delta=\sqrt{2r_1g_2/q_1} \epsilon_m^{\alpha}$ and obtain that with probability at least  $1- 4 \exp\left(- \frac{2r_1g_2q_2}{q_1} m \epsilon_m^{2\alpha}\right)\geq 1- 4\exp\left(- \frac{2r_1q_2}{q_1} n \epsilon_m^{2\alpha}\right)$,
\begin{align}\label{numJ}
  \sup_{\big\{ \theta\in \Theta:h_{mj}(\theta,\theta_0)\geq \sqrt{2r_1g_2/q_1} \epsilon_m^{\alpha} \big\}}  \prod_{i=1}^m \left[\frac{p(Y_{ji}|\theta)}{p(Y_{ji}|\theta_0)} \right]^{\gamma} \leq  \exp\left(- 2r_1 g_2 m \epsilon_m^{2\alpha}\right) \leq \exp\left(- 2r_1 n \epsilon_m^{2\alpha}\right)
\end{align}

Therefore, based on \eqref{J22}, \eqref{denomJ}, and \eqref{numJ}, we obtain that with probability at least $1-4\exp\left(-\frac{2r_1q_2}{q_1}  n \epsilon_m^{2\alpha}\right)-\exp(-r_2 m \epsilon_m^{2\alpha})$,
\begin{align*}
  \Pi_m \left(\theta\in \Theta: \rho(\theta,\theta_0)> c_1\epsilon_m ~\Big|~ Y_{[j]}\right)
  &\leq   \exp\left( - 2r_1 n \epsilon_m^{2\alpha} + r_1 n \epsilon_m^{2\alpha}  \right)\leq  \exp\left(- r_1 n \epsilon_m^{2\alpha}\right).
\end{align*}
Let $A_{\epsilon_m}$ be the event $\big\{ \theta \in \Theta: \Pi \left(\theta\in \Theta: \rho(\theta,\theta_0)> c_1\epsilon_m ~\Big|~ Y_{[j]}\right) \leq \exp\left(- r_1 n \epsilon_m^{2\alpha}\right)\big \}$. Then we can bound the second term in \eqref{J12} as
\begin{align*}
& E_{P_{\theta_0}} \Pi_m\left(\rho(\theta,\theta_0)> c_1\epsilon_m~\Big|~Y_{[j]}\right)  \\
\leq{}& E_{P_{\theta_0}} \left[I(A_{\epsilon_m})\Pi_m\left(\rho(\theta,\theta_0)>c_1\epsilon_m~\Big|~Y_{[j]}\right)\right]
+E_{P_{\theta_0}} \left[I(A^c_{\epsilon_m})\Pi_m\left(\rho(\theta,\theta_0)>c_1\epsilon_m~\Big|~Y_{[j]}\right)\right] \\
\leq{}& \exp\left(- r_1 n \epsilon_m^{2\alpha}\right) + P_{\theta_0}^{(n)}(A^c_{\epsilon_m})\cdot 1\\
\leq{} & \exp\left(- r_1 n \epsilon_m^{2\alpha}\right) + 4 \exp\left(-\frac{2r_1q_2}{q_1}  n \epsilon_m^{2\alpha}\right)+\exp(-r_2 m \epsilon_m^{2\alpha})\\
\leq{}& 6\exp\left(-c_2  m \epsilon_m^{2\alpha}\right),
\end{align*}
for $c_2=\min(r_1,r_2,2r_1q_2/q_1)$, as clearly the second term is dominating the other two given $m\lesssim n$.

Therefore, for \eqref{J12}, since $\epsilon_m=(m/\log^2 m)^{-1/(2\alpha)}$, as $m\to\infty$, an explicit bound will be
\begin{align*}
  &E_{P_{\theta_0}} W_2^2 \left( \Pi_m (\cdot \mid Y_{[j]}) , \delta_{\theta_0}(\cdot) \right) \leq c_1^2 \frac{\log^{2/\alpha} m}{m^{1/\alpha}}  + 6M_0^2\exp\left(-c_2  \log^2 m\right) \\
  &\leq c_1^2 \frac{\log^{2/\alpha} m}{m^{1/\alpha}}  + \frac{1}{m^{1+\frac{1}{\alpha}}} \leq C_1 \frac{\log^{2/\alpha} m}{m^{1/\alpha}}
\end{align*}
as $m$ becomes sufficiently large, where the constant $C_1$ depends on $\alpha,c_1,c_2$, which further depends on $g_1,g_2,q_1,q_2,r_1,r_2,C_L$. Since $q_1,q_2$ in Lemma 5 and $r_1,r_2$ in Lemma 6 depend on $g_1,g_2,D_1,D_2,\kappa,c_{\pi}$, it follows that $C_1$ depends on $g_1,g_2,C_L,D_1,D_2,\kappa,c_{\pi}$. $\Box$

Based on Lemma 7 in Supplementary Material, if the assumption (A5) holds, then we have
\begin{align*}
& E_{P_{\theta_0}^{(n)}} \left[ W^2_2 \left\{ \overline \Pi_n(\cdot \mid Y^{(n)}),\delta_{\theta_0}(\cdot)  \right\} \right] \leq E_{P_{\theta_0}^{(n)}} \left[ \frac{1}{k} \sum_{j=1}^{k} W_2 \left\{\Pi_m(\cdot \mid Y_{[j]}),\delta_{\theta_0}(\cdot)\right\} \right]^2 \\
&\leq \frac{1}{k} \sum_{j=1}^{k} E_{P_{\theta_0}^{(n)}} W^2_2 \left\{\Pi_m(\cdot \mid Y_{[j]}),\delta_{\theta_0}(\cdot)\right\} \leq C_1 \frac{\log^{2/\alpha} m}{m^{1/\alpha}},
\end{align*}
where the first inequality follows from Lemma 7 in Supplementary Material, the second inequality follows from the Cauchy-Schwarz inequality, and the third inequality follows from the subset bound \eqref{jbound}. $\Box$

\section{Univariate density estimation}
\label{dens-est}

Let $X_1, \ldots, X_n$ be $n$ copies of a scalar random variable $X$ that follows probability distribution $P_0$ with density $p_0$. The full data are randomly split into $k$ subsets and $X_{j1}, \ldots, X_{jm}$ represent the data on subset $j$ ($j=1, \ldots, k$). The hierarchical model for density estimation using the stick-breaking representation of
Dirichlet process mixtures is
\begin{align}
  \label{eq:dp-dens}
  &X_{ji} \mid z_{ji}, \{\mu_h\}_{h=1}^{\infty}, \{\sigma^2_h\}_{h=1}^{\infty} \sim \Ncal(\mu_{z_{ji}}, \sigma_{z_{ji}}^2), \quad z_{ji} \sim \sum_{h=1}^{\infty} \nu_h \delta_{h}, \quad \nu_h = V_h \prod_{l < h} (1 - V_l), \quad V_h \mid \alpha \sim \text{Beta}(1, \alpha), \nonumber\\
  &\quad \alpha \sim \text{Gamma}(a_{\alpha}, b_{\alpha}), \quad \mu_h \mid \sigma^2_h \sim \Ncal(0, \sigma^2_h), \quad \sigma^2_h \sim \text{Inverse-Gamma}(a_{\sigma}, b_{\sigma}),
\end{align}
where $a_{\sigma} > 2$ and Beta, Gamma, and Inverse-Gamma random variables have means $\frac{1}{1+\alpha}$, $\frac{a_{\alpha}}{b_{\alpha}}$, and $\frac{b_{\sigma}}{a_{\sigma}-1}$ and variances $\frac{\alpha}{(1+\alpha)^2 (2 + \alpha)}$, $\frac{a_{\alpha}}{b^2_{\alpha}}$, and $\frac{b^2_{\sigma}}{(a_{\sigma}-1)^2(a_{\sigma}-2)}$. If $l^*$ is the maximum number of atoms in the stick-breaking representation, then the prior density $\pi$ is in the form a discrete mixture. We cannot use existing sampling algorithms directly if $\pi$ is raised to a power of $1/k$, so it is unclear how to sample from the subset posterior density of competing approaches in Section \ref{sec:sub-post}.

We show that it is still possible to sample from the subset posterior density in \eqref{eqn:wasp-post} using data augmentation. Let $L_j$ be the likelihood given $X_{j1}, \ldots, X_{jm}$ and latent variables $z_{j1}, \ldots, z_{jm}$ in \eqref{eq:dp-dens}, then
\begin{align}
  \label{eq:dens1}
  L_j(\{\mu_h\}_{h=1}^{l^*}, \{\sigma^2_h\}_{h=1}^{l^*}, \{\nu_h\}_{h=1}^{l^*})
  &= \prod_{h=1}^{l^*} (2 \pi \sigma_h^2)^{-\frac{\sharp h_j}{2}} e^ {- \frac{1}{2 \sigma^2_h}   \sum_{i=1}^m 1(z_{ji} = h) \left(x_{ji} - \mu_h\right)^2} \nu_h^{\sharp h_j},
\end{align}
where $1(z_{ji} = h)$ is 1 if $z_{ji} = h$ and 0 otherwise and $\sharp h_j = \sum_{i=1}^m 1(z_{ji} = h)$. For stochastic approximation, we raise $L_j$ in \eqref{eq:dens1} to the power $\gamma$ and obtain
\begin{align}
  \label{eq:dens2}
    L_j^{\gamma}(\{\mu_h\}_{h=1}^{l^*}, \{\sigma^2_h\}_{h=1}^{l^*}, \{\nu_h\}_{h=1}^{l^*}) &= \prod_{h=1}^{l^*} (2 \pi \sigma_h^2)^{-\frac{\gamma \sharp h_j}{2}} e^{- \frac{\gamma }{2 \sigma^2_h}  \sum_{i=1}^m 1(z_{ji} = h)  \left(x_{ji} - \mu_h\right)^2 } \nu_h^{\gamma \sharp h_j }.
\end{align}
Standard arguments imply that the analytic form of full conditional densities of parameters are
\begin{align}
  \label{eq:dens3}
  \mu_h \mid \text{rest} &\propto e^{- \frac{\gamma \sharp h_j + 1}{2 \sigma_h^2} \left( \mu_h^2 - 2 \mu_h \gamma \frac{\sum_{i=1}^m 1(z_{ji} = h) x_{ji}} {\gamma \sharp h_j + 1} \right)}, \nonumber\\
  \sigma_h^2 \mid \text{rest} &\propto \sigma_h^{2^{-\frac{\gamma \sharp h_j}{2}}} e^{- \frac{\gamma} {2 \sigma^2_h} \sum_{i=1}^m 1(z_{ji} = h)  \left(x_{ji} - \mu_h\right)^2}
  \sigma_h^{2^{-\frac{1}{2}}} e^{- \frac{\mu_h^2}{2 \sigma_h^2}} \sigma_h^{2^{-a_{\sigma} - 1}} e^{-\frac{b_{\sigma}}{\sigma_h^2}},\nonumber\\
  V_h \mid \text{rest} &\propto V_h^{\gamma  \sum_{i = 1}^{m} 1(z_i = h)} (1 - V_h)^{\gamma \sum_{i = 1}^{m} 1(z_{ji} > h)} (1 - V_h)^{\alpha - 1},\nonumber\\
  \alpha \mid \text{rest} &\propto \alpha^{a_{\alpha} - 1} e^{- b_{\alpha} \alpha} \alpha^{l^*} \prod_{h=1}^{l^*} (1 - V_d)^{\alpha - 1}
\end{align}
for $h=1, \ldots, l^*$. Let
\begin{align}
  \label{eq:dens4}
  m_{jh} &= \frac{\gamma \sum_{i=1}^m 1(z_{ji} = h) x_{ji}}{\gamma \sharp h_j + 1}, \quad v_{jh} = \frac{\sigma_h^2}{\gamma \sharp h_j + 1}, \\
  a_{jh} &= \frac{\gamma \sharp h_j + 1}{2} + a_{\sigma}, \quad  b_{jh} = \frac{\gamma}{2} \sum_{i=1}^m 1(z_{ji} = h)  \left(x_{ji} - \mu_h\right)^2 + \frac{\mu_h^2}{2} + b_{\sigma}
\end{align}
for $h=1, \ldots, l^*$, then all full conditional densities are tractable in terms of standard distributions:
\begin{align}
  \label{eq:dens5}
  \mu_{jh} \mid \text{rest}  &\sim N(m_{jh}, v_{jh}), \quad \sigma_{jh}^2 \mid \text{rest}  \sim \text{Inverse-Gamma}(a_{jh}, b_{jh}), \nonumber\\
  V_{jh} \mid \text{rest}  &\sim \text{Beta}(1+ \gamma  \sum_{i = 1}^{m} 1(z_{ji} = h), \alpha + \gamma \sum_{i = 1}^{m} 1(z_{ji} > h)), \nonumber\\
  \alpha_{jh} \mid \text{rest}  &\sim \text{Gamma}(a_{\alpha} + l^*, b_{\alpha} - \sum_{h=1}^{l^*} \log (1 - V_{jh})).
\end{align}
Finally, posterior distribution of the latent variables is
\begin{align}
  z_{ji} \mid \text{rest} \sim \sum_{h=1}^{l^*} p_{jh} \delta_h, \quad p_{jh} = \frac{\nu_{jh} \Ncal(\mu_{jh}, \sigma_{jh}^2)}{\sum_{\tilde h = 1}^{l^*}\nu_{j \tilde h} \Ncal(\mu_{j \tilde h}, \sigma_{j \tilde h}^2)}, \quad (i = 1, \ldots, m),
\end{align}
where $\nu_{jh} = V_{jh} \prod_{l < h} (1 - V_{jl})$ and $\Ncal(m, v)$ is the Gaussian density with mean $m$ and variance $v$.

\section{Linear program}
\label{sec:linear-program}

\begin{align}
  & \underset{\ab, T_1, \ldots, T_k}{\text{minimize}} & &\sum_{j=1}^k \mathrm{trace}(T_j^T D_j) \nonumber\\
  & \text{subject to} &  \nonumber\\
  & & & 0 \leq a_i \leq 1,  \quad i = 1, \ldots, g, \nonumber\\
  & & & 0 \leq \left( T_j \right)_{uv} \leq 1,  \quad u = 1, \ldots, g, \quad v = 1, \ldots, s_j, \quad j = 1, \ldots, k, \nonumber\\
  &  & &\one^T \ab = 1, \nonumber\\
  &  & &T_j \one_{s_j} = \ab, \quad j = 1, \ldots, k, \nonumber\\
  &  & &T_j^T \one_{s} = \frac{\one_{s_j}}{s_j}, \quad j = 1, \ldots, k. \label{lpwasp}
\end{align}
This linear program can be solved using a variety of linear programming solvers in \texttt{Matlab} or \texttt{R}, including the algorithms of \citet{CutDou14}  and \citet{Srietal15}.

\bibliographystyle{Chicago}
\bibliography{papers}

\clearpage

\renewcommand\thesection{\arabic{section}}
\renewcommand\thesubsection{\thesection.\arabic{subsection}}
\renewcommand\thesubsubsection{\thesubsection.\arabic{subsubsection}}

\setcounter{section}{0}

\begin{center}
\textbf{\Large Supplementary Material for Scalable Bayes via Barycenter in Wasserstein Space}
\end{center}

\section{Technical Lemmas}
To show Theorem 3.1, we first introduce in Lemma \ref{wongshen} a generalized version of the concentration inequality in Theorem 1 of \citet{WonShe95}. The proof parallels the original proof in \citet{WonShe95}, with several adaptations for the $\inid$ setup. Let $Z_{ji}(\theta)=\log [p(Y_{ji}|\theta)/p(Y_{ji}|\theta_0)]$, and $\tilde Z_{ji}(\theta) = \max(Z_{ji}(\theta),-\tau)$ be the lower truncated version of $Z_{ji}(\theta)$ for some constant $\tau>0$ to be chosen later. Let $Z_j(\theta)=(Z_{j1}(\theta),\ldots,Z_{jm}(\theta))^T$ and $\tilde Z_j(\theta)=(\tilde Z_{j1}(\theta),\ldots,\tilde Z_{jm}(\theta))^T$.

\begin{lemma}\label{wslem4}
Let $c_{1\tau} = 2e^{-\tau/2}/(1-e^{-\tau/2})^2$. Then for any $\theta \in \Theta$,
\begin{equation}\label{wss1}
\frac{1}{m} \sum_{i=1}^m E \tilde Z_{ji}(\theta) \leq -(1-c_{1\tau})h_{mj}^2 (\theta,\theta_0).
\end{equation}
\end{lemma}

\noindent {\bf Proof of Lemma \ref{wslem4}:}\\
The proof is a simple adaptation of Lemma 2 and Lemma 4 in \citet{WonShe95}. First note that for every observation $Y_{ji}$ (which takes value $y_{ji}$), we can define the event $A_{ji}=\{y_{ji}:p(y_{ji})|\theta)/p(y_{ji}|\theta_0)<e^{-\tau}\}$. Their Lemma 2 implies that $P(A_{ji})\leq (1-e^{-\tau/2})^{-2} h^2(p_{ji}(\cdot|\theta),p_{ji}(\cdot|\theta_0))$, which further implies by simple averaging over $i=1,\ldots,m$ that
\begin{equation}\label{wss2}
\frac{1}{m} \sum_{i=1}^m P(A_{ji}) \leq (1-e^{-\tau/2})^{-2} h_{mj}^2 (\theta,\theta_0).
\end{equation}
Following the same derivation of their Lemma 4, we have for every individual $\tilde Z_{ji}$,
$$E\tilde Z_{ji}(\theta) \leq - h^2(p_{ji}(\cdot|\theta),p_{ji}(\cdot|\theta_0)) + 2e^{-\tau/2} P(A_{ji}).$$
Then a simple averaging over $i=1,\ldots,m$ together with \eqref{wss2} gives \eqref{wss1}. $\Box$

\begin{lemma}\label{wslem5}
Let $c_{2\tau} = (e^{\tau/2}-1-\tau/2)/(1-e^{-\tau/2})^2$. For any $t>0$, integer $\ell\geq 2$ and any $\theta\in \Theta$ that satisfies $h_{mj}(\theta,\theta_0)\leq r$,
\begin{align*}
& \frac{1}{m}\sum_{i=1}^m E_{P_{\theta_0}} \left|\frac{\tilde Z_{ji}(\theta)}{2\sqrt{2c_{2\tau}}r}\right|^{\ell} \leq \frac{\ell!}{2}\left(\frac{1}{\sqrt{2c_{2\tau}}r}\right)^{\ell-2}.
\end{align*}
\end{lemma}
\noindent {\bf Proof of Lemma \ref{wslem5}:}\\
Lemma 5 in \citet{WonShe95} is stated for every single observation $Y_{ji}$, so it still holds for individual $\tilde Z_{ji}(\theta)$. For all $i=1,\ldots,m$,
$$E_{P_{\theta_0}}\left[\exp\left(\left|\tilde Z_{ji}(\theta)/2\right|\right)-1-\left|\tilde Z_{ji}(\theta)/2\right| \right]\leq c_{2\tau} h^2(p_{ji}(\cdot|\theta),p_{ji}(\cdot|\theta_0)),$$
where $c_{2\tau}$ is as defined in the statement of the lemma. Averaging over $i=1,\ldots,m$ yields
$$\frac{1}{m}\sum_{i=1}^m E_{P_{\theta_0}}\left[\exp\left(\left|\tilde Z_{ji}(\theta)/2\right|\right)-1-\left|\tilde Z_{ji}(\theta)/2\right| \right]\leq c_{2\tau} h_{mj}^2(\theta,\theta_0) \leq c_{2\tau} r^2,$$
where the second inequality is from the condition $h_{mj}(\theta,\theta_0)\leq r$. Using $e^x-1-x\geq x^{\ell}/\ell!$ for all $x>0$, we have
$$\frac{1}{m}\sum_{i=1}^m E_{P_{\theta_0}}\left|\tilde Z_{ji}(\theta)\right|^{\ell}
\leq 2^{\ell} \ell! c_{2\tau} r.$$
Rearranging the terms and the conclusion follows. $\Box$

\begin{lemma}\label{wslem3}
Let $j\in \{1,\ldots,k\}$ be fixed. Suppose $\Xi$ is a subset of $\Theta$. Let $\tilde {\mathcal{Z}_j}(\Xi)=\left\{\tilde Z_j(\theta) , \theta\in \Xi \right\}.$ For any $u>0$,
\begin{align*}
H_{[]}\left(u,\tilde {\mathcal{Z}_j}(\Xi),\|\cdot\| \right)\leq H_{[]} \left(u/(2e^{\tau/2}),{\mathcal{P}_j}(\Xi),h_{mj} \right).
\end{align*}
\end{lemma}
\noindent {\bf Proof of Lemma \ref{wslem3}:}\\
The proof follows the argument in the proof of Lemma 3 in \citet{WonShe95}. We can derive that for each $i=1,\ldots,m$, for any $\theta_1,\theta_2\in \Xi$,
$$E_{P_{\theta_0}} \left[\tilde Z_{ji}(\theta_1)-\tilde Z_{ji}(\theta_2)\right]^2 \leq 4e^{\tau} h_{mj}^2(\theta_1,\theta_2).$$
Then averaging over $i=1,\ldots,m$ gives the relation between two norms
$$\left\|\tilde Z_{j}(\theta_1)-\tilde Z_{j}(\theta_2)\right\|\leq 2e^{\tau/2} h_{mj}(\theta_1,\theta_2),$$
which further implies the relation between the bracketing entropies. $\Box$

\begin{lemma}\label{vandergeer}
(\citet{GeeLed13} Theorem 8) Let $j\in \{1,\ldots,k\}$ be fixed. Suppose a class of functions $\mathcal{F}_j=\left\{\f(\bby) =(f_1(y_1),\ldots,f_m(y_m))^T, \bby=(y_1,\ldots,y_m)\in \otimes_{i=1}^m \Ycal_{ji} \right\}$ satisfies\\
\noindent (i) $\sup_{\f\in \mathcal{F}_j } \|\f\| \leq 1$;\\
\noindent (ii) For any integer $\ell\geq 2$, $\sup_{\f\in\mathcal{F}_j} |\f|_{\ell}^{\ell} \leq \ell! M^{\ell-2}/2$, for some constant $M>0$;\\
Then for any $t>0$,
\begin{align*}
P_{\theta_0} \left(\sup_{\f\in\mathcal{F}_j}\frac{1}{\sqrt{m}}\sum_{i=1}^m \left[f_{i}(Y_{ji})-E_{P_{\theta_0}}f_{i}(Y_{ji})\right] \geq \min_{S\in \mathbb{N}}R_S + \frac{36M(1+t)}{\sqrt{m}} + 24\sqrt{6t}\right) \leq 2e^{-t},
\end{align*}
where
\begin{align*}
R_S \equiv 2^{-S}\sqrt{m} + 14\sqrt{6} \sum_{s=0}^S 2^{-s} \sqrt{H_{[]}\left(2^{-s},\mathcal{F}_j,\|\cdot\|\right)} + \frac{36MH_{[]}\left(1,\mathcal{F}_j,\|\cdot\|\right)}{\sqrt{m}}.
\end{align*}
\end{lemma}
\noindent {\bf Proof of Lemma \ref{vandergeer}:}\\
The theorem we present here is slightly different from the original Theorem 8 in \citet{GeeLed13} in that we have used the generalized bracketing entropy in Definition A.2 in the manuscript. Although the original version is presented with the usual $L_2$-bracketing entropy for univariate functions, the whole ``chaining along a tree'' argument will still go through, if we replace the $\|\cdot\|$ and $|\cdot|_q$ norms and bracketing entropies in their proofs by our generalized versions to multivariate functions as in Definition A.2. $\Box$

\begin{lemma}\label{wongshen}
(Generalization of \citet{WonShe95} Theorem 1) Assume (A3) holds. Then for any $\delta >0$, there exist positive constants $q_1, q_2$ that depend on $D_1,D_2$, such that for all subsets $Y_{[j]}$ with $j=1,\ldots, k$ and all sufficiently large $m$,
\begin{align}
P_{\theta_0}^{(n)}\left(\sup_{h_{mj}(\theta,\theta_0)\geq \delta}\prod_{i=1}^m \frac{p(Y_{ji} \mid \theta)}{p(Y_{ji} \mid \theta_0)} \geq \exp(- q_1 m \delta^2)  \right) \leq 4\exp(-q_2 m\delta^2)
\end{align}
\end{lemma}

\noindent {\bf Proof of Lemma \ref{wongshen}:}\\
We consider the class of functions
$$\hat{\mathcal{Z}}_j(r)=\left\{\frac{\tilde Z_j(\theta)}{2\sqrt{2c_{2\tau}}r}:\theta\in \Theta \text{ satisfies } h_{mj}(\theta,\theta_0)\leq r\right\},$$
for a fixed $r>0$. This class is a rescaled version of
$$\tilde {\mathcal{Z}}_j(\{\theta\in \Theta: h_{mj}(\theta,\theta_0)\leq r\})=\left\{\tilde Z_j(\theta):\theta\in \Theta \text{ satisfies } h_{mj}(\theta,\theta_0)\leq r\right\}$$
as in Lemma \ref{wslem3}. By Lemma \ref{wslem5}, $\hat{\mathcal{Z}}_j(r)$ satisfies Conditions (i) and (ii) in Lemma \ref{vandergeer} with the constant $M=1/(\sqrt{2c_{2\tau}}r)$. Therefore the concentration inequality in Lemma \ref{vandergeer} holds for $\hat{\mathcal{Z}}_j(r)$.

We first simplify the term $R_S$ in the inequality. $R_S$ involves the $L_2$-bracketing entropy of the class $\hat{\mathcal{Z}}_j(r)$. Since $H_{[]}\left(u,\hat{\mathcal{Z}}_j(r),\|\cdot\|\right)$ is nonincreasing in $u$, we have
\begin{align}\label{entbound1}
& \sum_{s=0}^S 2^{-s} \sqrt{H_{[]}\left(2^{-s},\hat{\mathcal{Z}}_j(r),\|\cdot\|\right)}
\leq  2\sum_{s=0}^S \int_{2^{-(s+1)}}^{2^{-s}} \sqrt{H_{[]}\left(u,\hat{\mathcal{Z}}_j(r),\|\cdot\|\right)} du \nonumber \\
&=  2 \int_{2^{-(S+1)}}^1 \sqrt{H_{[]}\left(u,\hat{\mathcal{Z}}_j(r),\|\cdot\|\right)} du \nonumber \\
&\overset{(i)}{=} 2 \int_{2^{-(S+1)}}^1 \sqrt{H_{[]}\left(2\sqrt{2c_{2\tau}}ru,\tilde{\mathcal{Z}}_j(\{\theta\in \Theta: h_{mj}(\theta,\theta_0)\leq r\}),\|\cdot\|\right)} du \nonumber \\
&= \frac{1}{\sqrt{2c_{2\tau}}r} \int_{2^{-S}\sqrt{2c_{2\tau}}r}^{2\sqrt{2c_{2\tau}}r} \sqrt{H_{[]}\left(u,\tilde{\mathcal{Z}}_j(\{\theta\in \Theta: h_{mj}(\theta,\theta_0)\leq r\}),\|\cdot\|\right)} du \nonumber \\
&\overset{(ii)}{\leq}  \frac{1}{\sqrt{2c_{2\tau}}r} \int_{2^{-S}\sqrt{2c_{2\tau}}r}^{2\sqrt{2c_{2\tau}}r} \sqrt{H_{[]}\left(u/(2e^{\tau/2}),{\mathcal{P}}_j(\{\theta\in \Theta: h_{mj}(\theta,\theta_0)\leq r\}),h_{mj}\right)} du \nonumber \\
&=  \frac{\sqrt{2e^{\tau}}}{\sqrt{c_{2\tau}}r} \int_{2^{-(S+1)}\sqrt{2c_{2\tau}e^{-\tau}}r}^{\sqrt{2c_{2\tau}e^{-\tau}}r} \sqrt{H_{[]}\left(u,{\mathcal{P}}_j(\{\theta\in \Theta: h_{mj}(\theta,\theta_0)\leq r\}),h_{mj}\right)} du,
\end{align}
where (i) follows from the scaling relation between $\hat{\mathcal{Z}}_j(r)$ and $\tilde {\mathcal{Z}}_j(\{\theta\in \Theta: h_{mj}(\theta,\theta_0)\leq r\})$, and (ii) follows from Lemma \ref{wslem3}. We choose integer $S\geq 1$ such that $2^{-(S+2)}\leq \sqrt{2c_{2\tau}e^{-\tau}}r/2^{12} \leq 2^{-(S+1)}$. It is possible to do so because we only need to consider $r\leq \sqrt{2}$ (since the $h_{mj}$ distance is upper bounded by $\sqrt{2}$), $c_{2\tau}e^{-\tau}\leq 1/2$ for all $\tau\geq 0$, and it is guaranteed that $\sqrt{2c_{2\tau}e^{-\tau}}r/2^{12}\leq \sqrt{2}/2^{12} < 1/4$.

Now we can apply (A3) to \eqref{entbound1} and obtain that uniformly over all $j=1,\ldots,k$,
\begin{align}\label{RS2}
& \sum_{s=0}^S 2^{-s} \sqrt{H_{[]}\left(2^{-s},\hat{\mathcal{Z}}_j(r),\|\cdot\|\right)}
\leq \frac{\sqrt{2e^{\tau}}}{\sqrt{c_{2\tau}}r} \int_{\left(\sqrt{2c_{2\tau}e^{-\tau}}r\right)^2/2^{12}}^{\sqrt{2c_{2\tau}e^{-\tau}}r}  \sqrt{\Psi(u,r)}du \nonumber \\
&\leq \frac{\sqrt{2e^{\tau}}}{\sqrt{c_{2\tau}}r}\cdot D_2\sqrt{m}\left(\frac{\sqrt{2c_{2\tau}e^{-\tau}}}{D_1}r\right)^2
= \frac{2D_2\sqrt{2c_{2\tau}e^{-\tau}}}{D_1^2}\sqrt{m}r.
\end{align}
Furthermore, since (A3) says $\Psi(u,r)$ is nonincreasing in $u$, we can also derive that
\begin{align}\label{RS3}
& H_{[]}\left(1,\hat{\mathcal{Z}}_j(r),\|\cdot\|\right) = H_{[]}\left(2\sqrt{2c_{2\tau}}r,\tilde{\mathcal{Z}}_j(\{\theta\in \Theta: h_{mj}(\theta,\theta_0)\leq r\}),\|\cdot\|\right) \nonumber \\
&\leq H_{[]}\left(\sqrt{2c_{2\tau}e^{-\tau}}r, \mathcal{P}_j(\{\theta\in \Theta: h_{mj}(\theta,\theta_0)\leq r\}),h_{mj}\right)
\leq \Psi\left(\sqrt{2c_{2\tau}e^{-\tau}}r,r\right) \nonumber \\
& \leq \left[\frac{1}{\sqrt{2c_{2\tau}e^{-\tau}}r- \left(\sqrt{2c_{2\tau}e^{-\tau}}r\right)^2/2^{12}}
\int_{\left(\sqrt{2c_{2\tau}e^{-\tau}}r\right)^2/2^{12}}^{\sqrt{2c_{2\tau}e^{-\tau}}r} \sqrt{\Psi(u,r)} du \right]^2 \nonumber\\
&\leq  \frac{2 D_2^2 c_{2\tau}e^{-\tau} m r^2}{D_1^4\left(1-\sqrt{2c_{2\tau}e^{-\tau}}r/2^{12}\right)^2}
\leq \frac{8D_2^2 c_{2\tau}e^{-\tau} m r^2}{D_1^4},
\end{align}
where in the last step, we used the fact that $\sqrt{2c_{2\tau}e^{-\tau}}r/2^{12} <1/2$.

By our choice $2^{-(S+2)}\leq \sqrt{2c_{2\tau}e^{-\tau}}r/2^{12}$, it follows from \eqref{RS2} and \eqref{RS3} that
\begin{align}\label{RSfinal}
\min_{S\in \mathbb{N}} R_S & \leq \frac{\sqrt{2c_{2\tau}e^{-\tau}}}{2^{10}}\cdot \sqrt{m}r + 14\sqrt{6} \cdot \frac{2D_2\sqrt{2c_{2\tau}e^{-\tau}}}{D_1^2}\sqrt{m}r + \frac{36}{\sqrt{2c_{2\tau}m}r}\cdot \frac{8D_2^2 c_{2\tau}e^{-\tau} m r^2}{D_1^4}\nonumber \\
& \leq \left[\frac{\sqrt{2}}{2^{10}} + 56\sqrt{3}\frac{D_2}{D_1^2} + 144\sqrt{2}\frac{D_2^2}{D_1^4}e^{-\tau/2}\right] \sqrt{c_{2\tau}e^{-\tau}} \sqrt{m} r.
\end{align}
In the inequality of Lemma \ref{vandergeer}, we let $t=c_{3\tau} m r^2$ with integer $\ell\geq 1$ and constant $c_{3\tau}$ to be chosen later, then with probability at least $1-2e^{- c_{3\tau} m r^2}$, the empirical process
$$\sup_{\f\in \hat{\mathcal{Z}}_j(r)}\frac{1}{\sqrt{m}}\sum_{i=1}^m \left[f_{i}(Y_{ji})-E_{P_{\theta_0}}f_{i}(Y_{ji})\right]$$
will not exceed the following upper bound
\begin{align*}
& \min_{S\in \mathbb{N}} R_S  +  \frac{36M(1+t)}{\sqrt{m}} + 24\sqrt{6t} \\
& \leq \left[\frac{\sqrt{2}}{2^{10}} + 56\sqrt{3}\frac{D_2}{D_1^2} + 144\sqrt{2}\frac{D_2^2}{D_1^4}e^{-\tau/2}\right] \sqrt{c_{2\tau}e^{-\tau}} \sqrt{m} r + \frac{36(1+ c_{3\tau} m r^2)}{\sqrt{2c_{2\tau}m}r} + 24\sqrt{6c_{3\tau}m}r \\
& = c_{4\tau}\sqrt{m}r + \frac{36}{\sqrt{2c_{2\tau}m}r},
\end{align*}
where $c_{4\tau} = \left[\frac{\sqrt{2}}{2^{10}} + 56\sqrt{3}\frac{D_2}{D_1^2} + 144\sqrt{2}\frac{D_2^2}{D_1^4}e^{-\tau/2}\right] \sqrt{c_{2\tau}e^{-\tau}} + \frac{36c_{3\tau}}{\sqrt{2c_{2\tau}}} +  24\sqrt{6c_{3\tau}}.$
As a result, the empirical process on the class $\tilde{\mathcal{Z}}_j(\{\theta\in \Theta: h_{mj}(\theta,\theta_0)\leq r\})$ satisfies that with probability at least $1-2e^{-c_{3\tau} m r^2}$,
\begin{align}\label{conineq1}
&\sup_{\theta\in \Theta: h_{mj}(\theta,\theta_0)\leq r}\frac{1}{\sqrt{m}}\sum_{i=1}^m \left[\tilde Z_{ji}(\theta)-E_{P_{\theta_0}}\tilde Z_{ji}(\theta)\right] \leq 2\sqrt{2c_{2\tau}}r\cdot \left(c_{4\tau}\sqrt{m}r + \frac{36}{\sqrt{2c_{2\tau}m}r}\right) \nonumber \\
&\leq 2\sqrt{2c_{2\tau}}c_{4\tau} \cdot \sqrt{m}r^2 + \frac{72}{\sqrt{m}}.
\end{align}
On the set $\left\{\theta\in \Theta: r\leq h_{mj}(\theta,\theta_0)\leq 2r\right\}$, we have $h_{mj}^2(\theta,\theta_0)\geq r^2$. By Lemma \ref{wslem4}, it follows that
$$\frac{1}{m} \sum_{i=1}^m E \tilde Z_{ji}(\theta) \leq -(1-c_{1\tau})r^2.$$
This together with \eqref{conineq1} implies that with probability at least $1-2e^{-c_{3\tau} m r^2}$,
\begin{align*}
& \sup_{\theta\in \Theta: r\leq h_{mj}(\theta,\theta_0)\leq 2r} \frac{1}{m}\sum_{i=1}^m \tilde Z_{ji}(\theta)
\leq -\left(1-c_{1\tau}- 8\sqrt{2c_{2\tau}}c_{4\tau}\right) r^2 + \frac{72}{m}.
\end{align*}
Now we choose $\tau$ such that $e^{-\tau/2}=1/32$. Then $c_{1\tau} < 0.07$, $29 <c_{2\tau}<30$. Set $c_{3\tau}=1/2^{30}$. By (A3), $D_2\leq D_1^2/2^{12}$. So
\begin{align*}
&8\sqrt{2c_{2\tau}}c_{4\tau} \leq  8\sqrt{60} \cdot \left\{\left[\frac{\sqrt{2}}{2^{10}} +\frac{56\sqrt{3}}{2^{12}} + \frac{144\sqrt{2}}{16\cdot 2^{24}}\right] \sqrt{\frac{30}{2^{10}}} + \frac{36}{\sqrt{58}\cdot 2^{30}} +  24\sqrt{\frac{6}{2^{30}}}\right\} < 0.377,
\end{align*}
which implies that $1-c_{1\tau}- 8\sqrt{2c_{2\tau}}c_{4\tau}> 0.55$. Thus we have proved that for any $0<r\leq \sqrt{2}$, uniformly for all $j=1,\ldots,k$ and for all sufficiently large $m$,
\begin{align}\label{conineq2}
& P_{\theta_0}^{(n)}\left(\sup_{ \{\theta\in \Theta: r\leq h_{mj}(\theta,\theta_0)\leq 2r \}} \frac{1}{m}\sum_{i=1}^m Z_{ji}(\theta) \geq -0.55 r^2 + \frac{72}{m} \right) \nonumber \\
\leq{}& P_{\theta_0}^{(n)}\left(\sup_{ \{\theta\in \Theta: r\leq h_{mj}(\theta,\theta_0)\leq 2r \}} \frac{1}{m}\sum_{i=1}^m \tilde Z_{ji}(\theta) \geq -0.55r^2 + \frac{72}{m} \right) \leq 2\exp\left(- mr^2/2^{30}\right).
\end{align}
Finally for a given $\delta$, we set $r=2^{\ell} \delta$ for integers $\ell\geq 0$, $m > 72/0.05/\delta = 1440/\delta$, and let $L$ be the smallest integer such that $2^L \delta^2 >2$. It follows that
\begin{align*}
& P_{\theta_0}^{(n)}\left(\sup_{ \{ \theta\in \Theta: h_{mj}(\theta,\theta_0)\geq\delta \} }  \prod_{i=1}^m \frac{p(Y_{ji} \mid \theta)}{p(Y_{ji} \mid \theta_0)} \geq \exp(-0.5 m \delta^2) \right) \\
\leq{}& P_{\theta_0}^{(n)}\left(\sup_{ \{\theta\in \Theta: h_{mj}(\theta,\theta_0)\geq\delta \} } \frac{1}{m}\sum_{i=1}^m  Z_{ji}(\theta) \geq -0.55\delta^2 + \frac{72}{m} \right) \\
\leq{}& \sum_{\ell=0}^L P_{\theta_0}^{(n)}\left(\sup_{ \{ \theta\in \Theta: 2^{\ell}\delta \leq h_{mj}(\theta,\theta_0)\leq 2^{\ell+1}\delta \}} \frac{1}{m}\sum_{i=1}^m  Z_{ji}(\theta) \geq -0.55(2^{\ell}\delta)^2 + \frac{72}{m} \right) \\
\leq{}& \sum_{\ell=0}^L 2 \exp\left(- 2^{2\ell}m\delta^2/2^{30}\right) \leq 4  \exp\left(- m\delta^2/2^{30}\right).
\end{align*}
We set $q_1=0.5$, $q_2=1/2^{30}$ and complete the proof. $\Box$

\begin{lemma}\label{gv2000denom}
Assume (A3) holds. Then for any $\delta>0$, there exist positive constants $r_1, r_2$ that depend on $g_1,g_2,\kappa,c_{\pi}$, such that for every subset $Y_{[j]}$ ($j=1,\ldots, k$), for any $t\geq \epsilon_m^{2\alpha} $,
\begin{align*}
P_{\theta_0}^{(n)}\left(\int_{\Theta}\prod_{i=1}^m \frac{p(Y_{ji}|\theta)^{\gamma}}{p(Y_{ji}|\theta_0)^{\gamma}} \Pi(d \theta)
\leq \exp(- r_1 n t)  \right) \leq \exp(-r_2 m t).
\end{align*}
\end{lemma}

\noindent {\bf Proof of Lemma \ref{gv2000denom}:}\\
Define the event $\Theta_{\epsilon_m}$ as
$$\Theta_{\epsilon_m} = \left\{\theta\in \Theta: \frac{1}{m}\sum_{i=1}^m E_{P_{\theta_0}}\exp\left(\log_+\frac{p(Y_{ji}|\theta_0)}{p(Y_{ji}|\theta)} \right) -1 \leq \epsilon_m^{2\alpha} \right\}$$
then (A4) can be written as $\Pi(\Theta_{\epsilon_m})\geq \exp(-c_{\pi}n\epsilon_m^{2\alpha})$. For $A\subseteq \Theta$, let $\Pi_{\epsilon_m}(A) = \Pi(A\cap \Theta_{\epsilon_m})/\Pi(\Theta_{\epsilon_m})$ be the prior measure $\Pi$ restricted measure to the set $\Theta_{\epsilon_m}$. For the left-hand side of the conclusion, we have
\begin{align}\label{denom1}
& P_{\theta_0}^{(n)}\left(\int_{\Theta}\prod_{i=1}^m \frac{p(Y_{ji}|\theta)^{\gamma}}{p(Y_{ji}|\theta_0)^{\gamma}} \Pi(d \theta) \leq \exp(- r_1 n t)  \right) \nonumber \\
\overset{(i)}{\leq} {}& P_{\theta_0}^{(n)}\left(\int_{\Theta_{\epsilon_m}}\prod_{i=1}^m \frac{p(Y_{ji}|\theta)^{\gamma}}{p(Y_{ji}|\theta_0)^{\gamma}} \Pi(d \theta) \leq \exp(- r_1 n t)  \right) \nonumber\\
\overset{}{\leq} {}& P_{\theta_0}^{(n)}\left(\Pi(\Theta_{\epsilon_m})\cdot \int_{\Theta_{\epsilon_m}}\prod_{i=1}^m \frac{p(Y_{ji}|\theta)^{\gamma}}{p(Y_{ji}|\theta_0)^{\gamma}} \Pi_{\epsilon_m}(d \theta) \leq \exp(- r_1 n t)  \right) \nonumber\\
\overset{(ii)}{\leq} {}& P_{\theta_0}^{(n)}\left(\int_{\Theta_{\epsilon_m}}\prod_{i=1}^m \frac{p(Y_{ji}|\theta)^{\gamma}}{p(Y_{ji}|\theta_0)^{\gamma}} \Pi_{\epsilon_m}(d \theta) \leq \exp(- r_1 n t + c_{\pi}n\epsilon_m^{2\alpha})  \right) \nonumber\\
\overset{(iii)}{\leq} {}& P_{\theta_0}^{(n)}\left( \sum_{i=1}^m \int_{\Theta_{\epsilon_m}}\log \frac{p(Y_{ji}|\theta_0)}{p(Y_{ji}|\theta)} \Pi_{\epsilon_m}(d \theta) \geq  r_1 g_1 m t - c_{\pi} g_2 m \epsilon_m^{2\alpha}  \right) \nonumber\\
\overset{(iv)}{\leq} {}&P_{\theta_0}^{(n)}\Bigg( \sum_{i=1}^m\left[\int_{\Theta_{\epsilon_m}}\log \frac{p(Y_{ji}|\theta_0)}{p(Y_{ji}|\theta)}\Pi_{\epsilon_m}(d \theta) - E_{P_{\theta_0}}\int_{\Theta_{\epsilon_m}}\log \frac{p(Y_{ji}|\theta_0)}{p(Y_{ji}|\theta)}\Pi_{\epsilon_m}(d \theta)  \right] \nonumber \\
& \geq r_1 g_1 m t - (c_{\pi}g_2  + \kappa^{-1}) m \epsilon_m^{2\alpha}  \Bigg).
\end{align}
In the derivation above, (i) follows from making the region of the integral smaller; (ii) is from (A4); (iii) is an application of Jensen's inequality and uses the fact that $g_1 \gamma m \leq n \leq g_2 \gamma m$; (iv)
follows by using Fubini's theorem, the inequality $x\leq (e^{\kappa x}-1)/\kappa$ for $x \geq 0$, and (A4):
\begin{align*}
&\sum_{i=1}^m E_{P_{\theta_0}}\int_{\Theta_{\epsilon_m}}\log \frac{p(Y_{ji}|\theta_0)}{p(Y_{ji}|\theta)}\Pi_{\epsilon_m}(d \theta) \leq \int_{\Theta_{\epsilon_m}}\sum_{i=1}^m E_{P_{\theta_0}}\log_+ \frac{p(Y_{ji}|\theta_0)}{p(Y_{ji}|\theta)}\Pi_{\epsilon_m}(d \theta) \\
& \leq \int_{\Theta_{\epsilon_m}} \frac{1}{\kappa}\sum_{i=1}^m \left[E_{P_{\theta_0}}\exp\left(\kappa \log_+ \frac{p(Y_{ji}|\theta_0)}{p(Y_{ji}|\theta)}\right)-1\right]\Pi_{\epsilon_m}(d \theta) \leq \kappa^{-1} m \epsilon_m^{2\alpha}  .
\end{align*}
Now in \eqref{denom1}, if we choose $r_1$ large enough and let $Z_i= \int_{\Theta_{\epsilon_m}}\log \frac{p(Y_{ji}|\theta_0)}{p(Y_{ji}|\theta)}\Pi_{\epsilon_m}(d \theta) $, we can use the Bernstein's inequality (Corollary 2.10 in \citet{Mas03}) to control the large deviation for the sum of $Z_i$'s. By inspecting the conditions for this inequality, we can see that for any integer $\ell \geq 2$, by the inequality $(\kappa x)^\ell/\ell ! \leq e^{\kappa x}-1$, (A4), and the Jensen's inequality,
$$\sum_{i=1}^m E_{P_{\theta_0}} \left[(Z_i)_+^\ell\right ] \leq \frac{\ell!}{\kappa^\ell}\sum_{i=1}^m \left[\exp\left(\kappa E_{P_{\theta_0}}(Z_i)_+ \right)-1\right] \leq \frac{\ell!}{\kappa^\ell} m\epsilon_m^{2\alpha} .$$
Therefore in the Corollary 2.10, we can set $c=\kappa^{-1}$, $v=2\kappa^{-2} m\epsilon_m^{2\alpha}$, $x=r_1 g_1 m t-(c_{\pi} g_2 +\kappa^{-1}) m \epsilon_m^{2\alpha}$. If we choose $r_1>(c_{\pi} g_2 +3\kappa^{-1})/ g_1$, then since $t \geq \epsilon_m^{2\alpha}$, we have $x >2\kappa^{-1}mt \geq 2\kappa^{-1} m\epsilon_m^{2\alpha}$. Hence it follows that $v/c = 2\kappa^{-1} m\epsilon_m^{2\alpha} < x$. We apply the Bernstein's inequality to \eqref{denom1} and obtain that
\begin{align*}
& P_{\theta_0}^{(n)}\left( \sum_{i=1}^m\left[Z_i - E_{P_{\theta_0}}Z_i  \right]  \geq r_1 g_1 m t - (c_{\pi} g_2 + \kappa^{-1}) m \epsilon_m^{2\alpha}  \right) \\
\leq{}& \exp\left[-\frac{x^2}{2\left(v+cx\right)}\right]\leq \exp\left(-\frac{x}{4c}\right) \leq  \exp\left(-\frac{\kappa x}{4}\right) \leq \exp\left(-\frac{mt}{2}\right).
\end{align*}
We set $r_2=1/2$ and complete the proof. $\Box$

\begin{lemma}\label{w2ineq}
Let $\overline \nu$ denote the $W_2$ barycenter of $N$ measures $\nu_1,\ldots,\nu_N$ in $\Pcal_2(\Theta)$. Then for any $\theta_0\in \Theta$,
\begin{align*}
  W_2(\overline\nu,\delta_{\theta_0})\leq \frac{1}{N} \sum_{j=1}^N W_2(\nu_j,\delta_{\theta_0}).
\end{align*}
\end{lemma}

\noindent {\bf Proof of Lemma \ref{w2ineq}:}\\

Theorem 4.1 in \citet{AguCar11} shows that $\overline \nu = \overline T\sharp\nu_1$, where
\begin{align*}
  \overline T (\theta) = \frac{1}{N} \sum_{j=1}^{N}  T^{1}_{j}(\theta),
\end{align*}
for any $\theta\in \Theta$ and $T^{1}_{j}$ is the optimal transport map that pushes $\nu_1$ forward to $\nu_j$, i.e. $\nu_j = T^1_j\sharp\nu_1$ and $T^1_1$ is the identity operator. We use the property of the $\rho$ metric in (A5) and obtain that
\begin{align}\label{w2ineq1}
& W_2^2(\overline \nu,\delta_{\theta_0})  = \int_{\Theta} \rho^2\left( \frac{1}{N} \sum_{j=1}^{N}  T^{1}_{j}(\theta),\theta_0 \right )  \nu_1(d \theta) \leq \int_{\Theta} \left[\frac{1}{N}\sum_{j=1}^{N} \rho\left( T^{1}_{j}(\theta),\theta_0 \right )\right]^2  \nu_1(d \theta) \nonumber \\
& = \frac{1}{N^2} \sum_{j=1}^{N} \int_{\Theta} \rho^2 \left( T^{1}_{j}(\theta),\theta_0 \right )\nu_1(d \theta)  + \frac{1}{N^2} \sum_{j\neq l} \int_{\Theta} \rho\left( T^{1}_{j}(\theta),\theta_0 \right )  \rho\left( T^{1}_{l}(\theta),\theta_0 \right ) \nu_1(d \theta).
\end{align}
For each term in the second summation, we apply the Cauchy-Schwartz inequality and have
\begin{align*}
& \int_{\Theta} \rho\left( T^{1}_{j}(\theta),\theta_0 \right )  \rho\left( T^{1}_{l}(\theta),\theta_0 \right ) \nu_1(d \theta)\\
& \leq \sqrt{\int_{\Theta} \rho^2\left( T^{1}_{j}(\theta),\theta_0 \right )\nu_1(d \theta)} \cdot \sqrt{ \int_{\Theta}\rho^2\left( T^{1}_{l}(\theta),\theta_0 \right ) \nu_1(d \theta)} = W_2(\nu_j,\delta_{\theta_0})\cdot W_2(\nu_l,\delta_{\theta_0}).
\end{align*}
Therefore in \eqref{w2ineq1},
\begin{align*}
&W_2^2(\overline \nu,\delta_{\theta_0})\leq \frac{1}{N^2} \sum_{j=1}^{N}  W_2^2 (\nu_j,\delta_{\theta_0}) + \frac{1}{N^2} \sum_{j\neq l} W_2(\nu_j,\delta_{\theta_0})\cdot W_2(\nu_l,\delta_{\theta_0}) \\
& = \left[ \frac{1}{N}  \sum_{j=1}^{N}  W_2(\nu_j,\delta_{\theta_0}) \right]^2,
\end{align*}
and hence the conclusion follows. $\Box$

\section{Experiments}
\label{sec:experiments}

\subsection{Simulated data: finite mixture of Gaussians}
\label{sec:simul-data:-finite}

Consider the set of $L$ mixture of Gaussians. If $\yb_i \in \RR^p$ is the $i$th observation ($i=1, \ldots, n$) sampled from a mixture of $L$ Gaussians, then 
\begin{align}
  p(\yb_i  \mid \theta) = \sum_{l=1}^L \pi_l \Ncal_p(\yb \mid \mub_l, \Sigma_l), \label{mix-1}  
\end{align}
where $\pib = (\pi_1, \ldots, \pi_L)$ lies in the $(L-1)$-simplex, $\mu_l$ and $\Sigma_l$ ($l=1, \ldots, L$) are the mean and covariance parameters of a $p$-variate Gaussian distribution, and $\theta = \{\pib, \mub_1, \ldots, \mub_L, \Sigma_1, \ldots, \Sigma_L\}$. We cluster $\yb_1, \ldots, \yb_n$ into $L$ clusters using K-Means clustering and randomly split the members of every cluster into $k$ subsets. This ensures that full-data are split into $k$ subsets such that the mixture proportions are represented in every subset. Let $\yb_{j1}, \ldots, \yb_{jm}$ represent the data on subset $j$ ($j=1, \ldots, k$). The hierarchical model for the data on subset $j$ is
\begin{align}
  \label{eq:mix-2}
  &\yb_{ji} \mid z_{ji}, \theta \sim \Ncal_p(\mub_{z_{ji}}, \Sigma_{z_{ji}}), \quad z_{ji} \sim \sum_{h=1}^{L} \pi_h \delta_{h},  \\
  &\pib \sim \text{Dirichlet}(L^{-1}, \ldots, L^{-1}), \quad \mub_h \mid \Sigma_h \sim \Ncal_p(\zero, 100 \Sigma_h), \quad \Sigma_h \sim \text{Inverse-Wishart}(2, 4 I_p), \quad h = 1, \ldots, L, \nonumber
\end{align}
where 2 is the prior degrees of freedom and $4I_p$ is the scale matrix of the Inverse-Wishart distribution.

The posterior distribution of $\theta$ after stochastic approximation is derived using standard arguments for finite mixture of Gaussians. The likelihood given $\yb_{j1}, \ldots, \yb_{jm}$ and latent variables $z_{j1}, \ldots, z_{jm}$ is 
\begin{align}
  \label{eq:mix-3}
  L_j(\{\mub_h\}_{h=1}^{L}, \{\Sigma_h\}_{h=1}^{L}, \{\pi_h\}_{h=1}^{L})
  &= \prod_{h=1}^{L} (2 \pi^p |\Sigma_l|)^{-\frac{\sharp h_j} {2}} e^ {- \frac{1}{2}   \sum_{i=1}^m 1(z_{ji} = h) \left(\yb_{ji} - \mub_h\right)^T \Sigma_h^{-1} \left(\yb_{ji} - \mub_h\right)} \pi_h^{\sharp h_j},
\end{align}
where $1(z_{ji} = h)$ is 1 if $z_{ji} = h$ and 0 otherwise and $\sharp h_j = \sum_{i=1}^m 1(z_{ji} = h)$. For stochastic approximation, we raise $L_j$ in \eqref{eq:mix-3} to the power $\gamma$ and obtain
\begin{align}
  \label{eq:mix-4}
  L_j^{\gamma}(\{\mub_h\}_{h=1}^{L}, \{\Sigma_h\}_{h=1}^{L}, \{\pi_h\}_{h=1}^{L})
  &= \prod_{h=1}^{L} (2 \pi^p |\Sigma_l|)^{-\frac{\gamma \sharp h_j} {2}} e^ {- \frac{\gamma}{2}   \sum_{i=1}^m 1(z_{ji} = h) \left(\yb_{ji} - \mub_h\right)^T \Sigma_h^{-1} \left(\yb_{ji} - \mub_h\right)} \pi_h^{\gamma \sharp h_j}.
\end{align}
If we use $L_j^{\gamma}$ as the likelihood in \eqref{mix-1}, then the prior for $\theta$ in \eqref{mix-1} and simple extensions of standard arguments for finite mixture of Gaussians imply that the analytic form of full conditional densities of the parameters are as follows. Define 
\begin{align}
  \label{eq:mix-5}
  h_j = \{i : 1(z_{ji} = h) = 1, \; i = 1, \ldots, m\},  \quad \overline \yb_{h_j} = \frac{1}{\sharp h_j} \sum_{i \in h_j} \yb_{ji}  \quad (h = 1, \ldots, L),
\end{align}
and the complete conditionals of the parameters are
\begin{align}  
  &\pib_j  \mid \text{rest} \sim \text{Dirichlet} \left(\gamma \sum_{i=1}^m 1(z_{ji} = 1) + L^{-1}, \ldots, \gamma \sum_{i=1}^m 1(z_{ji} = L) + L^{-1} \right) \nonumber \\
  &\mub_{jh} \mid \text{rest} \sim \text{Normal} \left\{\frac{\gamma \sharp h_j} { 0.01 + \gamma \sharp h_j}  \overline \yb_{h_j}, \frac{1} { 0.01 + \gamma \sharp h_j} \Sigma_h \right\}, \nonumber \\
  &\Sigma_{jh} \mid \text{rest} \sim \text{Inverse-Wishart} \left\{ \gamma \sharp h_j + 3,
    \sum_{i \in h_j} (\yb_{ji} - \overline \yb_{h_j})(\yb_{ji} - \overline \yb_{h_j})^T + \frac{0.01 \cdot \gamma \sharp h_j} {0.01 + \gamma \sharp h_j} \overline \yb_{h_j} \overline \yb_{h_j}^T + 4 I_p \right\} \nonumber\\
  &z_{ji} \mid \text{rest} \sim \sum_{h=1}^{L} p_{jh} \delta_h, \quad p_{jh} = \frac{\pi_{jh} \Ncal_p(\yb_{ji} \mid \mub_{jh}, \Sigma_{jh})}{\sum_{\tilde h = 1}^{L} \pi_{j\tilde h} \Ncal_p(\yb_{ji} \mid \mub_{j \tilde h }, \Sigma_{j \tilde h})}    \label{eq:mix-6}
\end{align}
for $h=1, \ldots, L$ and $i = 1, \ldots, m$; see Chapter 9 in \citet{Bis06} for details.

All full conditionals are analytically tractable in terms of standard distributions. The Gibbs sampler iterates between the following four steps:
\begin{enumerate}
\item Sample $\pib_j$ from Dirichlet distribution in \eqref{eq:mix-6}.
\item Sample $\mub_{jh}$ from Normal distribution in \eqref{eq:mix-6} for $h=1, \ldots, L$.
\item Sample $\Sigma_{jh}$ from Inverse-Wishart distribution in \eqref{eq:mix-6} for $h=1, \ldots, L$.
\item Sample $z_{ji}$ from categorical distribution in \eqref{eq:mix-6} for $i = 1, \ldots, m$.  
\end{enumerate}

\subsection{Simulated data analysis: Linear mixed effects model}
\label{stoc-lme}

The conditional densities of $\betab$ and $\Sigma$ in two steps. First, the sampling model of the linear mixed effects model implies that the likelihood of $i$th observation in $j$th subset is
\begin{align}
  \label{eq:lme1}
  L_{ji} = \int_{\RR^q} \Ncal_{n_i}(\yb_{ji} \mid X_i \betab + Z_i \ub_i, \tau^2 I_{n_i}) \Ncal_q(\ub_i \mid \zero, \Sigma) d \ub_i =
  \Ncal_{n_i}(\yb_{ji} \mid X_i \betab, Z_i \Sigma Z_i^T + \tau^2 I_{n_i}).
\end{align}
This implies that likelihood of $\betab$ and $\Sigma$ after stochastic approximation is 
\begin{align}
  \label{eq:lme2}
  L^{\gamma}_{j} = \prod_{i=1}^m \left\{ L_{ji} \right\}^{\gamma}= \prod_{i=1}^m \left\{ \Ncal_{n_i} (\yb_{ji} \mid X_i \betab, Z_i \Sigma Z_i^T + \tau^2 I_{n_i}) \right\}^{\gamma}.  
\end{align}
Second, the $j$th subset posterior distribution of $\betab$ and $\Sigma$ after stochastic approximation is calculated using $L^{\gamma}_{j}$ as the likelihood and the same priors for $\betab$ and $\Sigma$ as in the sampling model of  the linear mixed effects model. Instead of finding the analytic form of the posterior density, we use the \texttt{increment\_log\_prob} function in Stan \citep{Sta14} to specify that the likelihood of $\yb_{ji}$ as $\left\{ L_{ji} \right\}^{\gamma}$ \eqref{eq:lme2} and use the default priors of $\betab$ and $\Sigma$ in Stan to obtain samples of $(\betab, \Sigma)$ from the $j$th subset posterior after stochastic approximation. For
automatic differentiation variational inference (ADVI; \citet{Kucetal15}), we used the \texttt{vb} function in Stan with default options and used the same model file that we used for sampling from the full data posterior distribution. Table \ref{tbl:acc_fixef} describes the accuracies of ADVI, CMC, SA, SDP, and WASP in approximating the full data marginal posterior distributions of fixed effects in the simulation.

\begin{table}[t]
  \caption{Accuracies of the approximate posteriors for $\betab$ in linear mixed effects model simulation. The accuracies are averaged over 10 simulation replications and across all elements of $\betab$. Monte Carlo errors are in parenthesis. ADVI, automatic differentiation variational inference; SA, streamlined algorithm; SGLD, stochastic gradient Langevin dynamics with batch size in parenthesis; CMC, consensus Monte Carlo; SDP, semiparametric density product; WASP, Wasserstein posterior
  }
  \label{tbl:acc_fixef}
  \centering
  {\tiny
    \begin{tabular}{rcccc}
      \hline
      & \multicolumn{2}{c}{$p = 4$} & \multicolumn{2}{c} {$p = 80$} \\ 
      \hline
      ADVI & \multicolumn{2}{c}{0.22 (0.26)} & \multicolumn{2}{c}{0.46 (0.13)} \\ 
      SA & \multicolumn{2}{c}{0.97 (0.01)} & \multicolumn{2}{c}{0.96 (0.01)} \\
      SGLD (2000) & \multicolumn{2}{c}{0.83 (0.01)} & \multicolumn{2}{c}{0.91 (0.01)} \\
      SGLD (4000) & \multicolumn{2}{c}{0.95 (0.01)} & \multicolumn{2}{c}{0.96 (0.01)} \\                  
      \hline
      & $k = 10$ & $k = 20$ & $k=10$ & $k=20$ \\ 
      \hline 
      CMC & 0.96 (0.01) & 0.95 (0.02) & 0.95 (0.02) & 0.94 (0.03) \\ 
      SDP & 0.96 (0.02) & 0.95 (0.02) & 0.90 (0.06) & 0.90 (0.06) \\ 
      WASP & 0.97 (0.01) & 0.96 (0.02) & 0.95 (0.02) & 0.94 (0.03) \\ 
      \hline
    \end{tabular}
  }
\end{table}

\subsection{Simulated data analysis: Probablistic parafac model}
\label{p-parafac}

The derivation of modified full conditional densities of unknown parameters involves one key modification in the Gibbs sampling algorithm of \citet{DunXin09}. Assuming $z_{ji}$ is given, the contribution of $i$th observation in $j$th subset to the likelihood after stochastic approximation is
\begin{align*}
  L_{ji}^{\gamma}((\psib_h^{(1)})_{h=1}^{l^*}, \ldots, (\psib_h^{(q)})_{h=1}^{l^*}, \ldots, (\psib_h^{(p)})_{h=1}^{l^*}) = \left( \prod_{h=1}^{l^*}\nu_h  \prod_{q=1}^p \prod_{l=1}^{d_q} \psi_{h l}^{(q)^{1(x_{jiq} = l, z_{ji}=h)}} \right)^{\gamma},
\end{align*}
where $1(x_{jiq} = l, z_{ji}=h)$ is 1 if both conditions are true and is 0 otherwise and $l^*$ is the maximum number of atoms in the
stick breaking representation for the distribution of $z_{ji}$. The conditional posterior density of $\psib_h^{(q)}$ after stochastic approximation is proportional to
\begin{align*}
  \prod_{l=1}^{d_q} \psi_{h l}^{(q)^{a_{jl} - 1}}  \prod_{i=1}^m \prod_{l=1}^{d_q} \psi_{h l}^{(q)^{\gamma 1(x_{jiq} = l, z_{ji}=h)}} = \prod_{l=1}^{d_q} \psi_{h l}^{(q)^{a_{jl} + \gamma \sum_{i=1}^m 1(x_{jiq} = l, z_{ji}=h) - 1}},
\end{align*}
which implies that
\begin{align}
  \psib_{jh}^{(q)} \mid \text{rest} \sim \text{Dirichlet} \left(a_{q1} + \gamma 1(x_{jiq} = 1, z_{ji}=h), \ldots, a_{q d_q} + \gamma 1(x_{jiq} = d_q, z_{ji}=h)\right) \label{postpsi}
\end{align}
for $q = 1, \ldots, p$ and $h=1, \ldots, l^*$. The conditional densities of remaining parameters follow from Section 3.1 of \citet{DunXin09}:
\begin{align}
  V_{jh} \mid \text{rest}  &\sim \text{Beta}(1 + \gamma \sum_{i=1}^m 1(z_{ji} = h), \alpha + \gamma \sum_{i=1}^m 1(z_{ji} > h)), \label{postv}\\
  \alpha_j \mid \text{rest} &\sim \text{Gamma}(a_{\alpha} + l^*, b_{\alpha} - \sum_{h=1}^{l^*} \log (1 - V_{jh})). \label{postalpha}
\end{align}
Finally, we update the posterior density of responsibility of every observation as
\begin{align}
  z_{ji} \mid \text{rest} \sim \sum_{h=1}^{l^*} p_{jh} \delta_h, \quad p_{jh} = \frac{\nu_{jh} \prod_{q=1}^{p}  \psi_{h x_{jiq}}^{(q)}}{\sum_{h=1}^{l^*} \nu_{jh} \prod_{q=1}^{p}  \psi_{h x_{jiq}    }^{(q)}}, \quad (h=1 ,\ldots, l^*; \; i=1, \ldots, m), \label{postz}
\end{align}
where $\nu_{jh} = V_{jh} \prod_{l < h} (1 - V_{jl})$. The conditional posterior densities without stochastic approximation are obtained by substituting $\gamma=1$ and $m = n$ in the full conditionals \eqref{postpsi} -- \eqref{postalpha}.

All full conditionals are analytically tractable in terms of standard distributions. The Gibbs sampler iterates between the following four steps:
\begin{enumerate}
\item Sample $\psib^{(q)}_{jh}$ from Dirichlet distribution in \eqref{postpsi} for $q=1, \ldots, p$ and $h=1, \ldots, l^*$.
\item Sample $V_{jh}$ from Beta distribution in \eqref{postv} for $h = 1, \ldots, l^*$.
\item Sample $\alpha_j$ from Gamma distribution in \eqref{postalpha}.
\item Sample $z_{ji}$ from categorical distribution in \eqref{postz} for $i = 1, \ldots, m$.  
\end{enumerate}
We fix $a_{ql} = 1 / d_q$ for $q = 1, \ldots, p$ and $l = 1, \ldots, d_q$.

\subsection{Real data analysis: MovieLens data}

Table \ref{tbl:acc_ml_fixef} describes the accuracies of ADVI, CMC, SA, SDP, SGLD, and WASP in approximating the full data marginal posterior distributions of fixed effects in the MovieLens data analysis.

\begin{table}[t]
  \caption{Accuracies of the approximate posteriors of the fixed effects in the linear mixed effects model for MovieLens data. The accuracies are averaged over 10 replications. Monte Carlo errors are in parenthesis. ADVI, automatic differentiation variational inference; SA, streamlined algorithm; SGLD, stochastic gradient Langevin dynamics with batch size in parenthesis; CMC, consensus Monte Carlo; SDP, semiparametric density product; WASP, Wasserstein posterior}
  \label{tbl:acc_ml_fixef}
  \centering
  {\tiny
  \begin{tabular}{rcccccc}
    \hline
    & $\beta_{\text{Action}}$ & $\beta_{\text{Children $-$ Action}}$ & $\beta_{\text{Comedy $-$ Action}}$ & $\beta_{\text{Drama $-$ Action}}$ & $\beta_{\text{Popularity}}$ & $\beta_{\text{Previous}}$ \\ 
    \hline    
    ADVI & 0.16 (0.24) & 0.21 (0.32) & 0.31 (0.31) & 0.27 (0.31) & 0.53 (0.20) & 0.39 (0.27) \\
    SA & 0.00 (0.00) & 0.34 (0.05) & 0.64 (0.05) & 0.15 (0.02) & 0.01 (0.00) & 0.00 (0.00) \\  
    SGLD (2000) & 0.79 (0.02) & 0.72 (0.02) & 0.78 (0.02) & 0.82 (0.01) & 0.86 (0.02) & 0.80 (0.02) \\ 
    SGLD (4000) & 0.78 (0.02) & 0.72 (0.02) & 0.76 (0.03) & 0.80 (0.03) & 0.84 (0.02) & 0.77 (0.02) \\ 
    CMC & 0.95 (0.02) & 0.93 (0.02) & 0.92 (0.02) & 0.95 (0.02) & 0.94 (0.02) & 0.95 (0.03) \\ 
    SDP & 0.93 (0.03) & 0.92 (0.03) & 0.92 (0.04) & 0.93 (0.04) & 0.92 (0.04) & 0.94 (0.02) \\ 
    WASP & 0.96 (0.01) & 0.95 (0.01) & 0.95 (0.02) & 0.96 (0.01) & 0.96 (0.01) & 0.96 (0.01) \\ 
    \hline
  \end{tabular}
  }%
\end{table}

\begin{figure}[h]
  \centering
  \subfloat[]{
    \includegraphics[scale=0.2]{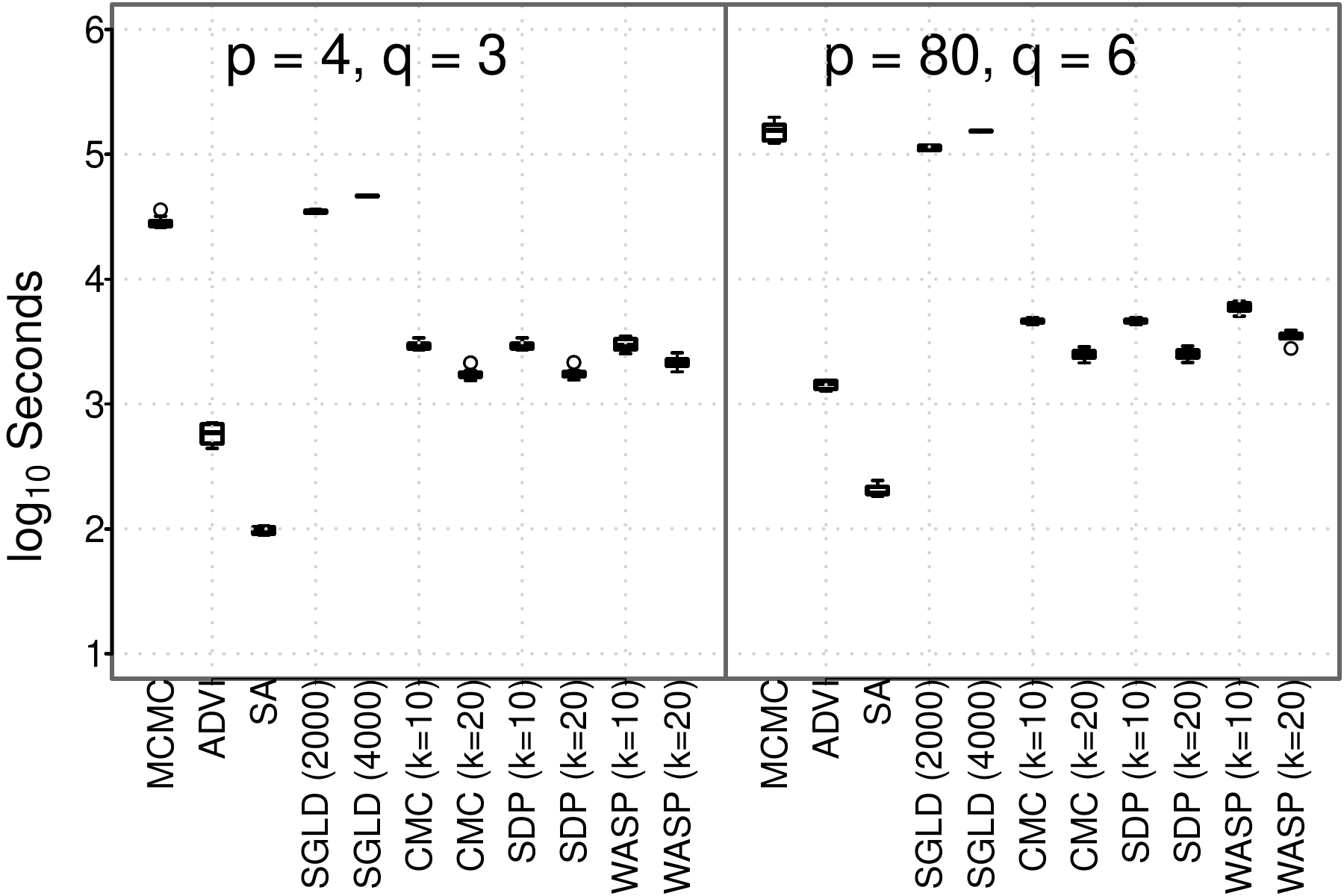}
    \label{fig:track}}
  \subfloat[]{
    \includegraphics[scale=0.2]{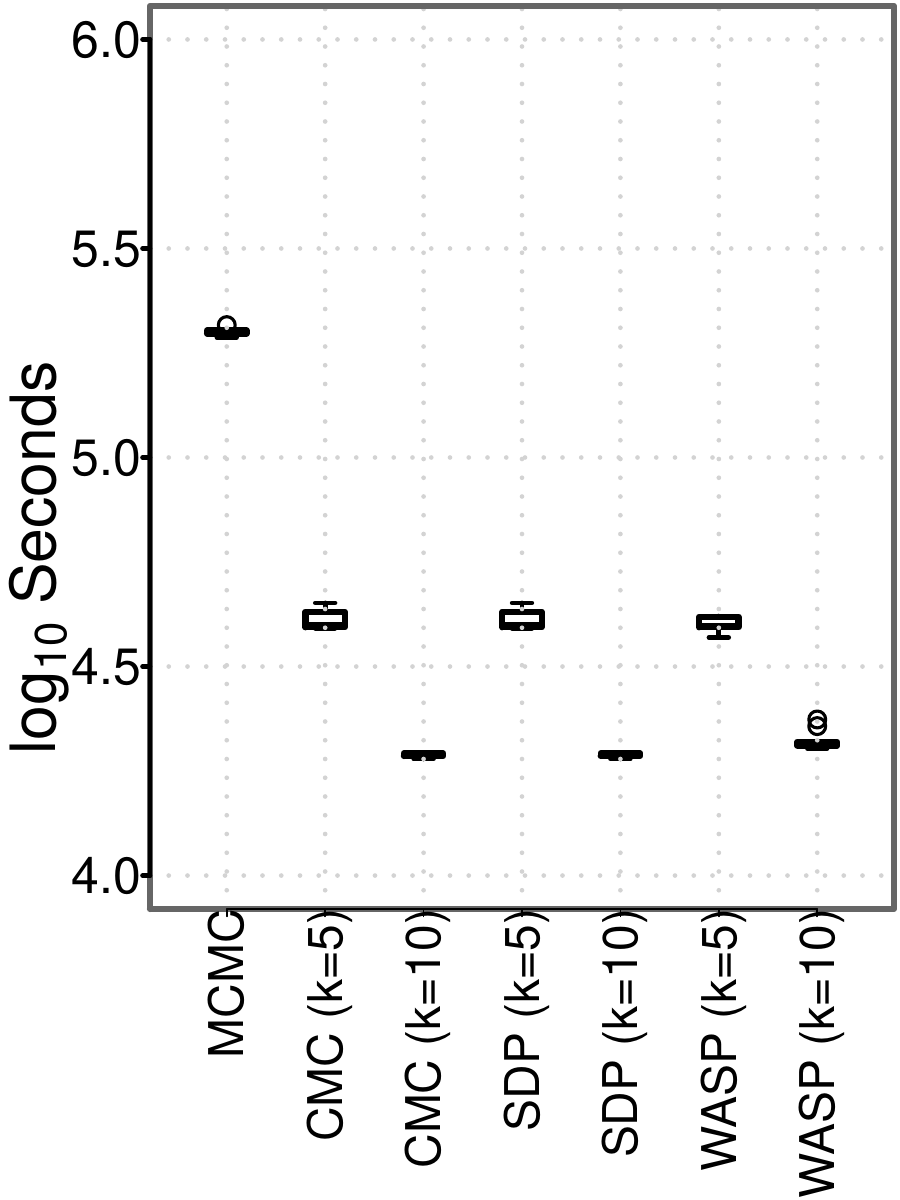}
    \label{fig:iters}}
  \subfloat[]{
    \includegraphics[scale=0.2]{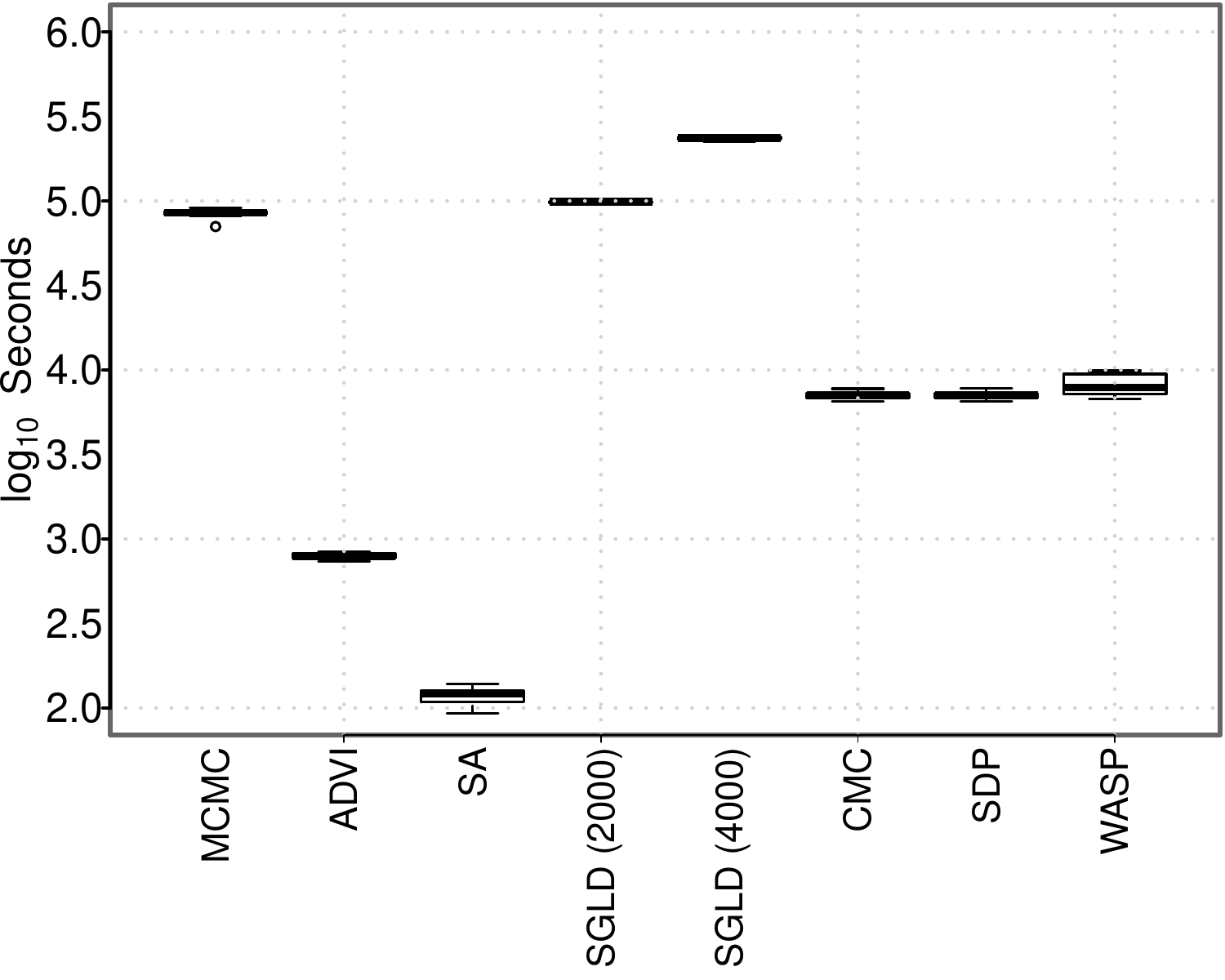}
    \label{fig:time}}
  \caption{Computation time for the methods used in (a) linear mixed effects model simulation, (b) probabilistic parafac model simulation, and (c) MovieLens data analysis. ADVI, automatic differentiation variational inference; SA, streamlined algorithm; SGLD, stochastic gradient Langevin dynamics with batch size in parenthesis; CMC, consensus Monte Carlo; SDP, semiparametric density product; WASP, Wasserstein posterior.}    
  \label{fig:dem-em}
\end{figure}

\end{document}